\documentclass{aa}
\usepackage{graphicx}
\usepackage{adjustbox}
\usepackage{natbib}
\usepackage{enumerate}
\usepackage{enumitem}
\usepackage{float}
\usepackage{siunitx}
\usepackage{txfonts}
\begin{document}
%

 \title{On attempting to automate the identification of mixed dipole modes for subgiant stars}

   \author{T.~Appourchaux}

   \institute{Universit\'e Paris-Saclay, Institut d'Astrophysique Spatiale, UMR 8617, CNRS, B\^atiment 121, 91405 Orsay Cedex, France
 }

   \date{Submitted July 3rd 2020.  Accepted August 24th 2020.}

 
  \abstract
  {The existence of mixed modes in stars is a marker of stellar evolution.  Their detection serves for a better determination of stellar age.}
   {The goal of this paper is to identify the dipole modes in an automatic manner without human intervention.}
   {I use the power spectra obtained by the {{\it Kepler}} mission for the application of the method.  I compute asymptotic dipole mode frequencies as a function of coupling factor and dipole period spacing, and other parameters.   { For each star, I collapse the power in an echelle diagramme aligned onto the monopole and dipole mixed modes.  The power at the null frequency is used as a figure of merit.  Using a genetic algorithm, I then optimise the figure of merit by adjusting the location of the dipole frequencies in the power spectrum}.  Using published frequencies, I compare the asymptotic dipole mode frequencies with published frequencies.  I also used published frequencies for deriving coupling factor and dipole period spacing using a non-linear least squares fit.  I use Monte-Carlo simulations of the non-linear least square fit for deriving error bars for each parameters.}
   {From the 44 subgiants studied, the automatic identification allows to retrieve within 3 $\mu$Hz at least 80\% of the modes for 32 stars, and within 6 $\mu$Hz at least 90\% of the modes for 37 stars.  The optimised and fitted gravity-mode period spacing and coupling factor agree with previous measurements.  Random errors for the mixed-mode parameters deduced from Monte-Carlo simulation are about 30-50 times smaller than previously determined errors, which are in fact systematic errors.}
   {The period spacing and coupling factors of mixed modes in subgiants are confirmed.  The current automated procedure will need to be improved using a more accurate asymptotic model and/or proper statistical tests}
\titlerunning{On attempting to automate the identification of mixed dipole modes for subgiant stars} 
   \maketitle
%
\section{Introduction}
The internal structure of stars has been derived with great details with the advent of the space missions CoRoT and {\it Kepler} \citep{Michel2008, Chaplin2010}.  The detection of pressure modes and mixed modes in red giants provided a wealth of information about the internal structure of these stars \citep{Mosser2014, Bedding2011a} and their internal dynamics \citep{Mosser2012}.  { Subgiant stars bridges the gap between the evolved red giants and the Main Sequence (MS) stars.  Subgiant stars being closer to MS stars  can help to infer informations about the evolution of their interior and dynamics}.  For subgiant stars, \citet{Deheuvels2014} also measured their core rotation using these mixed modes.  The properties of these mixed modes are such that they are easily detectable because they propagate like pressure modes until the surface of the star and are excited by turbulent convection just like the pressure modes { \citep{PG77, GH99, Dupret2009}}t; but these modes also propagate like gravity modes in the stellar core \citet{Unno89}.  Given these properties, these modes have also been searched in the Sun with no positive results \citep{Appourchaux2010}.  

The detection of these mixed modes in subgiants stars provides also an important marker of the age of these stars.  The subgiants evolve away from the MS by burning hydrogen in shells after exhaustion on the hydrogen in the core.  The time it take for these stars from the core exhaustion to the Helium burning phase is called the { First Giant Branch} \citep{VW93}.  The time spent on this branch last about 15\% to 30\% of the time spent on the MS \citep{VW93}.  In addition,  due to the presence of an Helium core, there is large increase in the Brunt-Va{\"{\i}}sala frequency which makes possible the existence of a propagation zone for gravity modes { \citep{Dziembowski2001, JCD2004a}.}  It leads to the existence of the so-called 
mixed modes that have a pressure mode character in the outer stellar regions, and a gravity mode character in the stellar core { \citep{Dziembowski2001, JCD2004a}.}  In addition, it is clear that since the change in the Brunt-Va{\"{\i}}sala frequency is so quick, the frequencies of the mixed modes change even faster because of their sensitivities to that frequency \citep{TA2003}.  Therefore the detection of mixed modes in subgiants is a powerful tool for getting a precise stellar age \citep{Lebreton2014_I}, even for getting a precise estimate of the mass \citep{Deheuvels2011}.

A  major challenge of the ESA PLATO mission is to derive stellar ages to better than 10\% \citep{Rauer2014} for the sample\footnote{known as the P1 sample} of stars brighter than $m_{\rm V}=11$ { magnitude} and with an instrumental noise lower than 50 ppm.hr$^{-1/2}$ provided with the 24 cameras \citep{Marchiori2019, Moreno2019}.  The derivation of the stellar age is obtained by comparing theoretical mode frequencies to observed mode frequencies \citep{Lebreton2014}.   The number of stars in the P1 sample for which an asteroseismic age with a precision can be derived to better than 10\% is larger than 8000.  The measurement of the observed mode frequencies can be achieved using the following steps: detecting the stellar oscillation mode envelope, inference of the large separation of the modes (related to the diameter of the star), tagging of the degree of the modes and mode frequency fitting using either Maximum Likelihood Estimators or Markov Chain Monte Carlo estimators.  All these steps can be automated for fitting a large number of stars \citep[See][and references therein]{BMTA2009, Hekker2010a}.  These automated method can be easily applied to MS stars having no mixed modes \citep{Fletcher2011}. 

It was already identified by \citet{TA2003} that the identification of mixed modes for a large amount of in subgiants could be a challenge.  For red giants having mixed modes, \citet{Vrard2016} devised a technique for automatically deriving the gravity mode period spacing allowing to fit the mixed modes.  Their technique is applied to stars for which the large separation is smaller than 18 $\mu$Hz but they use guess for the gravity mode period spacing derived from \citet{Mosser2014}; their technique is then not entirely automated.  { For subgiant stars, \citet{Bedding2012} devised a {\it replicated} echelle diagrams allowing to visually identify dipole mixed modes.} For subgiants stars with a large separation typically higher than 30 $\mu$Hz \citep{Mosser2014}, an automated detection procedure has been recently devised by \citet{Corsaro2020} for power spectra having the $l=1$ mode ridge slightly distorted.   For more complicated case, the mixed mode frequencies of the subgiant { stars} of \citet{Appourchaux2012} were all identified by hand.  {  In the case of PLATO, the manual identification becomes impractical for the mere 3000 subgiants of the P1 sample of PLATO, therefore requiring an automated procedure for that aim.}

The goal of this paper is to present a way to automatically provide guess for the frequencies of dipole mixed modes in subgiants. The first section recalls how the frequencies of these modes can be derived from an asymptotic expression based upon the work of \citet{Shibahashi1979} and of \citet{Mosser2015}.  The second section explains how one can build a diagnostic of the location of the mixed mode frequencies in a power spectrum derived from the Fourier transform of a stellar light curve (or from stellar radial velocity).  The third section details how the diagnostic can be optimised for finding the main characteristics of the parameters describing the location of the frequencies of the dipole mixed modes.  The next three sections provides the results for about 44 subgiant and a few early red giants using the optimisation procedure and the fitting of available frequencies.  I finally discuss the scientific implications and conclude.

\section{Asymptotic mixed dipole mode frequency}
The approach to the derivation of the asymptotic mixed mode frequency is based upon the work of \citep{Mosser2015}.  The methodology is repeated here for completeness.  The frequencies of the mixed dipole modes is asymptotic because the frequencies of the pressure modes and the gravity modes are assumed to follow the asymptotic description of \citet{TASSOUL80}.  A mixed mode has by definition a dual character of being both a pressure mode and a gravity mode.  The main equation describing the mixed mode character is related to the following continuity equation provided by \citet{Shibahashi1979}:
\begin{equation}
\tan \theta_p=q \tan \theta_g
\end{equation}
where $\theta_p$ and $\theta_g$ are the phase function of a dipole mixed mode frequency $\nu_{pg}$ with respect to the p modes and g modes frequency, and $q$ is the coupling factor.  
\\
\\
{\noindent \bf Pressure mode phase function:} The phase function $\theta_p$ is given by:
\begin{equation}
\theta_p=\pi \frac{(\nu_{pg}-\nu_{n_p,1})}{\Delta\nu(n_p)}
\end{equation} 
where $\nu_{n_p,1}$ is the frequency of the closest dipole pressure mode and $\Delta\nu(n_p)$ is the large separation at order $n_p$.    $\nu_{n_p,1}$ is given asymptotically by:
\begin{equation}
\nu_{n_p,1}=\nu_{n_p,0}+\left(\frac{1}{2}-d_{01}\right) \Delta\nu
\end{equation}
where $\nu_{n_p,0}$ is the radial mode frequency and $d_{01}$ is the relative small separation between $l=0$ and $l=1$.
In  \citet{Mosser2015}, they assume that the frequencies of the radial modes $\nu_{n_p,0}$ are given by:
\begin{equation}
\nu_{n_p,0}=\left(n_p+\epsilon_p+\frac{\alpha}{2}(n_p-n_{\rm max})^2\right) \Delta\nu
\end{equation}
where $n_p$ is the order of the mode, $\epsilon_p$ is the p-mode phase offset and $\alpha$ is the parameter describing the parabolic
departure of the frequency the asymptotic behaviour, $n_{\rm max}$ is the radial order at $\nu_{\rm max}$ and $\Delta\nu$ is the large 
separation.   As a consequence, the large separation varies from order to order as follows:
\begin{equation}
\Delta\nu(n_p)=(1+\alpha (n_p-n_{\rm max})) \Delta\nu
\end{equation} 
Here we note that the formulation of the $l=1$ mode frequencies with respect to the $l=0$ mode frequencies allows to compensate directly for the
surface effect.  Since this effect is of similar order for both, it is partly masked or even fully compensated by the parabolic term in $\alpha$.

{\noindent \bf Gravity mode phase function:} The phase function $\theta_g$ is given by:
\begin{equation}
\theta_g=\pi \frac{1}{P_1} \left(\frac{1}{\nu_{pg}}-\frac{1}{\nu_{n_g,1}}\right)
\end{equation}
where $\nu_{n_g,1}$ is the frequency of the closest dipole gravity mode, $n_g$ is the order of the mode and $P_1$
is the period spacing of the dipole gravity mode.  The frequency of the the dipole gravity mode is given by:
\begin{equation}
\frac{1}{\nu_{n_g,1}}=P_{n_g,1}=(n_g+\epsilon_g) P_1
\end{equation}
where $n_g$ is the g mode order (here it is positive) and $\epsilon_g$ is the g-mode phase offset.  $P_1$ is related to $P_0$ ($P_1=P_0/\sqrt{2}$).
\\
\\
{\noindent \bf Asymptotic mixed mode frequencies:} The frequencies $\nu_{pg}$ of the dipole mixed modes are then implicitly provided by solving the
following equation:
\begin{equation}
\tan \theta_p(\nu_{pg}, \Delta\nu,\alpha,\epsilon_p,n_p)-q \tan \theta_p(\nu_{pg}, P_1,\epsilon_g,n_g)=0
\end{equation}
When there is no coupling $q=0$, there is no mixed modes $\nu_{pg}=\nu_{n_p,1}$.  In practice, the solving of Eq.~(8) is done as follows:
\begin{enumerate}
\item Set $\Delta\nu$, $\epsilon_p$ and $\alpha$ for the $l=0$ mode frequencies (Eq.~4) 
\item Set $d_{01}$ for getting the $l=1$ mode frequencies (Eq.~3)
\item Set $P_1$ and $\epsilon_g$ (Eq.~7)
\item Set coupling factor $q$
\item Compute phases using Eq.~(2) and Eq.~(6).  Set mixed mode frequency $\nu_{n_g,1}$ as a running frequency $\nu$ with a resolution of 1 nHz.
\item Compute difference of phase tangent as: $\delta(\nu)=\tan \theta_p-q \tan \theta_g$  (Eq.~8)
\item Compute sign of $\delta$.  Since the derivative of $\delta(\nu)$ wrt to $\nu$ is always positive, $\delta(\nu)$ is an increasing function of $\nu$, then the only transition through zero is when $\delta$ changes sign from negative to positive.  When it occurs, it means that it crossed the zero line, i.e. we have a solution for a mixed-mode frequency.
\item The final detection of the location of mixed-mode frequency is done by computing the first difference of the sign of $\delta$. When there is a crossing from negative to positive
the first difference is then 2 (+1-(-1)).
\end{enumerate}
This pseudo code can easily be implemented in any language.



\section{Collapsing of mixed-mode peak power}
{ The identification of dipole mixed modes in subgiant stars rely on a visual assessment of the location of the mixed modes.  As mentioned earlier, the use of {\it replicated} echelle diagram \citep{Bedding2012} or just plain experience of a regular or irregular echelle diagram is sufficient for most application, i.e for tens of stars.  For the PLATO mission, I had to devise an automated procedure that would make the identification more immune to human intervention.  The novelty of the method developed in this section lies in the use of all the information available in the power spectrum for the monopole and dipole modes coupled with the asymptotic description introduced in the previous section.  The information used allows to derive a figure of merit that is related to the location of the dipole mode frequencies in the power spectrum.}

\begin{figure}[!tbp]
\centerline{\includegraphics[width=6 cm,angle=90]{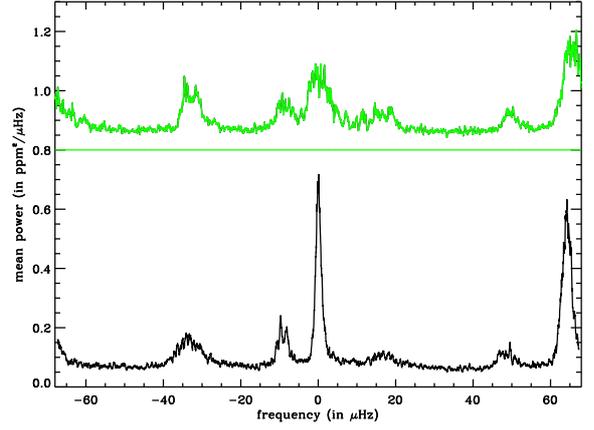}}
\caption{Superposed power for the LOI data: power averaged over 15 modes centered on the $l$=0 modes: (top, green) for a fixed large separation of 136 $\mu$Hz; (bottom, black) for an optimised large separation of 134.7 $\mu$Hz linearly dependent upon frequency.  The original power spectrum, smoothed to 10 bins, is taken from one year of data of the Luminosity Oscillations Imager seeing the Sun as a star \citep{TABA97}.}
\label{superposition}
\end{figure}

The idea behind the derivation of the guess frequencies for the dipole mixed modes is based upon the visualisation of the \'echelle diagram \citep{GG81} applied to a power spectrum of stellar time series. If one co-adds the power along each segment line of the \'echelle, one obtains the so-called {\it collapsed power}.  An example of collapsed power is shown on Fig.~\ref{superposition} where one can see two main doublets: one doublet for the $l=0-2$ modes located at 0 $\mu$Hz; one doublet for the $l=1-3$ modes located at +63 $\mu$Hz.  The other patches of power located at -35 $\mu$Hz and +18 $\mu$Hz seen in Fig.~\ref{superposition} are due to the $l=4$ modes and the $l=5$ modes, respectively.  Fig.~\ref{superposition} is computed using data from the Luminosity Oscillations Imager (LOI) seeing the Sun as a star \citep{TABA97}.  The additional sensitivity to the higher degree modes are due to the LOI pixels resolving the surface of the Sun (ibidem).

\begin{figure}[!tbp]
\includegraphics[width=6 cm,angle=90]{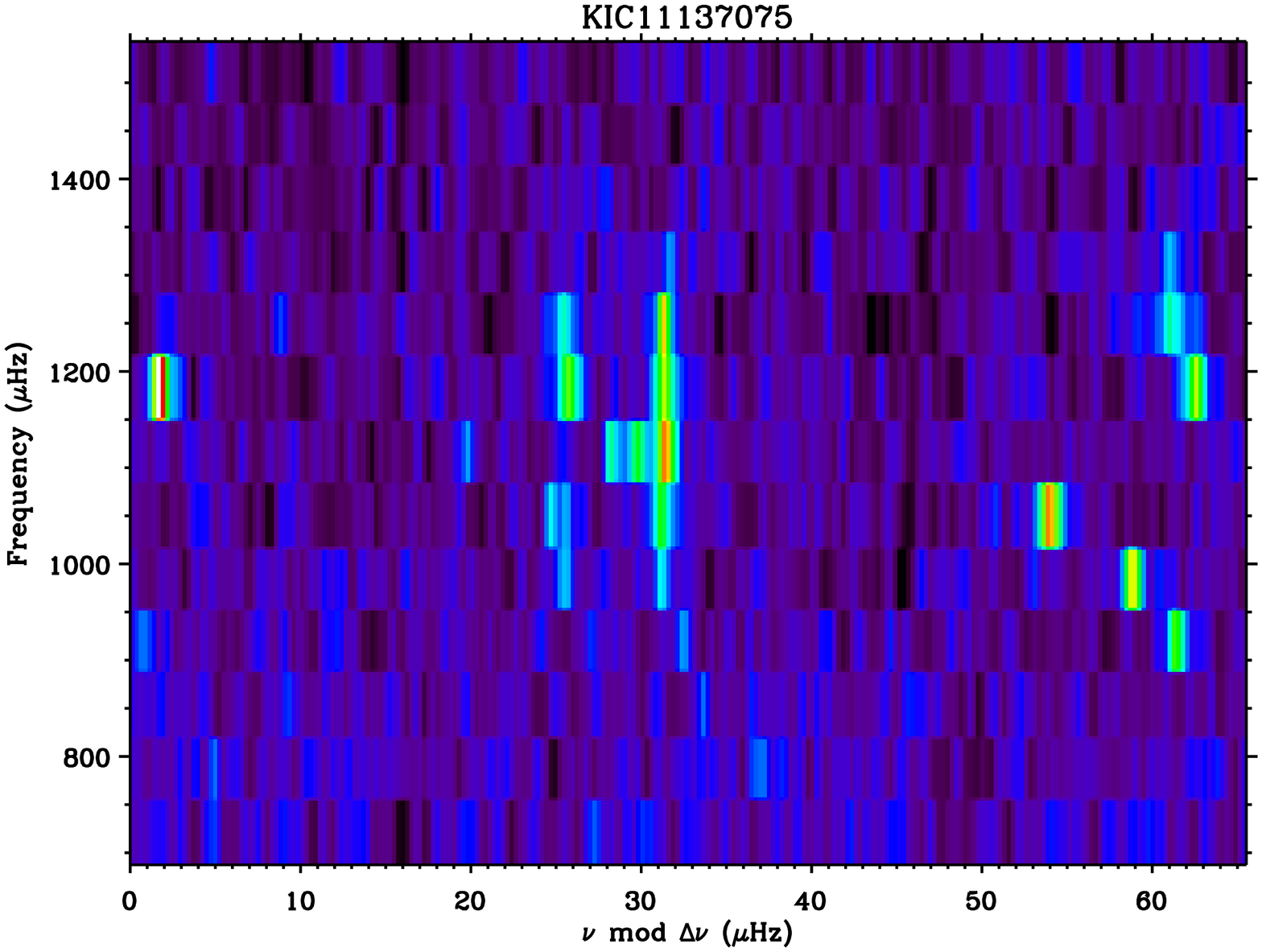}
\includegraphics[width=6 cm,angle=90]{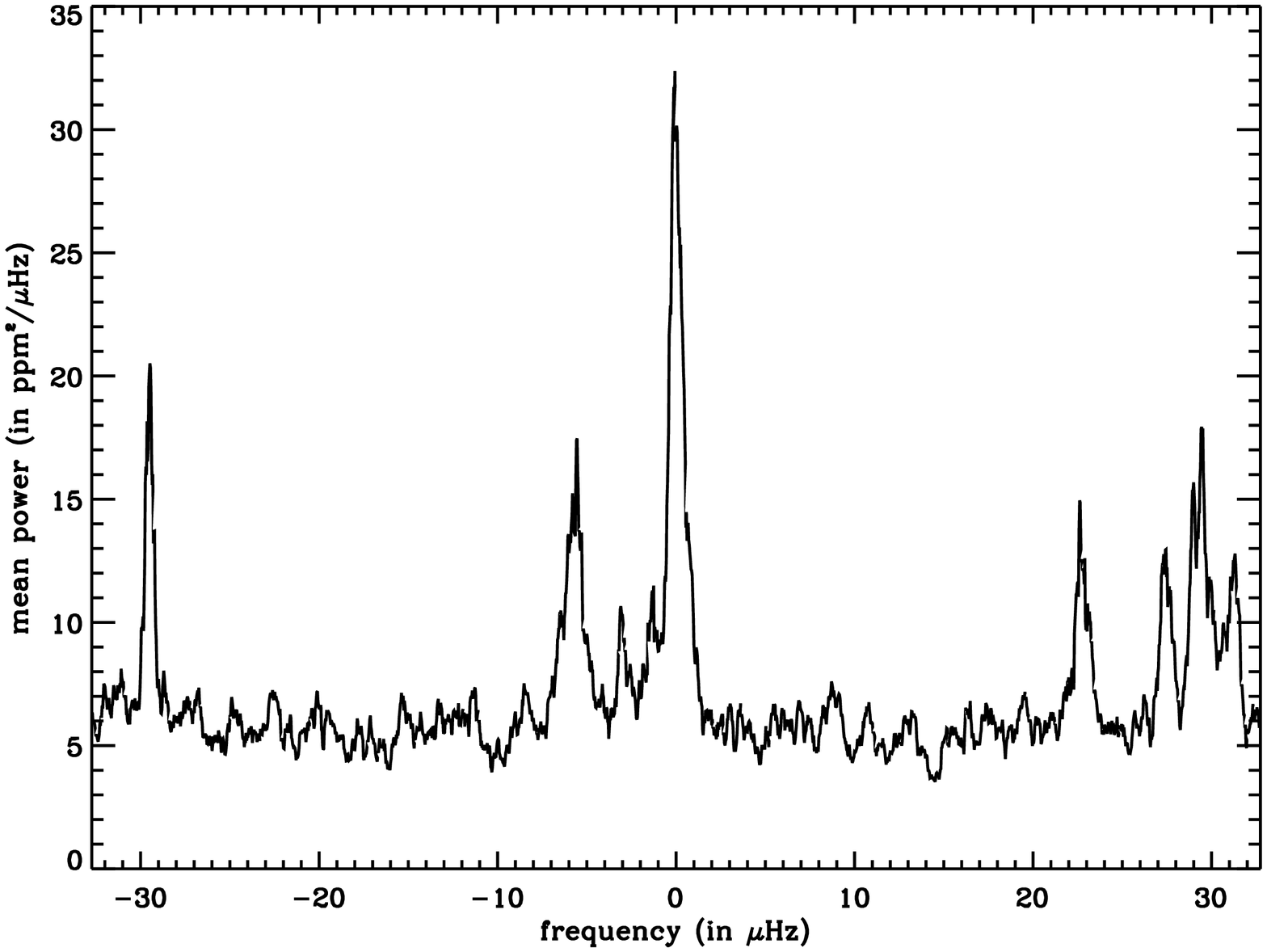}
\caption{(Top) Echelle diagram for KIC11137075 with a large separation of 65 $\mu$Hz.  (Bottom) Superposed power for KIC11137075 sta: power averaged over 8 modes centered on the $l$=0 modes.  The original power spectrum, smoothed to 10 bins, is taken from 6 months of {\it Kepler} data.}.
\label{superposition_kic}
\end{figure}

Fig.~\ref{superposition_kic} shows the collapsed for  a KIC11137075, a star with $l=1$ mixed modes already analysed by \citet{Tian2015}.  It is obvious that the power of the $l=1$ mixed modes is not concentrated at all compared to the $l=0-2$ mode pairs.  The idea is then to compute $l=1$ mixed mode frequencies resulting from solving Eq.~(8) and to collapse the power around these frequencies.  The power is then extracted around an $l=1$ mixed frequency using a window of the size of the large separation ($\pm \Delta \nu/2$), then the power is summed up over all $l=1$ mixed modes found from solving Eq.~(8).  Here the total power is {\it not} averaged over the number of mixed modes because the division would prevent to have a large total power when more mixed modes correspond to real peak of power.
In addition, since I am interested in the location of the dipole mode frequencies, I compensate for the intrinsic fall out of mode power away from $\nu_{\rm max}$ by dividing the original power spectrum by the inverse of a Gaussian mode envelope whose full width at half maximum is 2/3 of the full mode frequency range defined as  [$\nu_{\rm low}$,$\nu_{\rm high}$].  This compensation permits to give a lower weight to the modes close to $\nu_{\rm max}$, and a higher weight to the modes on the edge of the p-mode range.

\begin{table*}[!t]
\caption{Table of optimised parameters.  First column: Kepler Identification Catalog (KIC) number; second column: duration of observation; third column: percentage of mode within $\pm$3 microHz of the asymptotic dipole frequency; fourth column: source of available fitted frequencies; fifth column: frequency of maximum poser; sixth column: large separation; seventh column: parabolic coefficient of the $l=0$ mode frequencies; eighth column: small separation between the $l=0$ and the $l=1$ mode frequencies; ninth column: g-mode phase; tenth column: p-mode phase; eleventh column: coupling factor; twelfth column: dipole period spacing.}      
\label{Table_data} 
\centering          
\begin{tabular}{cccccccccccc}  
\hline
KIC&Duration&\%&Fitted frequencies&$\nu_{\rm max}$ &$\Delta\nu$  &$\alpha$&$d_{01}$&$\epsilon_g$&$\epsilon_p$&$q$&$P_1$\\
&(in months)&&&(in $\mu$Hz)&(in $\mu$Hz) &&&&&&(in s)\\
\hline
2991448 &3& 90 & \citet{Li2020} & 1080 & 61.5 & 0.00141 & +0.0548 & 0.39 & 1.45 & 0.07 & 516.69\\
4072740 &12& 100 &  & 260 & 18.4 & 0.00211 & +0.0109 & 0.32 & 1.39 & 0.20 & 88.64\\
4346201 &3& 100 & \citet{Li2020} & 950 & 55.5 & 0.00800 & +0.0580 & 0.24 & 1.31 & 0.08 & 707.61\\
5607242 &6& 88 & \citet{Appourchaux2012a} & 610 & 40.6 & 0.00138 & +0.0011 & 0.33 & 1.36 & 0.56 & 180.80\\
5689820 &9& 80 & \citet{Deheuvels2014} & 679 & 41.1 & 0.00111 & -0.0071 & 0.37 & 1.53 & 0.27 & 140.99\\
5723165 &12& 100 &  & 560 & 34.4 & 0.00182 & -0.0197 & 0.31 & 1.45 & 0.30 & 135.23\\
5955122 &6& 69 & \citet{Appourchaux2012a} & 826 & 49.6 & 0.00425 & +0.0202 & 0.45 & 1.19 & 0.14 & 416.22\\
6313425 &12& 100 &  & 600 & 37.0 & 0.00107 & -0.0608 & 0.69 & 1.41 & 0.76 & 208.73\\
6370489 &3& 80 & \citet{Li2020} & 889 & 51.8 & 0.00427 & +0.0379 & 0.24 & 1.17 & 0.11 & 389.34\\
6442183 &9& 83 & \citet{Tian2015} & 1125 & 64.9 & 0.00301 & +0.0659 & 0.56 & 1.43 & 0.08 & 481.28\\
6531928 &12& 100 &  & 450 & 31.3 & 0.00300 & +0.0033 & 0.21 & 1.45 & 0.31 & 114.88\\
6693861 &12& 50 & \citet{Li2020} & 775 & 47.3 & 0.00201 & -0.0621 & 0.22 & 1.39 & 0.28 & 231.42\\
6802438 &3& 100 &  & 567 & 34.5 & 0.00224 & -0.0240 & 0.23 & 1.44 & 0.38 & 137.97\\
6863041 &3& 85 &  & 600 & 42.5 & 0.00257 & -0.0451 & 0.33 & 1.41 & 0.19 & 280.00\\
7174707 &12& 82 & \citet{Li2020} & 825 & 47.2 & 0.00277 & -0.0580 & 0.48 & 1.50 & 0.27 & 182.05\\
7216846 &3& 100 &  & 500 & 28.8 & 0.00223 & -0.0306 & 0.67 & 1.36 & 0.23 & 120.67\\
7341231 &9& 100 & \citet{Appourchaux2012a} & 404 & 29.0 & 0.00119 & -0.0039 & 0.26 & 1.26 & 0.37 & 111.43\\
7747078 &9& 79 & \citet{Appourchaux2012a} & 977 & 53.7 & 0.00416 & +0.0305 & 0.54 & 1.33 & 0.10 & 319.31\\
7799349 &9& 100 & \citet{Deheuvels2014} & 560 & 33.2 & 0.00000 & -0.0244 & 0.23 & 1.49 & 0.23 & 118.58\\
7976303 &12& 80 & \citet{Appourchaux2012a} & 826 & 51.3 & 0.00410 & +0.0386 & 0.66 & 1.18 & 0.13 & 323.50\\
8026226 &6& 81 & \citet{Appourchaux2012a} & 520 & 34.4 & 0.00109 & +0.0880 & 0.51 & 0.97 & 0.43 & 350.56\\
8524425 &9& 77 & \citet{Appourchaux2012a} & 1078 & 59.7 & 0.00074 & +0.0572 & 0.42 & 1.39 & 0.10 & 325.61\\
8702606 &9& 100 & \citet{Deheuvels2014} & 626 & 39.7 & 0.00269 & +0.0143 & 0.24 & 1.35 & 0.43 & 174.11\\
8738809 &12& 80 & \citet{Li2020} & 850 & 49.7 & 0.00399 & +0.0229 & 0.47 & 1.10 & 0.06 & 396.94\\
8751420 &12& 100 & \citet{Deheuvels2014} & 570 & 34.6 & 0.00194 & -0.0192 & 0.33 & 1.45 & 0.31 & 132.30\\
9512063 &3& 73 & \citet{Li2020} & 850 & 49.5 & 0.00300 & -0.0197 & 0.47 & 1.35 & 0.26 & 271.89\\
9574283 &9& 100 & \citet{Deheuvels2014} & 448 & 29.7 & 0.00340 & -0.0124 & 0.37 & 1.47 & 0.27 & 114.56\\
10018963 &9& 93 & \citet{Appourchaux2012a} & 988 & 55.2 & 0.00393 & +0.0549 & 0.23 & 1.15 & 0.10 & 735.64\\
10147635 &12& 53 & \citet{Li2020} & 605 & 37.0 & 0.00146 & -0.0424 & 0.37 & 1.32 & 0.42 & 361.23\\
10273246 &12& 67 & \citet{Campante2011} & 800 & 48.0 & 0.00310 & -0.0077 & 0.42 & 1.38 & 0.18 & 487.02\\
10593351 &12& 100 & \citet{Li2020} & 500 & 31.3 & 0.00800 & +0.0603 & 0.52 & 1.25 & 0.48 & 360.89\\
10873176 &3& 80 & \citet{Li2020} & 790 & 49.0 & 0.00686 & +0.0049 & 0.51 & 1.12 & 0.33 & 300.39\\
10920273 &7.5& 86 & \citet{Campante2011} & 997 & 57.3 & 0.00396 & +0.0402 & 0.64 & 1.40 & 0.18 & 368.15\\
10972873 &12& 92 & \citet{Li2020} & 1000 & 58.3 & 0.00300 & +0.0495 & 0.70 & 1.40 & 0.12 & 300.09\\
11026764 &48& 73 & \citet{Appourchaux2012a} & 885 & 50.4 & 0.00558 & +0.0174 & 0.20 & 1.34 & 0.15 & 229.42\\
11137075 &12& 100 & \citet{Tian2015} & 1144 & 65.5 & 0.00228 & +0.0557 & 0.63 & 1.47 & 0.11 & 279.49\\
11193681 &9& 75 & \citet{Appourchaux2012a} & 752 & 42.9 & 0.00163 & +0.0018 & 0.63 & 1.33 & 0.14 & 308.74\\
11234888 &7.5& 80 &  & 675 & 41.3 & 0.00399 & -0.0697 & 0.64 & 1.25 & 0.27 & 273.04\\
11395018 &9& 58 & \citet{Appourchaux2012a} & 840 & 47.7 & 0.00217 & -0.0038 & 0.70 & 1.34 & 0.17 & 319.71\\
11414712 &9& 59 & \citet{Appourchaux2012a} & 750 & 43.7 & 0.00137 & -0.1808 & 0.54 & 1.40 & 0.34 & 217.79\\
11713510 &3& 100 &  & 1201 & 68.8 & 0.00040 & +0.0718 & 0.40 & 1.46 & 0.04 & 562.74\\
11717120 &9& 89 & \citet{Appourchaux2012a} & 555 & 37.6 & 0.00010 & -0.0067 & 0.34 & 1.54 & 0.34 & 128.65\\
11771760 &9& 88 & \citet{Appourchaux2012a} & 505 & 32.4 & 0.00299 & -0.0487 & 0.43 & 1.09 & 0.59 & 201.71\\
12508433 &9& 75 & \citet{Deheuvels2014} & 758 & 45.0 & 0.00305 & -0.1000 & 0.20 & 1.43 & 0.38 & 178.08\\
\hline
\end{tabular}
\end{table*}

\begin{figure*}[!tbp]
\centerline{\includegraphics[width=6 cm,angle=90]{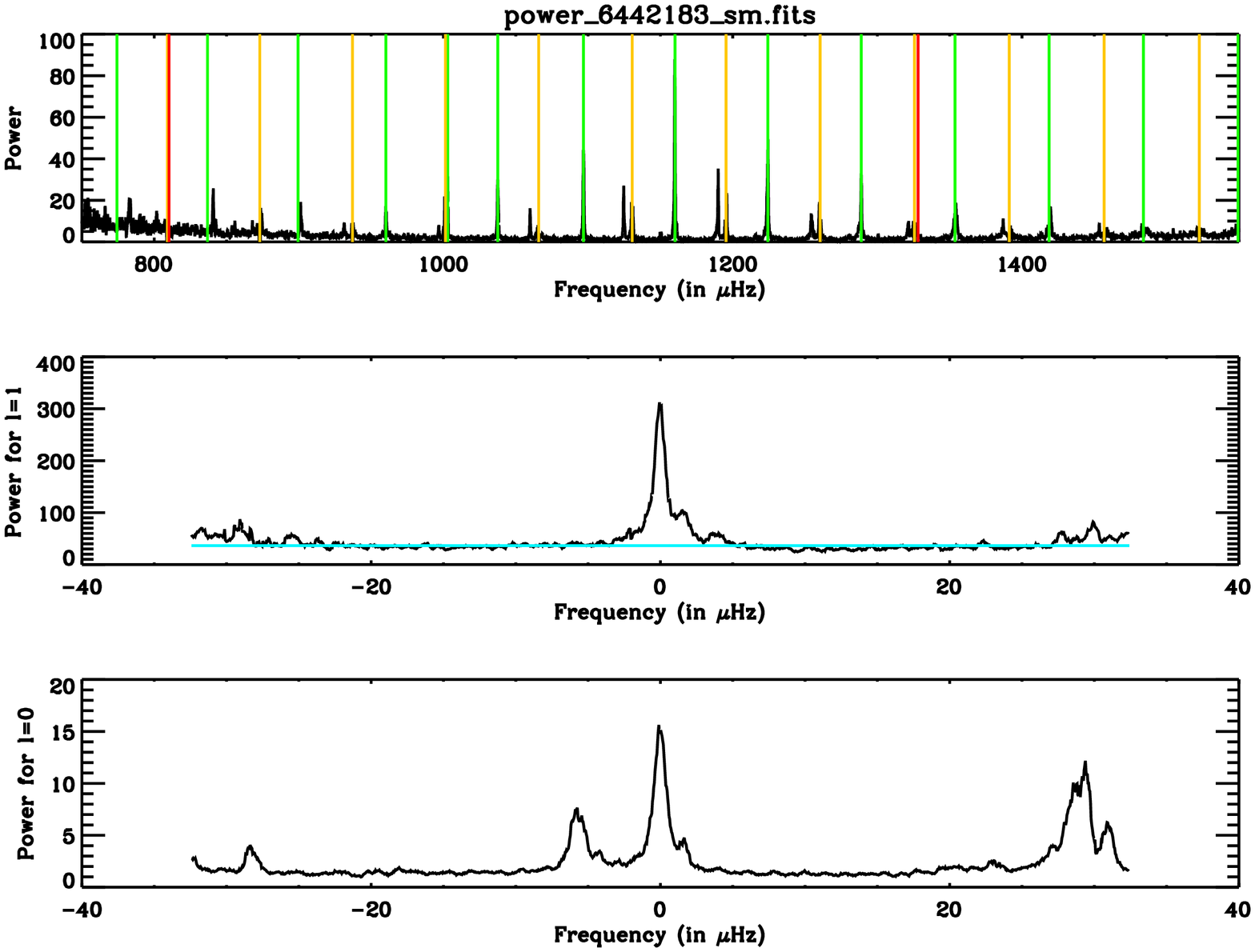}
\includegraphics[width=6 cm,angle=90]{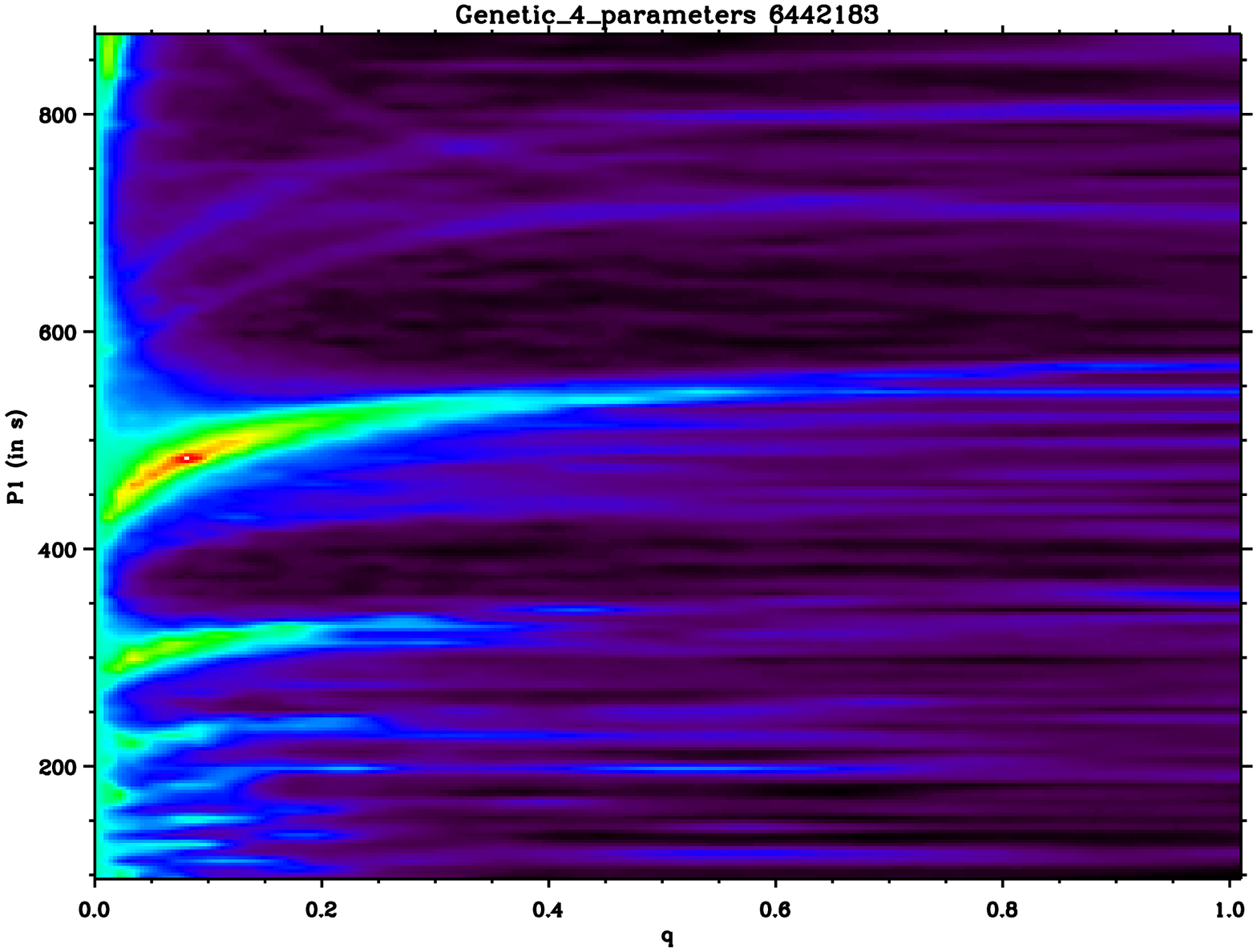}
}
\caption{(Left, top) Power spectrum as a function of frequency for KIC6442183.  The power spectrum is compensated for the p-mode envelope power.  The orange vertical lines indicate the location of the $l=0$ modes for which the power was put to zero.  The green vertical lines indicate the location of the dipole mixed modes.  The red vertical lines indicate the location of the gravity modes. (Left, Middle) Superposed power as a function of frequency for the dipole modes ($l=1$); the cyan horizontal lines indicates the level of noise.  (Left, Bottom) Superposed power as a function of frequency for the $l=0$ modes. (Right) Map of figure of merit for ($P_1, q$) with all other parameters fixed.}
\label{6442183}
\end{figure*}

\begin{figure*}[!tbp]
\centerline{\includegraphics[width=6 cm,angle=90]{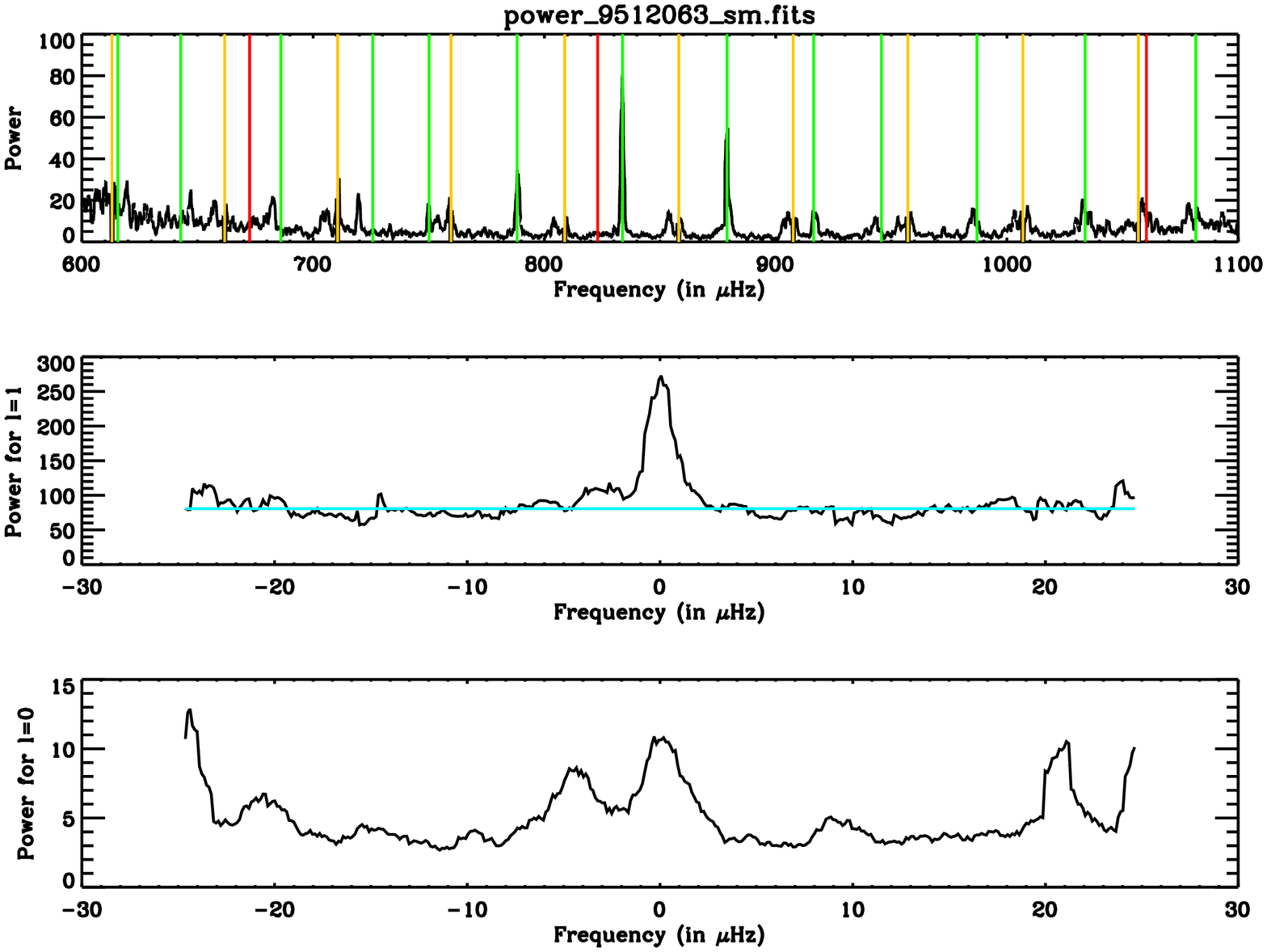}
\includegraphics[width=6 cm,angle=90]{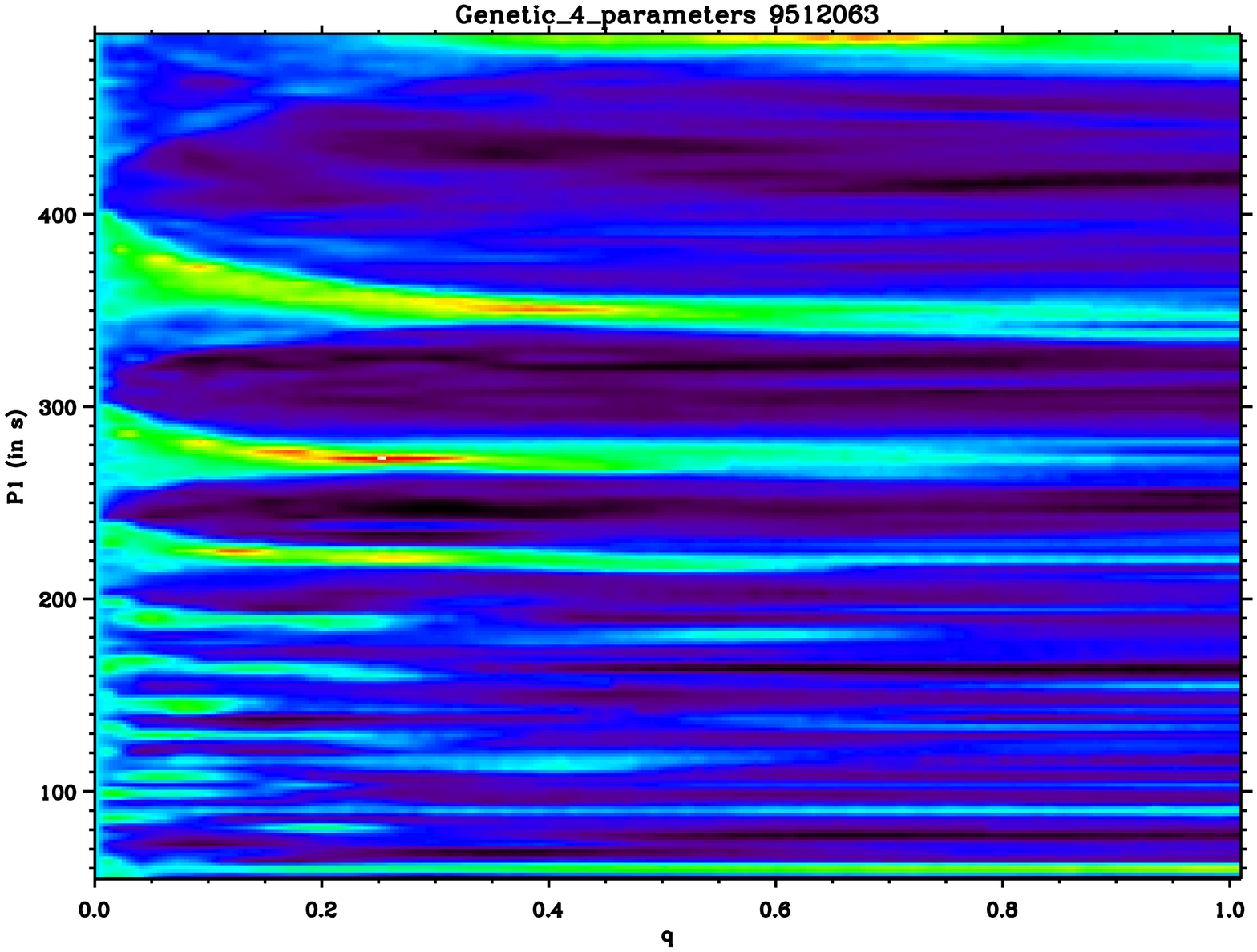}
}
\caption{(Left, top) Power spectrum as a function of frequency for KIC9512063.  The power spectrum is compensated for the p-mode envelope power.  The orange vertical lines indicate the location of the $l=0$ modes for which the power was put to zero.  The green vertical lines indicate the location of the dipole mixed modes.  The red vertical lines indicate the location of the gravity modes. (Left, middle) Superposed power as a function of frequency for the dipole modes ($l=1$); the cyan horizontal lines indicates the level of noise.  (Left, bottom) Superposed power as a function of frequency for the $l=0$ modes.  (Right) Map of figure of merit for ($P_1, q$) with all other parameters fixed.}
\label{9512063}
\end{figure*}

\begin{figure*}[!tbp]
\centerline{\includegraphics[width=6 cm,angle=90]{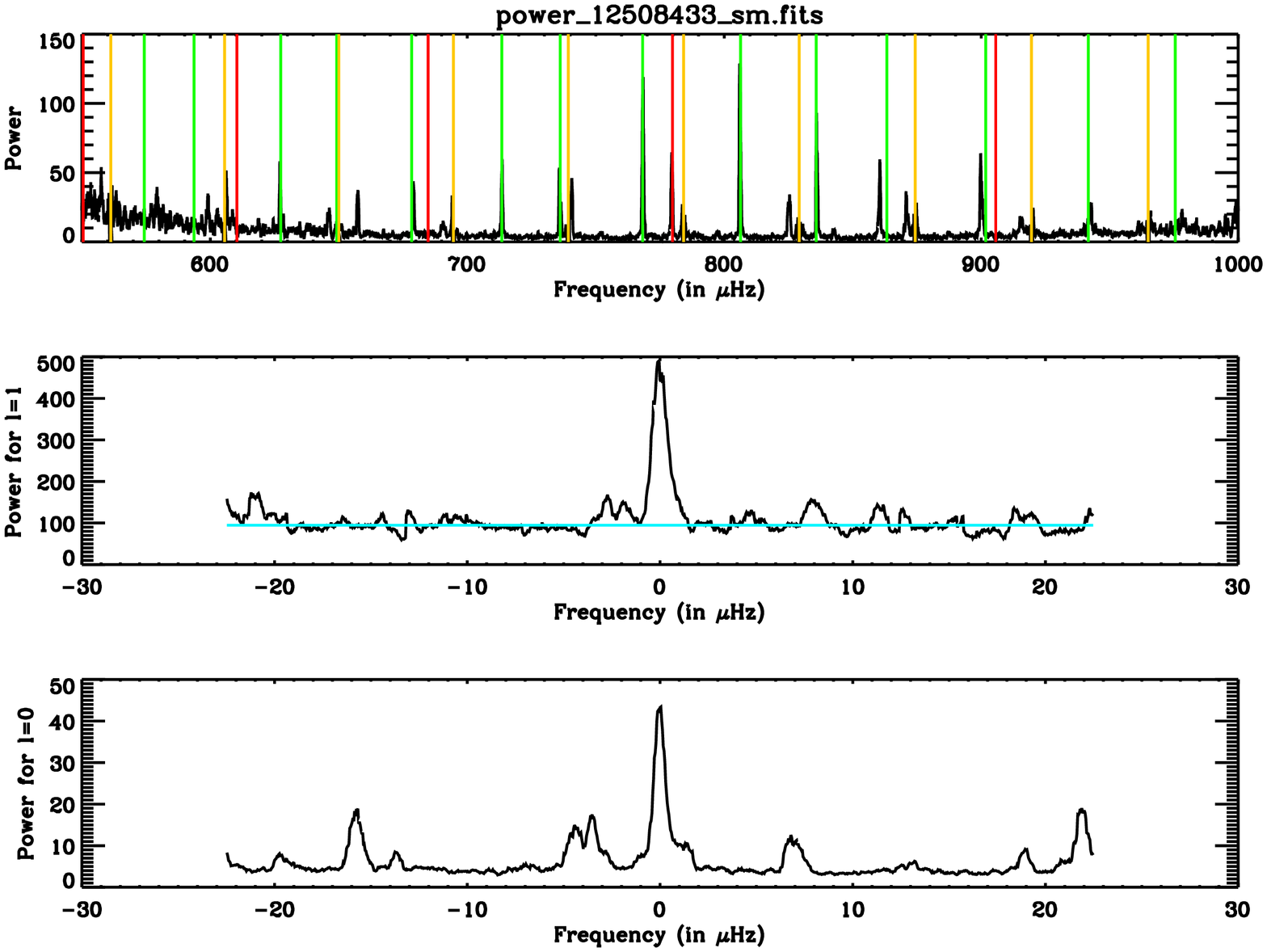}
\includegraphics[width=6 cm,angle=90]{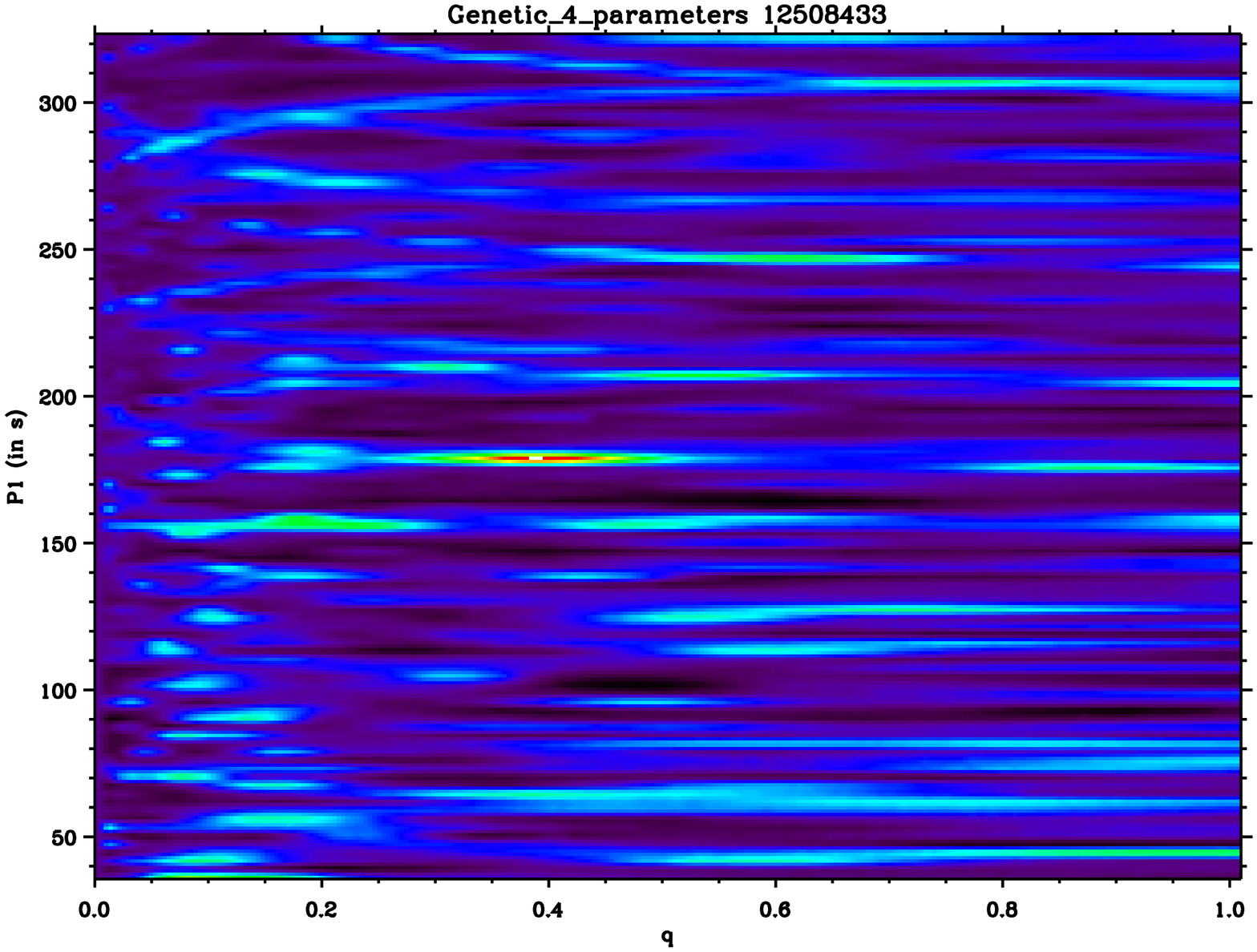}
}
\caption{(Left, top) Power spectrum as a function of frequency for KIC12508433.  The power spectrum is compensated for the p-mode envelope power.  The orange vertical lines indicate the location of the $l=0$ modes for which the power was put to zero.  The green vertical lines indicate the location of the dipole mixed modes.  The red vertical lines indicate the location of the gravity modes. (Left, middle) Superposed power as a function of frequency for the dipole modes ($l=1$); the cyan horizontal lines indicates the level of noise.  (Left, bottom) Superposed power as a function of frequency for the $l=0$ modes.  (Right) Map of figure of merit for ($P_1, q$) with all other parameters fixed.}
\label{12508433}
\end{figure*}

\section{Optimisation of collapsed power}
The computation of collapsed power is performed as explained in the previous section.  Then I need a way to optimise the location of the $l=0$ mode frequencies and of the $l=1$ mode frequencies.  The next idea is to compute the mean collapsed power within 1 $\mu$Hz of the center located at 0 $\mu$Hz.  I use this mean power as a figure of merit which is then maximised.  The procedure is done in the following steps:
\begin{enumerate}
\item Get a stellar power spectrum, smoothed it over 10 bins
\item Select a frequency range for the p-mode power as [$\nu_{\rm low}$,$\nu_{\rm high}$]
\item Divide the power spectrum by a proxy of the Gaussian envelope.
\item Optimise for the $l=0$ mode frequencies:
\begin{enumerate}
\item Set a starting value for $\Delta\nu$, $\epsilon_p$ and $\alpha$ for the $l=0$ mode frequencies (Eq.~4)
\item Compute collapsed power and get the figure of merit for $l=0$.
\item Optimise ($\Delta\nu$, $\epsilon_p$ and $\alpha$) for getting the highest figure of merit for $l=0$.
\end{enumerate}
\item Optimise for the $l=1$ mode frequencies:
\begin{enumerate}
\item Use the previously determined ($\Delta\nu$, $\epsilon_p$ and $\alpha$) as fixed value.
\item Set $d_{01}$ for getting the $l=1$ p-mode frequencies (Eq.~3)
\item Set starting values for $P_1$ and $\epsilon_g$ for getting the $l=1$ g-mode frequencies (Eq.~7)
\item Set starting value for $q$
\item Get $l=1$ mixed mode frequencies from solving Eq.~(8)
\item Compute collapsed power and get the figure of merit for $l=1$.
\item Optimise ($d_{01}$, $P_1$, $\epsilon_g$, $q$) for getting the highest figure of merit for $l=1$.
\end{enumerate}
\end{enumerate}

The maximisation of the $l=0$ figure of merit (Step 2c) is performed using a regular steepest-descent method since it is clear that there is only one minima that is very local and very close to the guess value of $\Delta\nu$, $\epsilon_p$ and $\alpha$.

The maximisation of the $l=1$ figure of merit (Step 5g) is a lot more complicated.  Regular algorithms based on steepest descent fails because of the many local extrema.  An algorithm that is able to search for the extremum of local extrema must be used for finding a proper solution.  Different algorithms were tried such as the following ones:
\begin{itemize}
\item Random search
\item Gaussian process prediction
\item Genetic algorithm
\end{itemize}
The random search is a brute force method which consist in doing 1 million evaluation of the $l=1$ figure of merit taken from the four parameters taken at random in a hypercube for the 4 parameters to be searched.  This algorithm was used for a similar problem involving solar g-mode detection  \citep{Appourchaux2019}.  The algorithm do find the global extremum but is very slow due to the large number of function evaluation.  

Another approach tested was to use to Gaussian Processes (GP) for evaluating / forecasting the figure of merit in the hypercube \citep[See][for a tutorial]{Frazier2018}.  An instructive application of GP can be found in \citet{Jones1998}.  The application of GP is most useful when the figure of merit to be computed takes a large amount of time in the hyperspace.  Instead of computing the figure of merit on a very fine grid of the hyperspace for finding the extremum, the idea is to forecast with a small number of calculation in hyperspace the shape of the figure of merit, and from there to sample the next point in hyperspace for checking the extremum of the forecasted figure of merit.  Usually the sampling of the next point in the hyperspace is done by computing another figure of merit called {\it the Expected Improvement} (EI).  The EI function can be much faster to compute on a fine grid of the hyperspace that the original figure of merit.  Although this is a promising approach, the transfer of the optimisation problem from the figure of merit to EI works on a small number of parameters (say less than 3), but fails for 4 parameters because the EI has local maxima that are Dirac like which are then easily missed unless the computing grid is extremely fine.  The original figure of merit  has already large variation and does not slowly vary; in hyperspace the EI is then similar to an empty space with small hypersphere sparsely distributed.

The most useful algorithm for the current problem is based upon a genetic approach.  The optimisation is based upon the Shuffled
Complex Evolution (SCE) algorithm, which mixes evolutionary and simplex algorithm \citet{Duan1993}.  This algorithm allows to search globally for the best maximum, and then search locally for the
highest maximum.   This algorithm was used by \citet{Thi2010} for fitting absorption profile observed in the infrared with European Southern Observatory's Very Large Telescope.  The fit employs an Interactive Data Language (IDL) software, that can be found at Thi's home page (www.astrowing.eu), which I use for the optimisation of the figure of merit.  Owing to the random character of the genetic algorithm, I also choose to run the SCE in several steps as follows:
\begin{enumerate}
\item Optimisation 10 times using SCE with the following parameters: $d_{01}=[-0.2,0.1]$, $\epsilon_g=[0.2,0.7], $q=[0.,0.8], $P_1=[100.,1000.]$
\item Determine the minimum and maximum of the optimised $P_1$ amongst the 10.  Compute $P_1^{\rm mean}=(P_1^{\rm max}+P_1^{\rm min})/2$, $\Delta_{P1}=(P_1^{\rm max}-P_1^{\rm min})/2$
\item Optimisation 10 times using SCE with the following parameters: same range for $d_{01}, \epsilon_g, q$, $P_1=[P_1^{\rm mean}-1.1\Delta_{P1}, P_1^{\rm mean}+1.1\Delta_{P1}$]
\end{enumerate}
This is more efficient than to run a million evaluation of the figure of merit since about it results in 10 times less evaluations in total.  

Figures~\ref{6442183}, \ref{9512063} and \ref{12508433} show results of the optimisation for three typical cases.  The ($P_1, q$) maps show the difficulties for finding the maximum of the figure of merit.  The typical cases are as follows:
\begin{itemize}
\item For period spacing larger than 200-300 s: a maximum clearly located for $q$ below 0.2.  Lower local maxima are present but can clearly excluded (Fig~\ref{6442183})
\item For $q$ larger than 0.2: Several local maxima located along parallel ridges along $q$.
\item For $P_1$ less than 200 s: several isolated local maxima.
\end{itemize}
The boundaries for these categories are by no means really strict but just rough indication.  Some stars may fall in one category or the other while still being out of the boundaries defined.
Still the three maps are typical of the various maps observed.  From these maps, it is obvious that the optimisation is not a simple problem as many local minima may prevent to find the global minimum.

Table~\ref{Table_data} gives the parameters resulting from the optimisation for each star.

\section{Results from optimisation: period spacing and coupling factor}
 I used two criteria that are useful to separate the red giants from the subgiants: the evolution criteria and the density of mixed modes.  \citet{Mosser2014} provided a boundary for the evolution criteria that subgiants fulfill $(\Delta\nu/36.5 \mu$Hz)~($P_1$/126 s) $> 1$.  The density of mixed modes is to the first order equivalent to the ratio of gravity-mode order to the pressure mode-order (${\cal N} \sim n_{\rm g}/n_{\rm p}$); \citet{Mosser2012a} gave this density as ${\cal N}=\Delta \nu/(P_1 \nu_{\rm max}^2)$.  
 
 Figure~\ref{P1_results} shows the comparison with the measurement made using the asymptotic relation as in \citet{Mosser2014}, and using avoided crossing as in \citet{Li2020} for the period spacing.   I note that the period spacing as derived from the asymptotic relation of \citet{Shibahashi1979} cannot be directly compared with the formulation used by \citet{Deheuvels2011} based on avoided crossing for harmonic oscillators.  For example, \citet{Benomar2013} and \citet{Li2020} provide several examples of echelle diagrams providing the location of the frequency of the gravity modes ($\gamma$ modes) using the formalism of \citet{Deheuvels2011}.  It is clear that the resulting frequencies of the mixed modes are greatly affected close to the frequency location of the gravity $\gamma$  modes, i.e. the mixed modes are further away than the regular $l=1$ ridge; on the other hand the mixed modes are closer to the original ridge in between the gravity $\gamma$ mode frequencies. In the formalism of \citet{Shibahashi1979}, the situation is opposite: close to a gravity mode, the mixed mode frequency is close to the original dipole mode frequency since $\theta_g \sim 0$, then $\theta_p \sim 0$; but when $\theta_g \sim \pi/2$, then $\theta_p \sim \pi/2$ which means that the mixed mode frequency can be anywhere within a $\Delta\nu$.  In the formalism of \citet{Shibahashi1979}, the gravity modes have a g-mode phase offset of 1/2 compared to the gravity $\gamma$ mode of  \citet{Benomar2013}.
 
 Figure~\ref{q_results} shows the comparison with the measurement made by \citet{Mosser2017} for the coupling factor (supplemented by data shown in \citet{Mosser2017} but not published; Mosser, 2020, private communication).  The comparison shows very good agreement with the work of \citet{Mosser2017} and \citet{Mosser2014}.  The period spacings from the optimisation also agree with the results of \citet{Deheuvels2014}.  The increase of the coupling factor for subgiant stars is due to a very thin zone where mixed modes are evanescent \citep{Pincon2020}.  This is the strong coupling assumption leading high transmission between the gravity mode cavity and that of the pressure mode cavity \citep{Takata2016}.

 
 It is also quite clear that there is a simple relation between the evolution criteria and the density of mixed modes.  Since $\nu_{\rm max}$ depends on the large separation, it can be shown that the density of mixed modes is inversely proportional to power of the evolution criteria; depending the state of evolution the exponent factor varies from -1 (for subgiants) to -1.5 (for red giants).  A better relation can be derived but the end result is that the evolution criteria as devised by \citet{Mosser2014} is sufficient for understanding the complexity of the mixed-mode pattern.

\begin{figure*}[!tbp]
\centerline{\includegraphics[width=6 cm,angle=90]{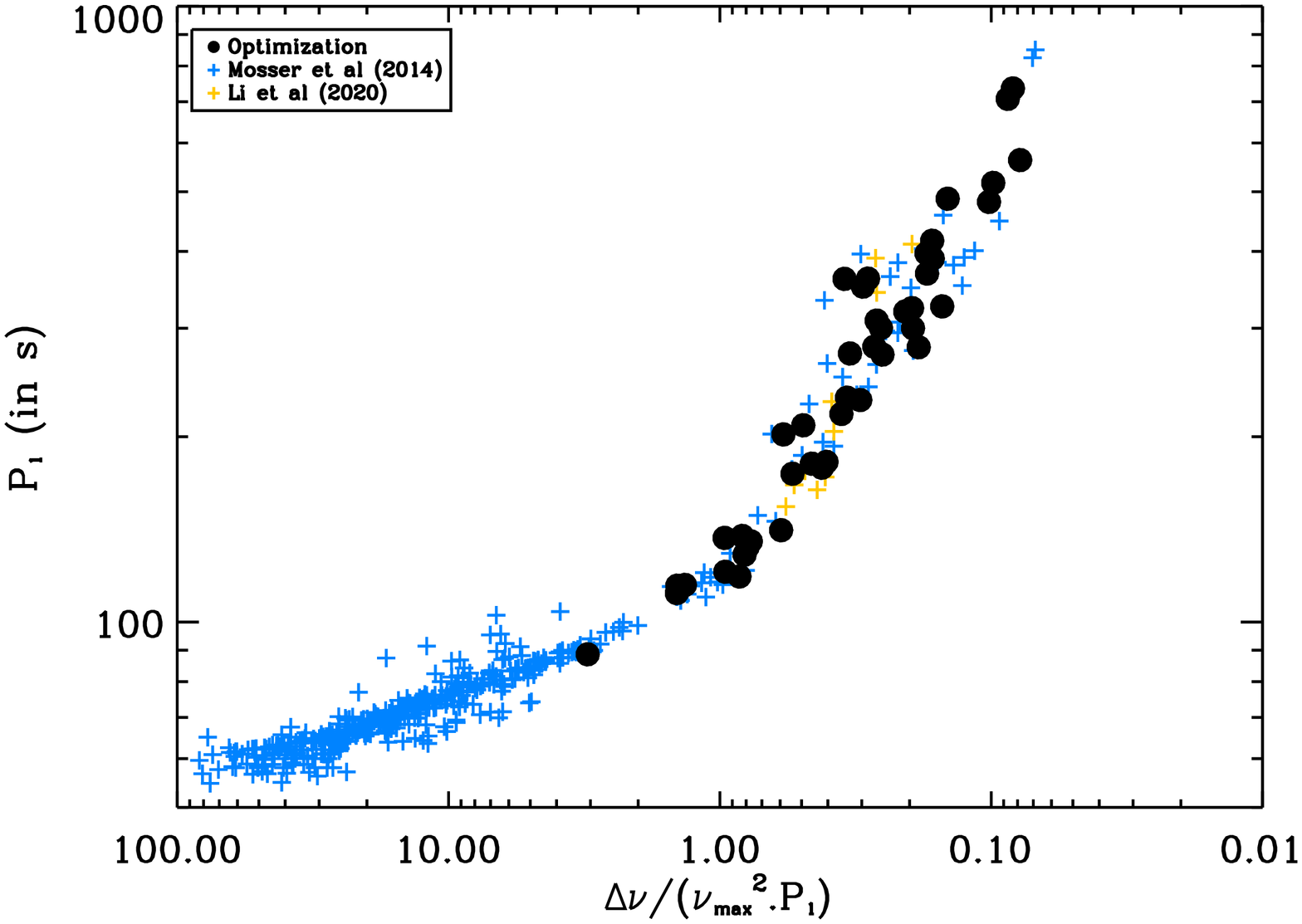}
\includegraphics[width=6 cm,angle=90]{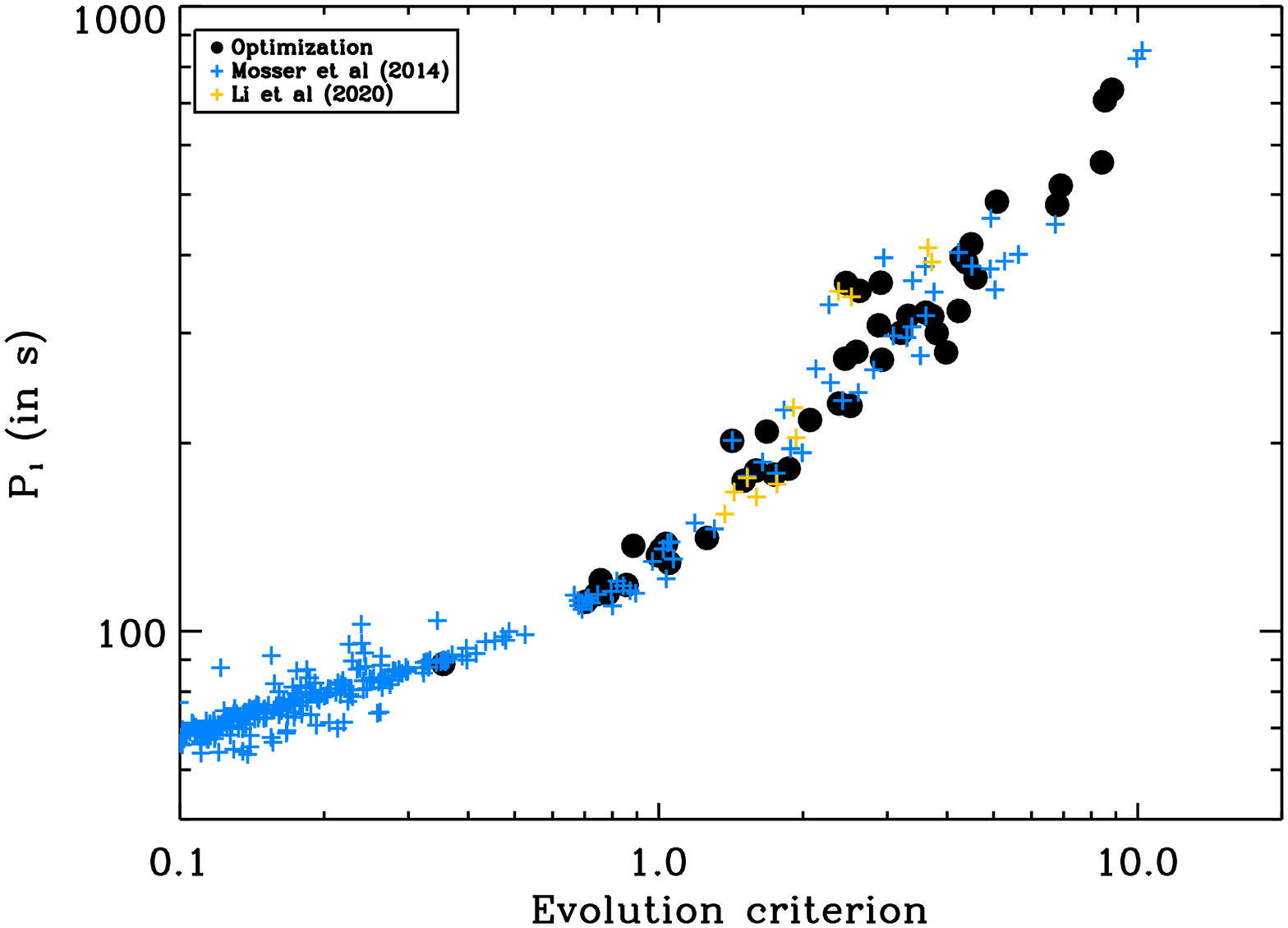}}
\caption{(Left)  Period spacing of the dipole modes as a function of the number of g modes. (Right) Period spacing of the dipole modes as a function of the evolution criteria. (Black disks) From the optimisation procedure; (Blue crosses) From \citet{Mosser2014}; (Orange crosses) From \citet{Li2020}}
\label{P1_results}
\end{figure*}

\begin{figure*}[!tbp]
\centerline{\includegraphics[width=6 cm,angle=90]{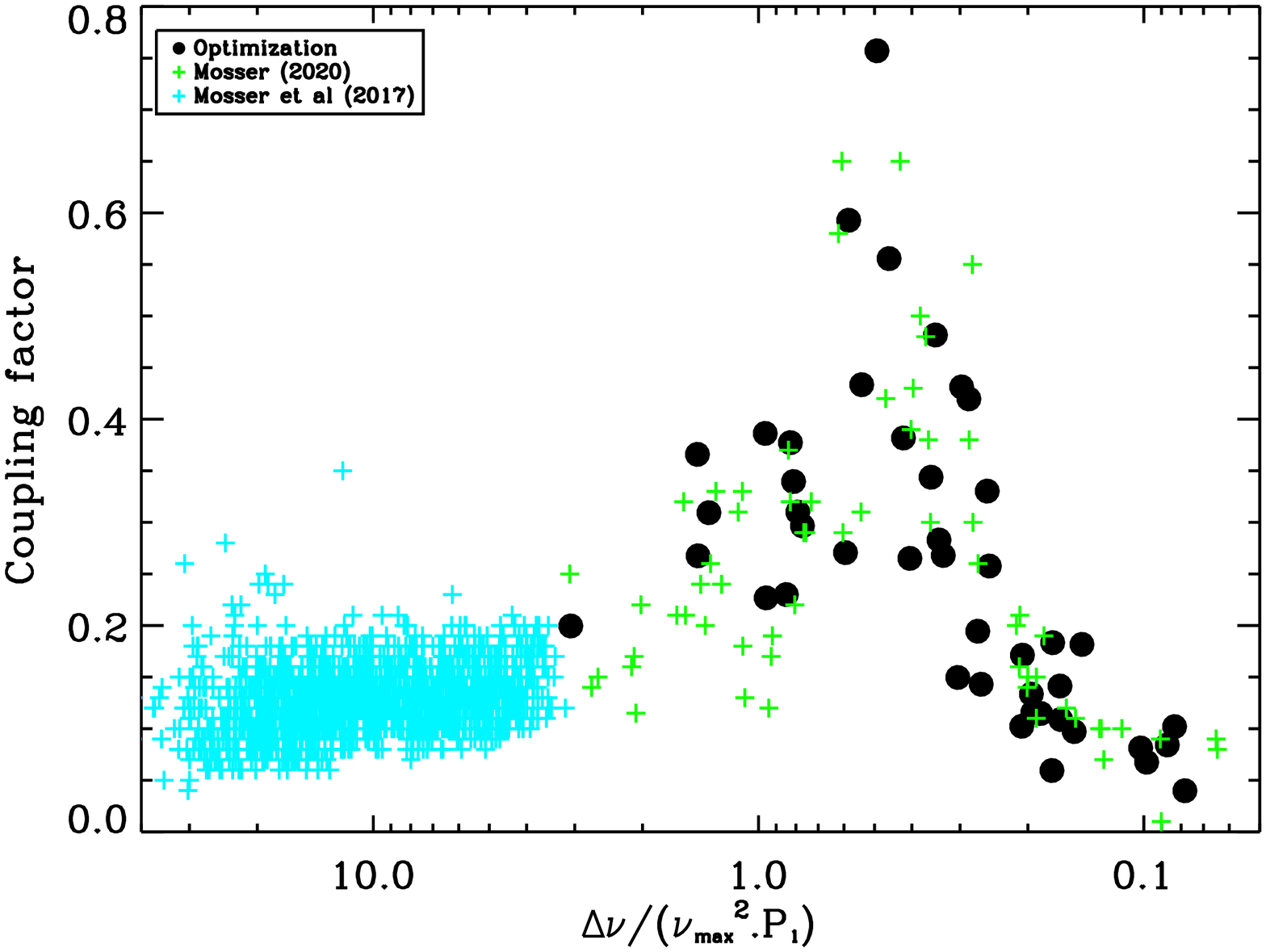}
\includegraphics[width=6 cm,angle=90]{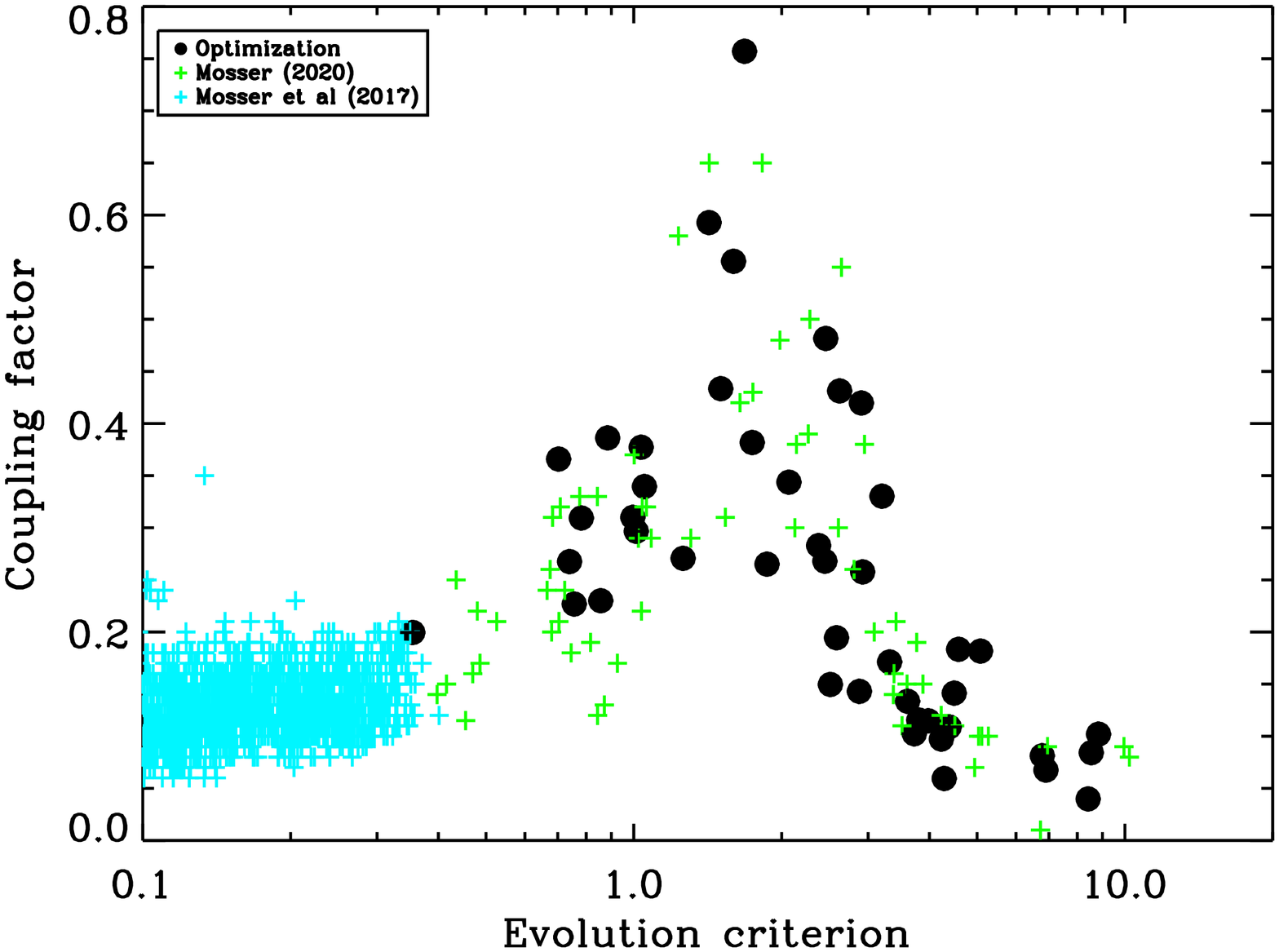}}
\caption{(Left)  Coupling factor as a function of the number of g modes. (Right) Coupling factor as a function of the evolution criteria. (Black disks) From the optimisation procedure, (Blue crosses) From \citet{Mosser2017}; (Green crosses) From Mosser (2020, private communication) for $\Delta\nu >$ 30 $\mu$Hz but unpublished in \citet{Mosser2017}.}
\label{q_results}
\end{figure*}

\section{Results from optimisation: $d_{01}$ and $\epsilon_p$}
\begin{figure}[!tbp]
\centerline{\includegraphics[width=6 cm,angle=90]{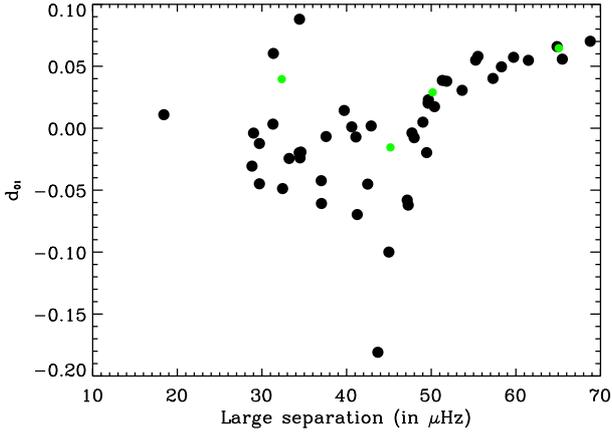}}
\caption{Relative small separation $d_{01}$ as a function of the large separation. (Black disks) From the optimisation procedure; (Green disks) From \citet{Benomar2013}  }.
\label{d_01}
\end{figure}

Figure~\ref{d_01} shows the relative small separation obtained from the optimisation compared to previous measurements obtained by \citet{Benomar2013}.  The results are comparable with the relative small separation provided by \citet{Stello2012} for a 1 $M_{\odot}$ star.  The small separation $d_{01}$ is much less powerful in terms of asteroseismic diagnostic compared to the small separation $d_{02}$ that can provide estimate or proxy of stellar ages \citep{Lebreton2009, Appourchaux2015}.

\begin{figure}[!tbp]
\centerline{\includegraphics[width=6 cm,angle=90]{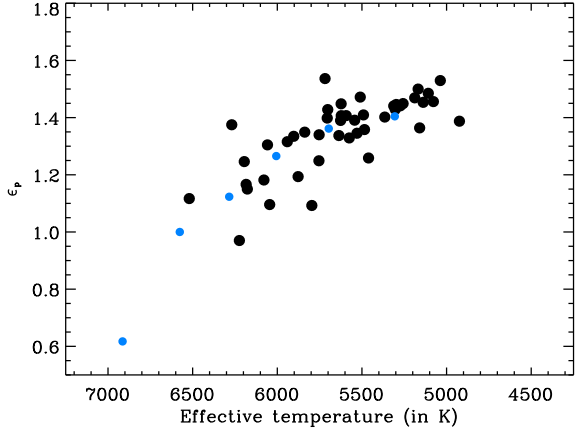}}
\caption{$\epsilon_p$ as a function of the effective temperature. (Black disks) From the optimisation procedure; (Blue disks) From the theoretical model of zero age MS stars given by \citet{Li2020}.}
\label{epsilon_p}
\end{figure}

Figure~\ref{epsilon_p} shows $\epsilon_p$ as a function of effective temperature compared to a theoretical model from \citet{Li2020}.  This ($\epsilon_p, T_{\rm eff}$) diagram is
used for as a diagnostic for identifying the $l=0-2$ modes vs the $l=1$ modes for stars with high effective temperature for which the mode linewidth is larger than the small separation $\delta_{02}$ \citep{White2011}.  In the case of subgiant stars, the identification is eased because of the presence of mixed modes.  The dependence of $\epsilon_p$ with high effective temperature for the subgiant stars  is similar to that of  \citet{White2011}.

\begin{figure*}[!tbp]
\centerline{\includegraphics[width=6 cm,angle=90]{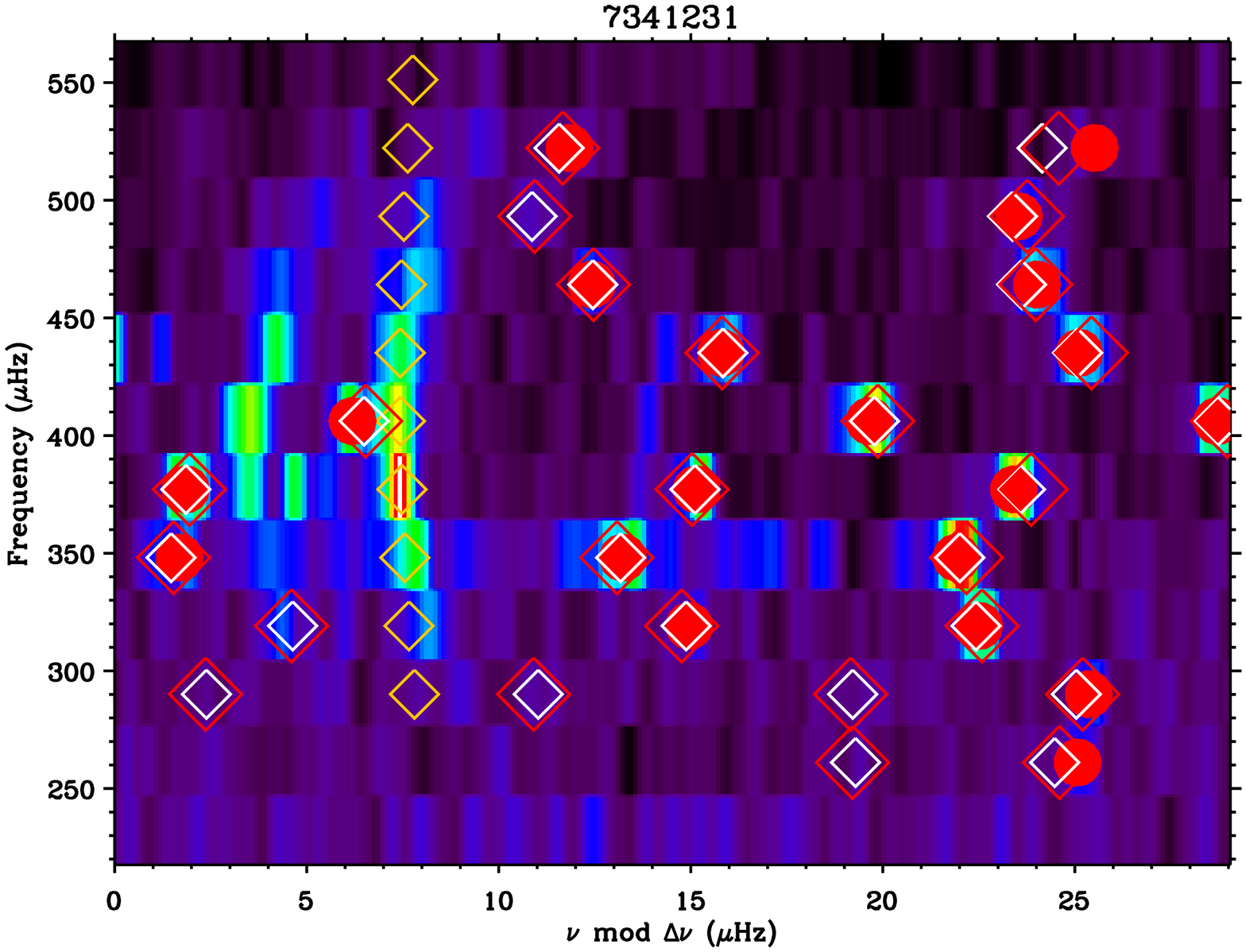}
\includegraphics[width=6 cm,angle=90]{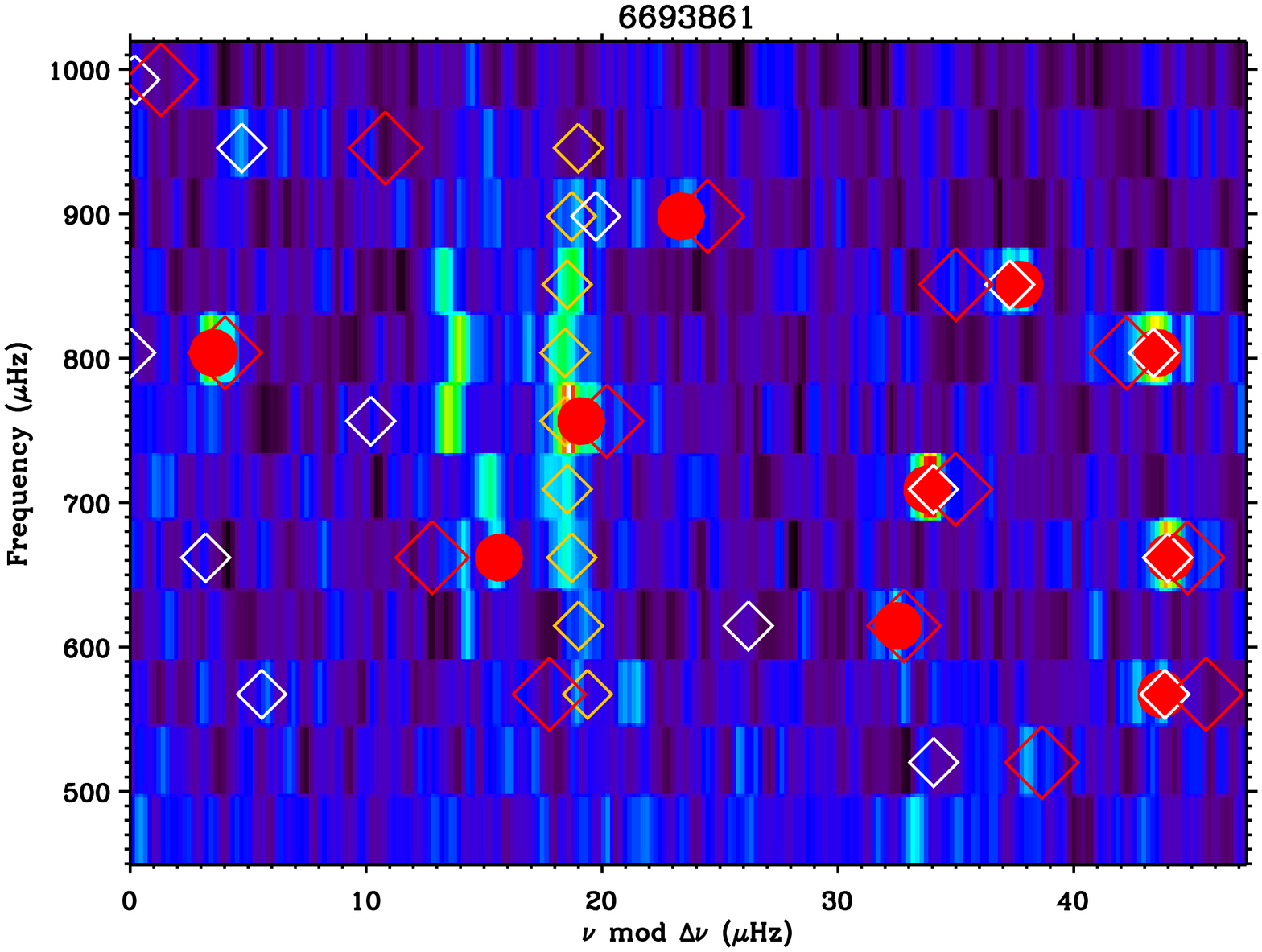}}
\caption{Echelle diagram of the amplitude spectra with: dipole frequencies fitted on the power spectra (Red circles), $l=0$ frequencies from the optimisation (Orange diamonds); asymptotic dipole frequencies from the optimisation (White diamonds) and dipole frequencies from fitting the asymptotic model on the fitted frequencies (Red diamonds). (Left) for a case with 100\% of asymptotic frequencies matching fitted frequencies.  (Right) for a case with 55\% of asymptotic frequencies matching fitted frequencies.}
\label{ech_with_freq}
\end{figure*}

\begin{figure*}[!tbp]
\centerline{\includegraphics[width=6 cm,angle=90]{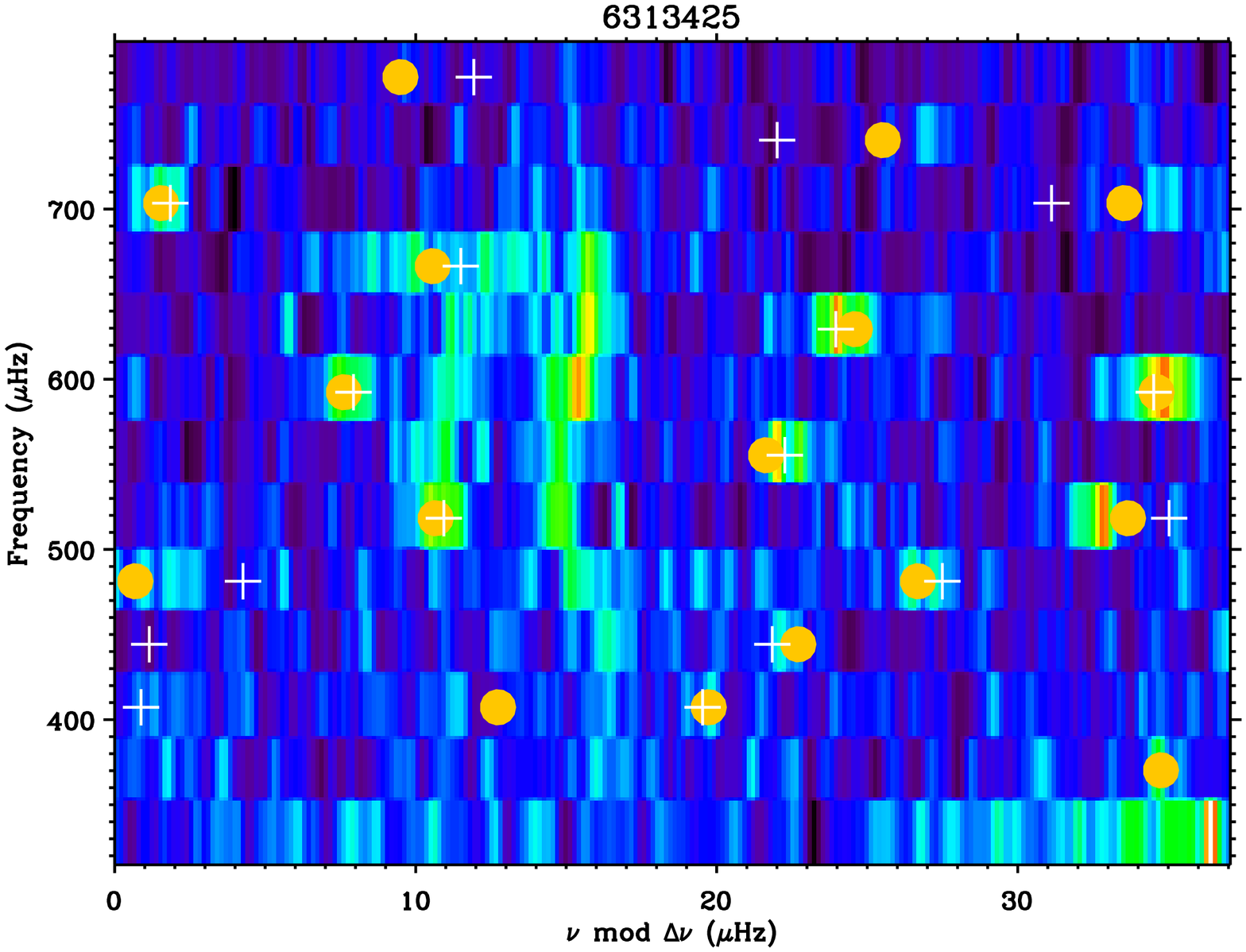}
\includegraphics[width=6 cm,angle=90]{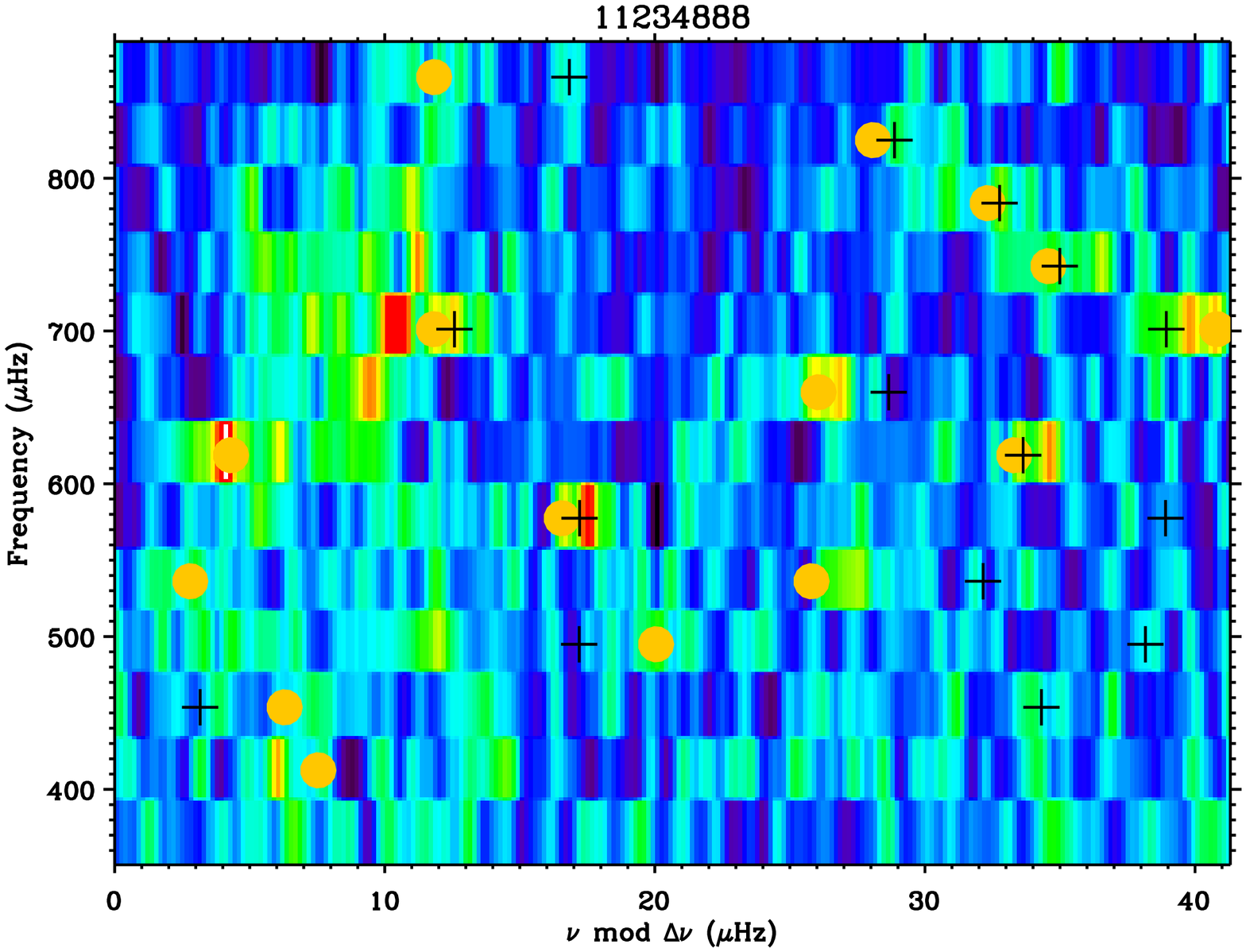}}
\caption{Echelle diagram of the amplitude spectra with visual frequencies for the dipole modes (Orange circles) and the asymptotic dipole frequencies from the optimisation (White / Black crosses).  (Left) for a case with 100\% of asymptotic frequencies matching visual frequencies.  (Right) for a case with 80\% of asymptotic frequencies matching visual frequencies..}
\label{ech_no_freq}
\end{figure*}

\subsection{Results: Asymptotic frequencies vs fitted frequencies}
For 35 stars out of the 44 stars, we have fitted frequencies available from \citet{Campante2011} , \citet{Appourchaux2012a},  \citet{Deheuvels2014},  \citet{Tian2015} and \citet{Li2020}.  We use these frequencies for comparing with the location of the asymptotic dipole frequencies that could have been used as guess.   For the remaining 9 stars, I visually checked whether a mode was present close to the asymptotic dipole frequency.  Table~\ref{Table_data} provides the percentage of frequencies (fitted or visual) that are within 3 $\mu$Hz of the asymptotic frequencies.   The automatic identification allows to retrieve within 3 $\mu$Hz at least 80\% of the modes for 32 stars, and.within 6 $\mu$Hz at least 90\% of the modes for 37 stars.  Figure~\ref{ech_with_freq} shows the best and worst matching for stars with fitted frequencies.  Figure~\ref{ech_no_freq} shows the best and worst matching for stars with visual frequencies.  Figures 15 to 20 give the comparison for all the other stars with fitted frequencies.

It is known that the asymptotic approximation is not applicable to mixed modes in subgiants \citep{Deheuvels2014}.  This is mainly due to the dipole mode frequency being of the same order of magnitude that of the Brunt-Va{\"{\i}}sala frequency.  In addition, \citet{Ong2020} mentioned that the asymptotic approximation fails for low $n_g$ and high $n_p$ which is the case when the density of mixed mode ${\cal N}$ is less than 1.  For most of the stars, the agreement is quite remarkable (Left of Figure~\ref{ech_with_freq}).  Nevertheless a success rate of 80\% is not satisfactory for an automatic guess determination.  It means for instance that out of the 3000 subgiants from P1 sample of PLATO, 900 stars will have between 50\% and 80\% of their dipole mode frequencies properly fitted.

\section{Results from least square fit: period spacing, coupling factor and $d_{01}$}
The optimisation procedure does not provide a direct access to the error bars for each parameter.  I could have derived the error bars from a Monte Carlo simulation of the optimisation process but this would have taken a vary long time; in addition it would have been not really pertinent because the original goal of the optimisation was to obtain guess frequencies for proper extraction of the dipole mode frequencies.  

Therefore I choose to fit directly the asymptotic frequencies to the fitted frequencies.  In that case, it is feasible to make Monte Carlo simulation of the fitting by using the error bars provided for the fitted frequencies by several authors.  As outline previously, the asymptotic model is far from perfect.  For several stars, the deviation of the asymptotic frequencies to the fitted frequencies can be very large amounting to more than 1000 $\sigma$ !  In other words, the fit is plagued with systematic errors due to the model itself.  Therefore, for the fitting the model using the fitted frequencies, I choose to minimise unweighted least square fit.  For computing error bars on the 4 parameters defining the location of the dipole mixed modes, I ran 100 Monte-Carlo simulations of the least square fitting using the fitted frequencies and their error bars as input randomised input.  For the least square fit, I also use the SCE algorithms with the following ranges: $d_{01}=[-0.2,0.2]$, $\epsilon_g=[0.,2.0],  $q=[0.,1.0], $P_1=[0.6P_{\rm optim},1.4P_{\rm optim}]$; where $P_{\rm optim}$ is the period spacing from the optimisation procedure.  In contrast, \citet{Buysschaert2016} used also non-linear least square fit but using a grid search on three parameters ($\epsilon_g$, $q$ and $P_1$)


Table~\ref{Table_data_fit} provides results of the fit with their error bars.  Note that only the dipole mode frequencies were fitted in the process, the parameters defining the $l=0$ mode frequencies were assumed to be the same as for the optimisation.  

Figure~\ref{P1_results} provides the result of the fit compared to the optimisation for the coupling factor and the period spacing, which shows a good agreement between the two determinations.  

Figure~\ref{error_P1_results} provides the errors from the Monte-Carlo simulation for the coupling factor and the period spacing.  In this figure, I added the error bars derived by \citet{Mosser2014} which are rather large compared to the error bars derived from the Monte-Carlo simulation.  A simple explanation to these large error bars is that the many solutions providing the period spacing can be mistreated as an {\it error} while it is a plain systematic error.  Multiple solutions for the period spacing arise when for given g-mode frequency at $n_g$,  $P_1$ can be expressed such that:
\begin{equation}
(n_g+\epsilon_g) P_1 = (n_g +\epsilon_g\pm1) (P_1+\Delta P_1)
\end{equation}
leading to
\begin{equation}
\Delta P_1 = \pm \frac{P_1}{n_g+1+\epsilon_g}
\end{equation}
I can replace ($n_g+1+\epsilon_g$) by $P_{\rm max}/P_1$, where $P_{\rm max}$ is the period corresponding to $\nu_{\rm max}$. Then I have:
\begin{equation}
\Delta P_1 = \pm P_1^2 \nu_{\rm max}
\label{sys}
\end{equation}
This is a rather crude approximation since the next g-mode frequency ($n_g$+1) should have also been taken into account.  It just gives a simple explanation for multiple solutions.  The same equation as Eq.~(\ref{sys}) was given by \citet{Vrard2016} but with a different meaning.  As a matter of fact, this equation does not provide an estimate of an error, as stated by \citet{Vrard2016}, but the separation between multiple solutions, or pseudo-periodic solutions.   Figures~\ref{6442183}, \ref{9512063} and \ref{12508433}  give good examples of the pseudo-periodic solutions that are roughly separated according to Eq.~(\ref{sys}).  As shown by Figure~\ref{error_P1_results}, the error derived by \citet{Mosser2014} are very close to the systematic error provided by Eq.~(\ref{sys}).  This explains why my error bars for the period spacing are about 30--50 times smaller than these systematics errors.

Figure~\ref{d_01_fit} shows the relative small separation obtained from the frequency fit compared to the optimisation and to previous measurements obtained by \citet{Benomar2013}.  The results with the fit are not very different from the optimisation procedure.

\begin{figure*}[!tbp]
\centerline{\includegraphics[width=6 cm,angle=90]{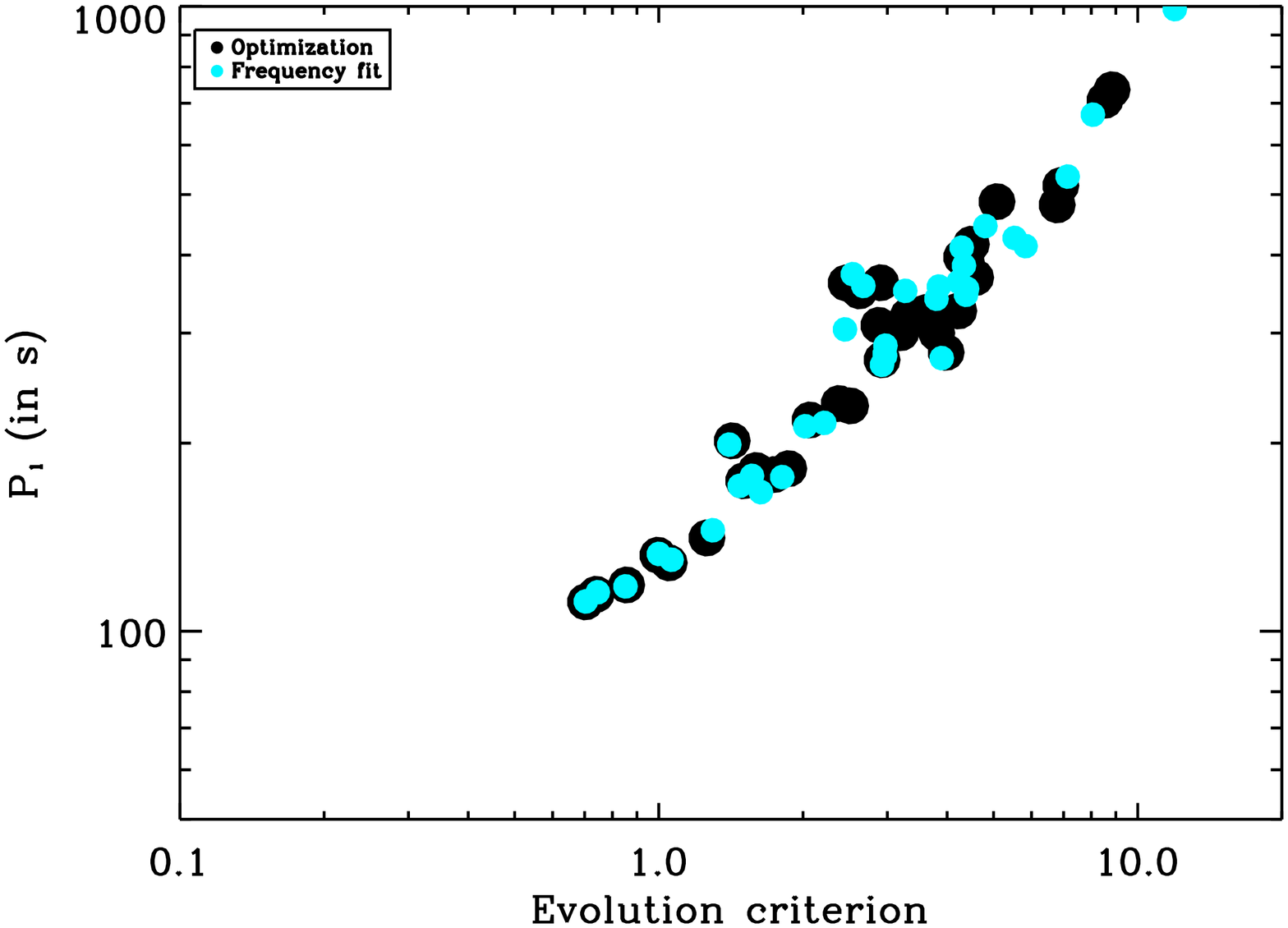}
\includegraphics[width=6 cm,angle=90]{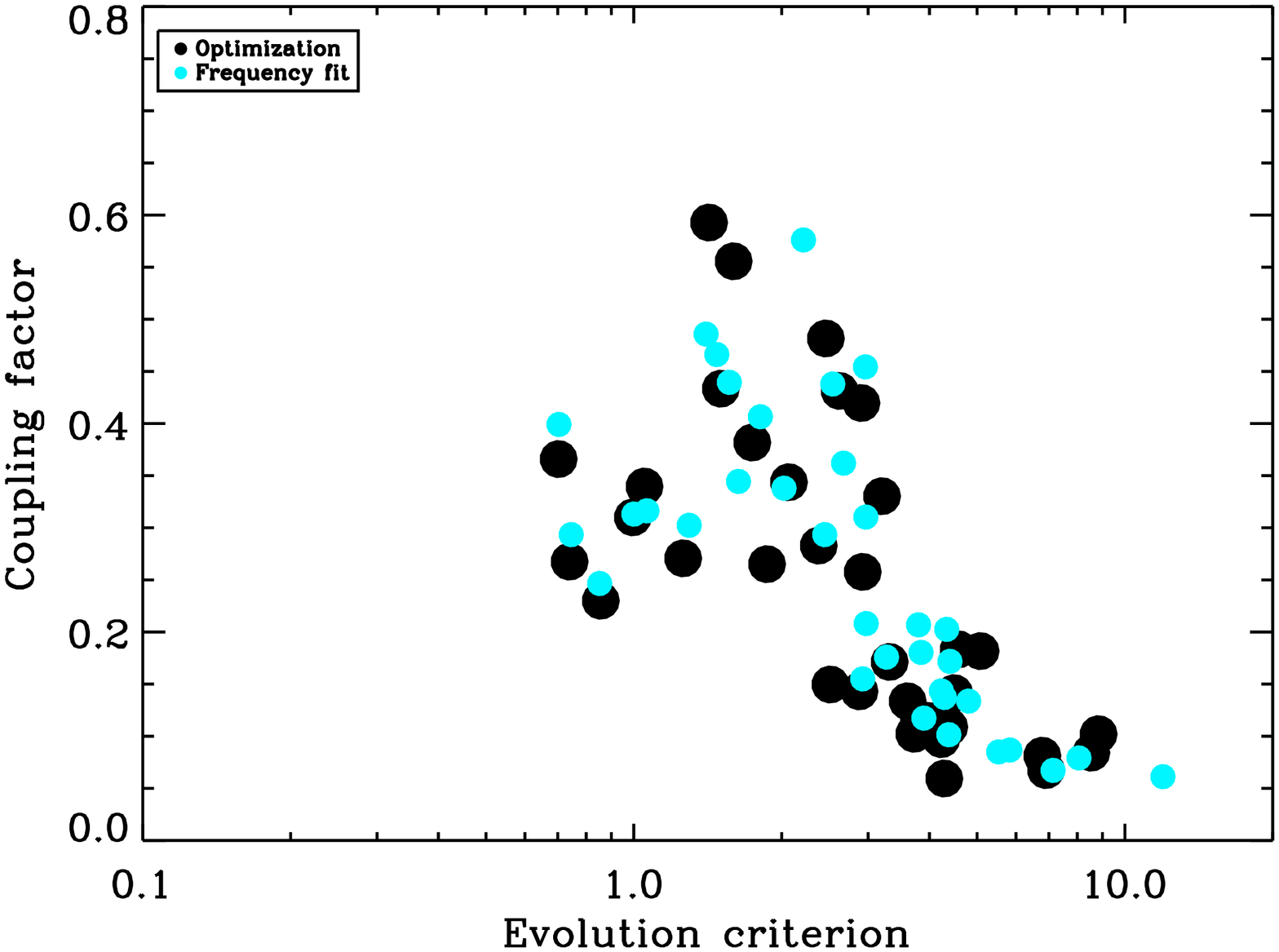}}
\caption{(Left)  Period spacing of the dipole modes as a function of the number of g modes. (Right) Period spacing of the dipole modes as a function of the evolution criteria. (Black disks) From the optimisation procedure; (Cyan disks) From the fit of the frequencies.}
\label{P1_results}
\end{figure*}

\begin{figure*}[!tbp]
\centerline{\includegraphics[width=6 cm,angle=90]{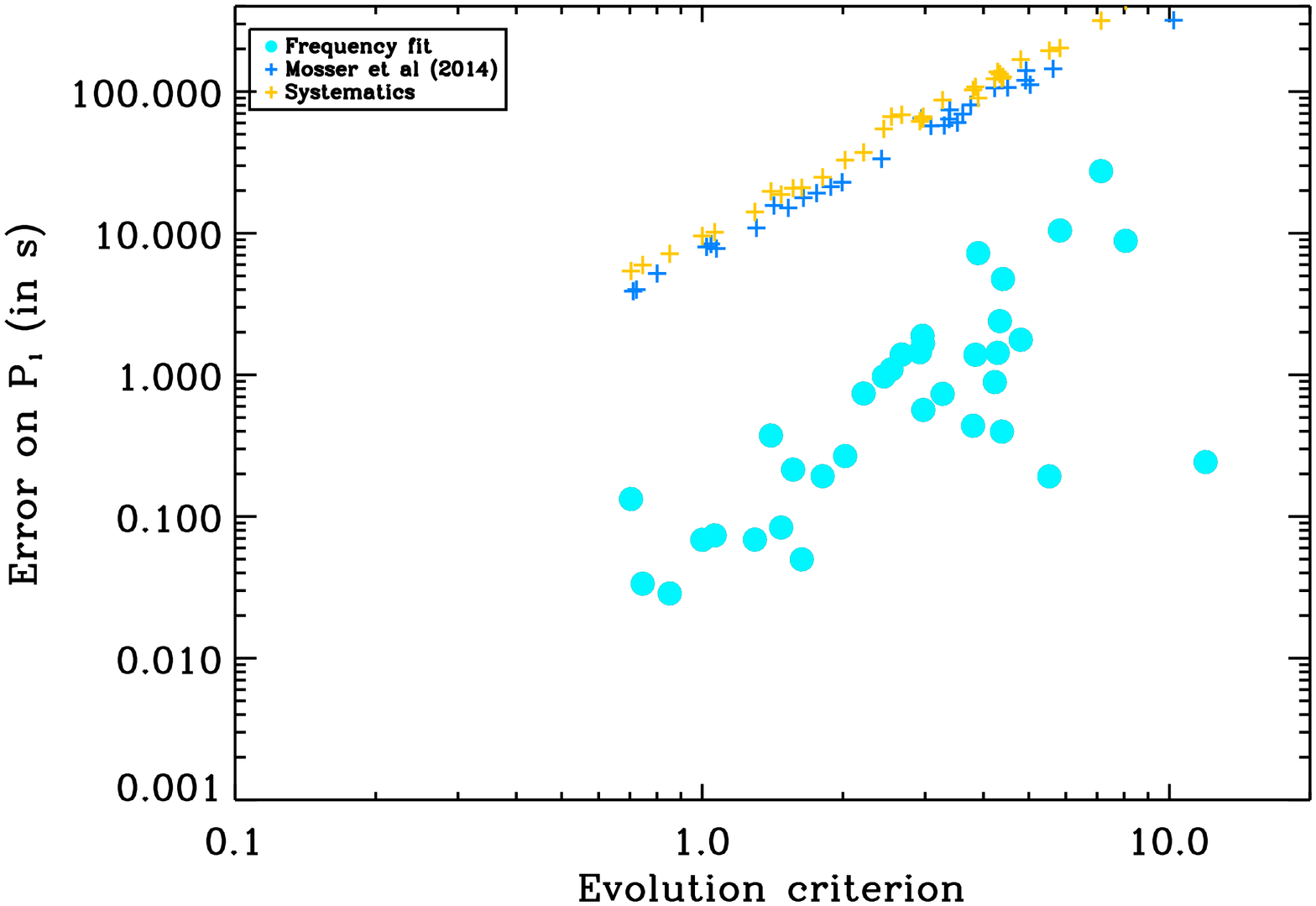}
\includegraphics[width=6 cm,angle=90]{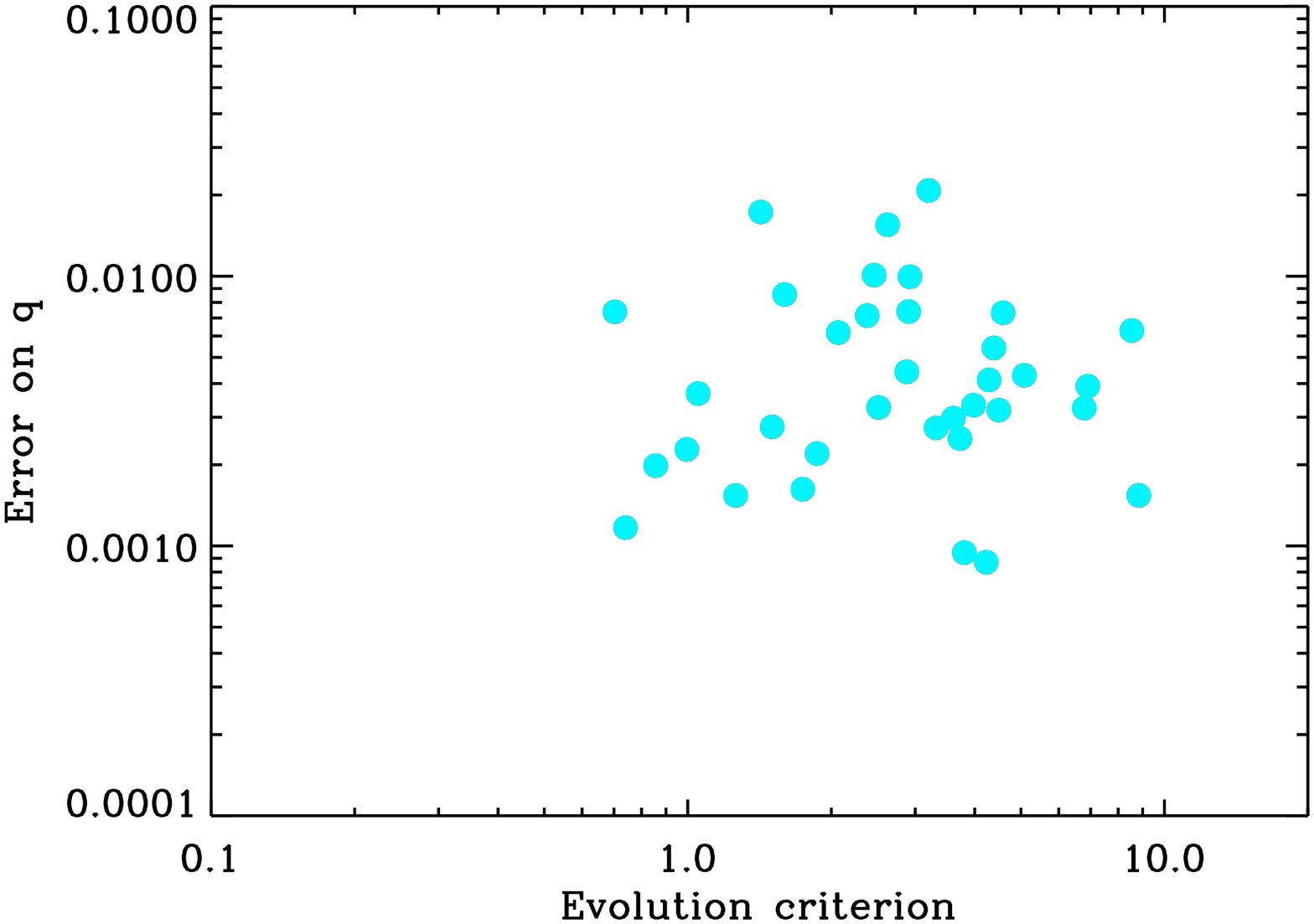}}
\caption{(Left)  Error on the period spacing of the dipole modes as a function of the number of g modes. (Right) Error on the coupling factor of the dipole modes as a function of the evolution criteria. (Black disks) From the optimisation procedure; (Cyan disks) From the fit of the frequencies; (Blue crosses) Error from \citet{Mosser2014};  (Orange crosses) Systematics from Eq~(\ref{sys}).}
\label{error_P1_results}
\end{figure*}

\begin{figure}[!tbp]
\centerline{\includegraphics[width=6 cm,angle=90]{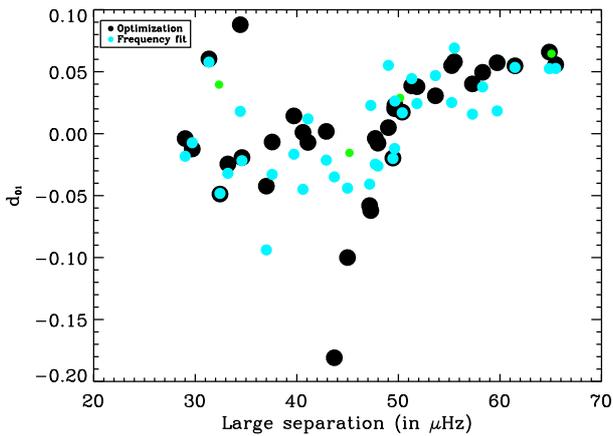}}
\caption{Relative small separation $d_{01}$ as a function of the large separation. (Black disks) From the optimisation procedure; (Cyan disks) From the fit of the frequencies; (Green disks) From \citet{Benomar2013}  }.
\label{d_01_fit}
\end{figure}

\section{Discussion}
{ Several alleys for improvement of the method are envisaged that use a combination of an improved asymptotic description together with detection scheme similar to that of \citet{Corsaro2020}.  The improvement should lead to an automatic identifying guess with a success rate better than 99\%.  The application of the optimisation procedure for the $l=2$ and $l=3$ mixed modes can be envisaged in a very similar fashion.  

The current computation of the figure of merit does not take into account the variation in mode height due to the mode inertia, as was observed by \citet{Benomar2014} for a couple of subgiant stars.
The extension of the method to red giants would require to take into account the variation of the mode height with the dipole mode frequency.  \citet{Dupret2009} showed that many dipole mixed modes are strongly attenuated, and would not contribute to any power in the figure of merit.  \citet{Benomar2014} and \citet{Mosser2015} provided an asymptotic expression for the mixed mode inertia
that can be used for both subgiant and red giants.  Last but not least, the rotational splitting is not included in the current model but would be necessary as the rotation in red giants can be similar to the mixed mode spacing \citet{Mosser2015}}.

\section{Conclusion}
I presented a new way of deriving the coupling factors and period spacing of dipole mixed modes.  The results obtained agree with previous measurements made by \citet{Mosser2014,Mosser2017}.  
I also obtained error bars on the parameters defining the dipole mixed modes using Monte-Carlo simulations.
The automated procedure is far from being perfect as it can provide for 32 stars out of 44 stars, 80\% of the dipole mode frequencies to within $\pm$~3 $\mu$Hz of the dipole peak power.  
{ The improvements of the method for subgiant stars, and the extension to red giant stars are discussed.}

\begin{table*}[!t]
\caption{Table of fitted parameters.  First column: KIC number; second column: relative small separation between the $l=0$ and the $l=1$ mode frequencies, and its error bars; third column: g-mode phase; and its error bar; fourth column: coupling factor and its error bar; fifth column: dipole period spacing and is error bar; sixth column: rms distance between the fit and the observations.}     
\label{Table_data_fit} 
\centering          
\begin{tabular}{cccccc}  
\hline
KIC&$d_{01}$&$\epsilon_g$&$q$& $P_1$&$<\delta_\nu>$\\
&&&& (in s)&in $\mu$Hz\\
\hline
2991448   &     +0.0535  $\pm$    0.0014&       0.3306  $\pm$     0.0966&      0.0672  $\pm$    0.0039 &      534.19  $\pm$       27.45&1.19\\
4346201   &     +0.0691  $\pm$    0.0034&       0.7409  $\pm$     0.0033  &    0.0612  $\pm$    0.0063 &      990.20  $\pm$      0.24&0.97\\
5607242   &    -0.0449  $\pm$      0.0042&       0.5308  $\pm$     0.0097  &    0.4395  $\pm$    0.0085 &      177.26  $\pm$      0.21&1.11\\
5689820   &     +0.0119  $\pm$    0.0008&      0.0239  $\pm$      0.0045  &    0.3024 $\pm$     0.0015 &      145.07  $\pm$     0.07&0.60\\
5955122   &    -0.0120  $\pm$      0.0015&       0.9456  $\pm$     0.0133&      0.1804  $\pm$    0.0032 &      356.22  $\pm$       1.39&2.29\\
6370489   &     +0.0243  $\pm$    0.0027&       0.2793  $\pm$     0.0182&      0.2026  $\pm$    0.0054 &      384.78  $\pm$       2.40&2.32\\
6442183   &     +0.0525  $\pm$    0.0009&       0.9068  $\pm$     0.0723&      0.0868  $\pm$    0.0032 &      413.41  $\pm$       10.45&2.17\\
6693861   &     +0.0227  $\pm$    0.0061&       0.4402  $\pm$     0.0227&      0.5761  $\pm$    0.0071 &      215.49  $\pm$      0.74&1.59\\
7174707   &    -0.0408  $\pm$      0.0023&       0.6549  $\pm$     0.0071  &    0.4065  $\pm$    0.0022 &      176.48  $\pm$      0.19&1.42\\
7341231   &    -0.0182  $\pm$      0.0040&       0.2348  $\pm$     0.0275&      0.3991  $\pm$    0.0074 &      111.59  $\pm$      0.13&0.31\\
7747078   &     +0.0469  $\pm$    0.0033&       0.1737  $\pm$     0.0068  &    0.1435  $\pm$    0.0025 &      362.53  $\pm$      0.89&2.06\\
7799349   &    -0.0320  $\pm$      0.0007&       0.3051  $\pm$     0.0041  &    0.2466  $\pm$    0.0020 &      117.96  $\pm$     0.029&0.36\\
7976303   &     +0.0444  $\pm$    0.0023&       0.5098  $\pm$     0.0044  &    0.2069  $\pm$    0.0030 &      340.51  $\pm$      0.44&2.89\\
8026226   &     +0.0180  $\pm$    0.0074&       0.4959  $\pm$     0.0152&      0.3620  $\pm$    0.0155&      357.04  $\pm$       1.39&1.69\\
8524425   &     +0.0183  $\pm$    0.0008&       0.7403  $\pm$     0.0011  &    0.0850  $\pm$    0.0009 &      426.03  $\pm$      0.19&3.44\\
8702606   &    -0.0166  $\pm$      0.0009&       0.4533  $\pm$     0.0048  &    0.4662  $\pm$    0.0028 &      170.85  $\pm$     0.084&0.53\\
8738809   &     +0.0263  $\pm$    0.0025&       0.1360  $\pm$     0.0105&      0.1337  $\pm$    0.0041 &      445.11  $\pm$       1.77&2.45\\
8751420   &    -0.0218  $\pm$      0.0013&       0.2585  $\pm$     0.0086  &    0.3131  $\pm$    0.0023 &      132.94  $\pm$     0.07&0.19\\
9512063   &    -0.0201  $\pm$      0.0047&       0.4262  $\pm$     0.0250&      0.3103  $\pm$    0.0100&      275.96  $\pm$       1.67&2.39\\
9574283   &   -0.0073  $\pm$       0.0006&       0.1996  $\pm$     0.0057  &    0.2934  $\pm$    0.0012 &      115.44  $\pm$     0.03&0.16\\
10018963 &     +0.0251  $\pm$    0.0009&       0.4137  $\pm$     0.0257&      0.0792  $\pm$    0.0015 &      671.05  $\pm$       8.84&1.82\\
10147635 &    -0.0938  $\pm$      0.0041&       0.2712  $\pm$     0.0139&      0.2934  $\pm$    0.0074 &      304.13  $\pm$      0.97&3.42\\
10273246 &    -0.0261  $\pm$      0.0027&       0.9828  $\pm$     0.0109&      0.1372  $\pm$    0.0043 &      411.07  $\pm$       1.43&1.92\\
10593351 &     +0.0578  $\pm$    0.0040&       0.3401  $\pm$     0.0153&      0.4379  $\pm$    0.0101&      372.79  $\pm$       1.09&1.21\\
10873176 &     +0.0552  $\pm$    0.0082&       0.7937  $\pm$     0.0340&      0.4544  $\pm$    0.0208&      277.90  $\pm$       1.89&2.26\\
10920273 &     +0.0157  $\pm$    0.0032&       0.8229  $\pm$     0.0444&      0.1718  $\pm$    0.0073 &      353.34  $\pm$       4.75&0.61\\
10972873 &     +0.0378  $\pm$    0.0016&       0.2816  $\pm$     0.0031  &    0.1015  $\pm$    0.0009 &      345.55  $\pm$      0.40&0.91\\
11026764 &     +0.0169  $\pm$    0.0015&       0.5427  $\pm$     0.0208&      0.1550 $\pm$     0.0033 &      266.92  $\pm$       1.45&1.42\\
11137075 &     +0.0526  $\pm$    0.0025&       0.6959  $\pm$     0.0880&      0.1174 $\pm$      0.0033 &      273.61  $\pm$       7.23&0.75\\
11193681 &    -0.0214  $\pm$      0.0023&       0.1512  $\pm$     0.0073  &    0.1759 $\pm$     0.0044 &      350.57  $\pm$      0.73&1.97\\
11395018 &    -0.0246  $\pm$      0.0020&       0.1522  $\pm$     0.0081  &    0.2081 $\pm$     0.0027 &      286.47  $\pm$      0.57&3.12\\
11414712 &    -0.0349  $\pm$      0.0027&       0.4476  $\pm$     0.0065  &    0.3380 $\pm$     0.0062 &      212.83  $\pm$      0.27&2.94\\
11717120 &    -0.0329  $\pm$      0.0017&       0.1992  $\pm$     0.0073  &    0.3163 $\pm$     0.0037 &      130.14  $\pm$     0.07&2.91\\
11771760 &    -0.0482  $\pm$      0.0051&       0.5693  $\pm$     0.0196&      0.4858 $\pm$     0.0173&      198.92  $\pm$      0.37&1.37\\
12508433 &    -0.0440  $\pm$     0.0006&       0.6047  $\pm$      0.0022  &    0.3444  $\pm$    0.0016 &      166.95  $\pm$     0.05&1.30\\
\hline
\end{tabular}
\end{table*}

\begin{acknowledgements}
Many thanks to my family for the long standing support during the exploration, questioning, computing and writing phase of this paper.
Thanks also to the Covid19, I had ample time to finish this cursed paper that merely took 18 months to complete.  Thanks to the {\it Kepler} gang for
allowing us to play with such wonderful data.  This paper benefitted from discussions with Othman Benomar, Patrick Gaulme, Benoit Mosser, Charly Pin{\c{c}}on and Mathieu Vrard.
Several of my forefathers and colleagues contributed in absentia to this paper: JEB, PC, VD, CF and DMR. 

\end{acknowledgements}

\bibliographystyle{aa}
\bibliography{thierrya}

\begin{thebibliography}{54}
\expandafter\ifx\csname natexlab\endcsname\relax\def\natexlab#1{#1}\fi

\bibitem[{{Appourchaux} {et~al.}(1997){Appourchaux}, {Andersen}, {Fr\"ohlich},
  {Jim\'enez}, {Telljohann}, \& {Wehrli}}]{TABA97}
{Appourchaux}, T., {Andersen}, B.~N., {Fr\"ohlich}, C., {et~al.} 1997,
  \solphys, {\bf 170}, 27

\bibitem[{{Appourchaux} {et~al.}(2015){Appourchaux}, {Antia}, {Ball},
  {Creevey}, {Lebreton}, {Verma}, {Vorontsov}, {Campante}, {Davies}, {Gaulme},
  {R{\'e}gulo}, {Horch}, {Howell}, {Everett}, {Ciardi}, {Fossati}, {Miglio},
  {Montalb{\'a}n}, {Chaplin}, {Garc{\'\i}a}, \& {Gizon}}]{Appourchaux2015}
{Appourchaux}, T., {Antia}, H.~M., {Ball}, W., {et~al.} 2015, \aap, 582, A25

\bibitem[{{Appourchaux} {et~al.}(2010){Appourchaux}, {Belkacem}, {Broomhall},
  {Chaplin}, {Gough}, {Houdek}, {Provost}, {Baudin}, {Boumier}, {Elsworth},
  {Garc{\'{\i}}a}, {Andersen}, {Finsterle}, {Fr{\"o}hlich}, {Gabriel}, {Grec},
  {Jim{\'e}nez}, {Kosovichev}, {Sekii}, {Toutain}, \&
  {Turck-Chi{\`e}ze}}]{Appourchaux2010}
{Appourchaux}, T., {Belkacem}, K., {Broomhall}, A.-M., {et~al.} 2010, \aapr,
  {\bf 18}, 197

\bibitem[{{Appourchaux} {et~al.}(2012{\natexlab{a}}){Appourchaux}, {Benomar},
  {Gruberbauer}, {Chaplin}, {Garcia}, {Handberg}, {Verner}, {Antia},
  {Campante}, {Davies}, {Deheuvels}, {Hekker}, {Howe}, {Salabert}, {Bedding},
  {White}, {Houdek}, {Silva Aguirre}, {Elsworth}, {Van Cleve}, {Clarke},
  {Hall}, \& {Kjeldsen}}]{Appourchaux2012}
{Appourchaux}, T., {Benomar}, O., {Gruberbauer}, M., {et~al.}
  2012{\natexlab{a}}, \aap, {\bf 537}, A134

\bibitem[{{Appourchaux} {et~al.}(2012{\natexlab{b}}){Appourchaux}, {Chaplin},
  {Garc{\'\i}a}, {Gruberbauer}, {Verner}, {Antia}, {Benomar}, {Campante},
  {Davies}, {Deheuvels}, {Handberg}, {Hekker}, {Howe}, {R{\'e}gulo},
  {Salabert}, {Bedding}, {White}, {Ballot}, {Mathur}, {Silva Aguirre},
  {Elsworth}, {Basu}, {Gilliland }, {Christensen-Dalsgaard}, {Kjeldsen},
  {Uddin}, {Stumpe}, \& {Barclay}}]{Appourchaux2012a}
{Appourchaux}, T., {Chaplin}, W.~J., {Garc{\'\i}a}, R.~A., {et~al.}
  2012{\natexlab{b}}, \aap, 543, A54

\bibitem[{{Appourchaux} \& {Corbard}(2019)}]{Appourchaux2019}
{Appourchaux}, T. \& {Corbard}, T. 2019, \aap, 624, A106

\bibitem[{{Appourchaux} {et~al.}(2004){Appourchaux}, {Moreira}, {Berthomieu},
  \& {Toutain}}]{TA2003}
{Appourchaux}, T., {Moreira}, O., {Berthomieu}, G., \& {Toutain}, T. 2004, in
  ESA Special Publication, Vol. 538, Stellar Structure and Habitable Planet
  Finding, ed. F.~{Favata}, S.~{Aigrain}, \& A.~{Wilson}, 109--115

\bibitem[{{Bedding}(2012)}]{Bedding2012}
{Bedding}, T.~R. 2012, in Astronomical Society of the Pacific Conference
  Series, Vol. 462, Progress in Solar/Stellar Physics with Helio- and
  Asteroseismology, ed. H.~{Shibahashi}, M.~{Takata}, \& A.~E. {Lynas-Gray},
  195

\bibitem[{{Bedding} {et~al.}(2011){Bedding}, {Mosser}, {Huber},
  {Montalb{\'a}n}, {Beck}, {Christensen-Dalsgaard}, {Elsworth},
  {Garc{\'{\i}}a}, {Miglio}, {Stello}, {White}, {De Ridder}, {Hekker}, {Aerts},
  {Barban}, {Belkacem}, {Broomhall}, {Brown}, {Buzasi}, {Carrier}, {Chaplin},
  {di Mauro}, {Dupret}, {Frandsen}, {Gilliland}, {Goupil}, {Jenkins},
  {Kallinger}, {Kawaler}, {Kjeldsen}, {Mathur}, {Noels}, {Aguirre}, \&
  {Ventura}}]{Bedding2011a}
{Bedding}, T.~R., {Mosser}, B., {Huber}, D., {et~al.} 2011, \nat, 471, 608

\bibitem[{{Benomar} {et~al.}(2013){Benomar}, {Bedding}, {Mosser}, {Stello},
  {Belkacem}, {Garcia}, {White}, {Kuehn}, {Deheuvels}, \&
  {Christensen-Dalsgaard}}]{Benomar2013}
{Benomar}, O., {Bedding}, T.~R., {Mosser}, B., {et~al.} 2013, \apj, 767, 158

\bibitem[{{Benomar} {et~al.}(2014){Benomar}, {Belkacem}, {Bedding}, {Stello},
  {Di Mauro}, {Ventura}, {Mosser}, {Goupil}, {Samadi}, \&
  {Garcia}}]{Benomar2014}
{Benomar}, O., {Belkacem}, K., {Bedding}, T.~R., {et~al.} 2014, \apjl, 781, L29

\bibitem[{{Buysschaert} {et~al.}(2016){Buysschaert}, {Beck}, {Corsaro},
  {Christensen-Dalsgaard}, {Aerts}, {Arentoft}, {Kjeldsen}, {Garc{\'\i}a},
  {Silva Aguirre}, \& {Degroote}}]{Buysschaert2016}
{Buysschaert}, B., {Beck}, P.~G., {Corsaro}, E., {et~al.} 2016, \aap, 588, A82

\bibitem[{{Campante} {et~al.}(2011){Campante}, {Handberg}, {Mathur},
  {Appourchaux}, {Bedding}, {Chaplin}, {Garc{\'{\i}}a}, {Mosser}, {Benomar},
  {Bonanno}, {Corsaro}, {Fletcher}, {Gaulme}, {Hekker}, {Karoff}, {R{\'e}gulo},
  {Salabert}, {Verner}, {White}, {Houdek}, {Brand{\~a}o}, {Creevey}, {Do{\v
  g}an}, {Bazot}, {Christensen-Dalsgaard}, {Cunha}, {Elsworth}, {Huber},
  {Kjeldsen}, {Lundkvist}, {Molenda-{\.Z}akowicz}, {Monteiro}, {Stello},
  {Clarke}, {Girouard}, \& {Hall}}]{Campante2011}
{Campante}, T.~L., {Handberg}, R., {Mathur}, S., {et~al.} 2011, \aap, {\bf
  534}, A6

\bibitem[{{Chaplin} {et~al.}(2010){Chaplin}, {Appourchaux}, {Elsworth},
  {Garc{\'{\i}}a}, {Houdek}, {Karoff}, {Metcalfe}, {Molenda-{\.Z}akowicz},
  {Monteiro}, {Thompson}, {Brown}, {Christensen-Dalsgaard}, {Gilliland},
  {Kjeldsen}, {Borucki}, {Koch}, {Jenkins}, {Ballot}, {Basu}, {Bazot},
  {Bedding}, {Benomar}, {Bonanno}, {Brand{\~a}o}, {Bruntt}, {Campante},
  {Creevey}, {Di Mauro}, {Do{\u g}an}, {Dreizler}, {Eggenberger}, {Esch},
  {Fletcher}, {Frandsen}, {Gai}, {Gaulme}, {Handberg}, {Hekker}, {Howe},
  {Huber}, {Korzennik}, {Lebrun}, {Leccia}, {Martic}, {Mathur}, {Mosser},
  {New}, {Quirion}, {R{\'e}gulo}, {Roxburgh}, {Salabert}, {Schou}, {Sousa},
  {Stello}, {Verner}, {Arentoft}, {Barban}, {Belkacem}, {Benatti}, {Biazzo},
  {Boumier}, {Bradley}, {Broomhall}, {Buzasi}, {Claudi}, {Cunha}, {D'Antona},
  {Deheuvels}, {Derekas}, {Garc{\'{\i}}a Hern{\'a}ndez}, {Giampapa}, {Goupil},
  {Gruberbauer}, {Guzik}, {Hale}, {Ireland}, {Kiss}, {Kitiashvili},
  {Kolenberg}, {Korhonen}, {Kosovichev}, {Kupka}, {Lebreton}, {Leroy},
  {Ludwig}, {Mathis}, {Michel}, {Miglio}, {Montalb{\'a}n}, {Moya}, {Noels},
  {Noyes}, {Pall{\'e}}, {Piau}, {Preston}, {Roca Cort{\'e}s}, {Roth}, {Sato},
  {Schmitt}, {Serenelli}, {Silva Aguirre}, {Stevens}, {Su{\'a}rez}, {Suran},
  {Trampedach}, {Turck-Chi{\`e}ze}, {Uytterhoeven}, {Ventura}, \&
  {Wilson}}]{Chaplin2010}
{Chaplin}, W.~J., {Appourchaux}, T., {Elsworth}, Y., {et~al.} 2010, \apjl, {\bf
  713}, L169

\bibitem[{{Christensen-Dalsgaard}(2004)}]{JCD2004a}
{Christensen-Dalsgaard}, J. 2004, \solphys, 220, 137

\bibitem[{{Corsaro} {et~al.}(2020){Corsaro}, {McKeever}, \&
  {Kuszlewicz}}]{Corsaro2020}
{Corsaro}, E., {McKeever}, J.~M., \& {Kuszlewicz}, J.~S. 2020, arXiv e-prints,
  arXiv:2006.08245

\bibitem[{{Deheuvels} {et~al.}(2014){Deheuvels}, {Do{\u{g}}an}, {Goupil},
  {Appourchaux}, {Benomar}, {Bruntt}, {Campante}, {Casagrande}, {Ceillier},
  {Davies}, {De Cat}, {Fu}, {Garc{\'\i}a}, {Lobel}, {Mosser}, {Reese},
  {Regulo}, {Schou}, {Stahn}, {Thygesen}, {Yang}, {Chaplin},
  {Christensen-Dalsgaard}, {Eggenberger}, {Gizon}, {Mathis},
  {Molenda-{\.Z}akowicz}, \& {Pinsonneault}}]{Deheuvels2014}
{Deheuvels}, S., {Do{\u{g}}an}, G., {Goupil}, M.~J., {et~al.} 2014, \aap, 564,
  A27

\bibitem[{{Deheuvels} \& {Michel}(2011)}]{Deheuvels2011}
{Deheuvels}, S. \& {Michel}, E. 2011, \aap, 535, A91

\bibitem[{Duan {et~al.}(1993)Duan, Gupta, \& Sorooshian}]{Duan1993}
Duan, Q.~Y., Gupta, V.~K., \& Sorooshian, S. 1993, Journal of Optimization
  Theory and Applications, 76, 501

\bibitem[{{Dupret} {et~al.}(2009){Dupret}, {Belkacem}, {Samadi}, {Montalban},
  {Moreira}, {Miglio}, {Godart}, {Ventura}, {Ludwig}, {Grigahc{\`e}ne},
  {Goupil}, {Noels}, \& {Caffau}}]{Dupret2009}
{Dupret}, M.~A., {Belkacem}, K., {Samadi}, R., {et~al.} 2009, \aap, 506, 57

\bibitem[{{Dziembowski} {et~al.}(2001){Dziembowski}, {Gough}, {Houdek}, \&
  {Sienkiewicz}}]{Dziembowski2001}
{Dziembowski}, W.~A., {Gough}, D.~O., {Houdek}, G., \& {Sienkiewicz}, R. 2001,
  \mnras, 328, 601

\bibitem[{{Fletcher} {et~al.}(2011){Fletcher}, {Broomhall}, {Chaplin},
  {Elsworth}, {Hekker}, \& {New}}]{Fletcher2011}
{Fletcher}, S.~T., {Broomhall}, A.~M., {Chaplin}, W.~J., {et~al.} 2011, \mnras,
  413, 359

\bibitem[{{Frazier}(2018)}]{Frazier2018}
{Frazier}, P.~I. 2018, arXiv e-prints, arXiv:1807.02811

\bibitem[{{Goldreich} \& {Keeley}(1977)}]{PG77}
{Goldreich}, P. \& {Keeley}, D.~A. 1977, \apj, {\bf 212}, 243

\bibitem[{{Grec}(1981)}]{GG81}
{Grec}, G. 1981, PhD thesis, Universit\'e de Nice

\bibitem[{{Hekker} {et~al.}(2010){Hekker}, {Broomhall}, {Chaplin}, {Elsworth},
  {Fletcher}, {New}, {Arentoft}, {Quirion}, \& {Kjeldsen}}]{Hekker2010a}
{Hekker}, S., {Broomhall}, A.~M., {Chaplin}, W.~J., {et~al.} 2010, \mnras, 402,
  2049

\bibitem[{{Houdek} {et~al.}(1999){Houdek}, {Balmforth},
  {Christensen-Dalsgaard}, \& {Gough}}]{GH99}
{Houdek}, G., {Balmforth}, N.~J., {Christensen-Dalsgaard}, J., \& {Gough},
  D.~O. 1999, \aap, {\bf 351}, 582

\bibitem[{Jones {et~al.}(1998)Jones, Schonlau, \& Welch}]{Jones1998}
Jones, D.~R., Schonlau, M., \& Welch, W.~J. 1998, Journal of Global
  Optimization, 13, 455

\bibitem[{{Lebreton} {et~al.}(2014{\natexlab{a}}){Lebreton}, {Goupil}, \&
  {Montalb{\'a}n}}]{Lebreton2014_I}
{Lebreton}, Y., {Goupil}, M.~J., \& {Montalb{\'a}n}, J. 2014{\natexlab{a}}, in
  EAS Publications Series, Vol.~65, EAS Publications Series, 177--223

\bibitem[{{Lebreton} {et~al.}(2014{\natexlab{b}}){Lebreton}, {Goupil}, \&
  {Montalb{\'a}n}}]{Lebreton2014}
{Lebreton}, Y., {Goupil}, M.~J., \& {Montalb{\'a}n}, J. 2014{\natexlab{b}}, in
  EAS Publications Series, Vol.~65, EAS Publications Series, 177--223

\bibitem[{{Lebreton} \& {Montalb{\'a}n}(2009)}]{Lebreton2009}
{Lebreton}, Y. \& {Montalb{\'a}n}, J. 2009, in IAU Symposium, Vol. 258, The
  Ages of Stars, ed. E.~E. {Mamajek}, D.~R. {Soderblom}, \& R.~F.~G. {Wyse},
  419--430

\bibitem[{{Li} {et~al.}(2020){Li}, {Bedding}, {Li}, {Bi}, {Stello}, {Zhou}, \&
  {White}}]{Li2020}
{Li}, Y., {Bedding}, T.~R., {Li}, T., {et~al.} 2020, \mnras, 495, 2363

\bibitem[{{Marchiori} {et~al.}(2019){Marchiori}, {Samadi}, {Fialho}, {Paproth},
  {Santerne}, {Pertenais}, {B{\"o}rner}, {Cabrera}, {Monsky}, \&
  {Kutrowski}}]{Marchiori2019}
{Marchiori}, V., {Samadi}, R., {Fialho}, F., {et~al.} 2019, \aap, 627, A71

\bibitem[{{Michel} {et~al.}(2008){Michel}, {Baglin}, {Auvergne}, {Catala},
  {Samadi}, {Baudin}, {Appourchaux}, {Barban}, {Weiss}, {Berthomieu},
  {Boumier}, {Dupret}, {Garcia}, {Fridlund}, {Garrido}, {Goupil}, {Kjeldsen},
  {Lebreton}, {Mosser}, {Grotsch-Noels}, {Janot-Pacheco}, {Provost},
  {Roxburgh}, {Thoul}, {Toutain}, {Tiph{\`e}ne}, {Turck-Chieze}, {Vauclair},
  {Vauclair}, {Aerts}, {Alecian}, {Ballot}, {Charpinet}, {Hubert},
  {Ligni{\`e}res}, {Mathias}, {Monteiro}, {Neiner}, {Poretti}, {Renan de
  Medeiros}, {Ribas}, {Rieutord}, {Cort{\'e}s}, \& {Zwintz}}]{Michel2008}
{Michel}, E., {Baglin}, A., {Auvergne}, M., {et~al.} 2008, Science, {\bf 322},
  558

\bibitem[{{Moreno} {et~al.}(2019){Moreno}, {Vielba}, {Manj{\'o}n}, {Motos},
  {V{\'a}zquez}, {Rodr{\'\i}guez}, {Saez}, {Sengl}, {Fern{\'a}ndez}, {Campos},
  {Mu{\~n}oz}, {Mas}, {Balado}, {Ramos}, {Cerruti}, {Pajas}, {Catal{\'a}n},
  {Alcacera}, {Valverde}, {Pflueger}, \& {Vera}}]{Moreno2019}
{Moreno}, J., {Vielba}, E., {Manj{\'o}n}, A., {et~al.} 2019, in Society of
  Photo-Optical Instrumentation Engineers (SPIE) Conference Series, Vol. 11180,
  \procspie, 111803N

\bibitem[{{Mosser} \& {Appourchaux}(2009)}]{BMTA2009}
{Mosser}, B. \& {Appourchaux}, T. 2009, \aap, 508, 877

\bibitem[{{Mosser} {et~al.}(2014){Mosser}, {Benomar}, {Belkacem}, {Goupil},
  {Lagarde}, {Michel}, {Lebreton}, {Stello}, {Vrard}, {Barban}, {Bedding},
  {Deheuvels}, {Chaplin}, {De Ridder}, {Elsworth}, {Montalban}, {Noels},
  {Ouazzani}, {Samadi}, {White}, \& {Kjeldsen}}]{Mosser2014}
{Mosser}, B., {Benomar}, O., {Belkacem}, K., {et~al.} 2014, \aap, 572, L5

\bibitem[{{Mosser} {et~al.}(2012{\natexlab{a}}){Mosser}, {Goupil}, {Belkacem},
  {Marques}, {Beck}, {Bloemen}, {De Ridder}, {Barban}, {Deheuvels}, {Elsworth},
  {Hekker}, {Kallinger}, {Ouazzani}, {Pinsonneault}, {Samadi}, {Stello},
  {Garc{\'{\i}}a}, {Klaus}, {Li}, {Mathur}, \& {Morris}}]{Mosser2012}
{Mosser}, B., {Goupil}, M.~J., {Belkacem}, K., {et~al.} 2012{\natexlab{a}},
  \aap, {\bf 548}, A10

\bibitem[{{Mosser} {et~al.}(2012{\natexlab{b}}){Mosser}, {Goupil}, {Belkacem},
  {Michel}, {Stello}, {Marques}, {Elsworth}, {Barban}, {Beck}, {Bedding}, {De
  Ridder}, {Garc{\'{\i}}a}, {Hekker}, {Kallinger}, {Samadi}, {Stumpe},
  {Barclay}, \& {Burke}}]{Mosser2012a}
{Mosser}, B., {Goupil}, M.~J., {Belkacem}, K., {et~al.} 2012{\natexlab{b}},
  \aap, {\bf 540}, A143

\bibitem[{{Mosser} {et~al.}(2017){Mosser}, {Pin{\c{c}}on}, {Belkacem},
  {Takata}, \& {Vrard}}]{Mosser2017}
{Mosser}, B., {Pin{\c{c}}on}, C., {Belkacem}, K., {Takata}, M., \& {Vrard}, M.
  2017, \aap, 600, A1

\bibitem[{{Mosser} {et~al.}(2015){Mosser}, {Vrard}, {Belkacem}, {Deheuvels}, \&
  {Goupil}}]{Mosser2015}
{Mosser}, B., {Vrard}, M., {Belkacem}, K., {Deheuvels}, S., \& {Goupil}, M.~J.
  2015, \aap, 584, A50

\bibitem[{{Ong} \& {Basu}(2020)}]{Ong2020}
{Ong}, J.~M.~J. \& {Basu}, S. 2020, arXiv e-prints, arXiv:2006.13313

\bibitem[{{Pin{\c{c}}on} {et~al.}(2020){Pin{\c{c}}on}, {Goupil}, \&
  {Belkacem}}]{Pincon2020}
{Pin{\c{c}}on}, C., {Goupil}, M.~J., \& {Belkacem}, K. 2020, \aap, 634, A68

\bibitem[{{Rauer} {et~al.}(2014){Rauer}, {Catala}, {Aerts}, {Appourchaux},
  {Benz}, {Brandeker}, {Christensen-Dalsgaard}, {Deleuil}, {Gizon}, {Goupil},
  {G{\"u}del}, {Janot-Pacheco}, {Mas-Hesse}, {Pagano}, {Piotto}, {Pollacco},
  {Santos}, {Smith}, {Su{\'a}rez}, {Szab{\'o}}, {Udry}, {Adibekyan}, {Alibert},
  {Almenara}, {Amaro-Seoane}, {Eiff}, {Asplund}, {Antonello}, {Barnes},
  {Baudin}, {Belkacem}, {Bergemann}, {Bihain}, {Birch}, {Bonfils}, {Boisse},
  {Bonomo}, {Borsa}, {Brand {\~a}o}, {Brocato}, {Brun}, {Burleigh}, {Burston},
  {Cabrera}, {Cassisi}, {Chaplin}, {Charpinet}, {Chiappini}, {Church},
  {Csizmadia}, {Cunha}, {Damasso}, {Davies}, {Deeg}, {D{\'\i}az}, {Dreizler},
  {Dreyer}, {Eggenberger}, {Ehrenreich}, {Eigm{\"u}ller}, {Erikson}, {Farmer},
  {Feltzing}, {de Oliveira Fialho}, {Figueira}, {Forveille}, {Fridlund},
  {Garc{\'\i}a}, {Giommi}, {Giuffrida}, {Godolt}, {Gomes da Silva}, {Granzer},
  {Grenfell}, {Grotsch-Noels}, {G{\"u}nther}, {Haswell}, {Hatzes},
  {H{\'e}brard}, {Hekker}, {Helled}, {Heng}, {Jenkins}, {Johansen},
  {Khodachenko}, {Kislyakova}, {Kley}, {Kolb}, {Krivova}, {Kupka}, {Lammer},
  {Lanza}, {Lebreton}, {Magrin}, {Marcos-Arenal}, {Marrese}, {Marques},
  {Martins}, {Mathis}, {Mathur}, {Messina}, {Miglio}, {Montalban}, {Montalto},
  {Monteiro}, {Moradi}, {Moravveji}, {Mordasini}, {Morel}, {Mortier},
  {Nascimbeni}, {Nelson}, {Nielsen}, {Noack}, {Norton}, {Ofir}, {Oshagh},
  {Ouazzani}, {P{\'a}pics}, {Parro}, {Petit}, {Plez}, {Poretti}, {Quirrenbach},
  {Ragazzoni}, {Raimondo}, {Rainer}, {Reese}, {Redmer}, {Reffert},
  {Rojas-Ayala}, {Roxburgh}, {Salmon}, {Santerne}, {Schneider}, {Schou},
  {Schuh}, {Schunker}, {Silva-Valio}, {Silvotti}, {Skillen}, {Snellen}, {Sohl},
  {Sousa}, {Sozzetti}, {Stello}, {Strassmeier}, {{\v{S}}vanda}, {Szab{\'o}},
  {Tkachenko}, {Valencia}, {Van Grootel}, {Vauclair}, {Ventura}, {Wagner},
  {Walton}, {Weingrill}, {Werner}, {Wheatley}, \& {Zwintz}}]{Rauer2014}
{Rauer}, H., {Catala}, C., {Aerts}, C., {et~al.} 2014, Experimental Astronomy,
  38, 249

\bibitem[{{Shibahashi}(1979)}]{Shibahashi1979}
{Shibahashi}, H. 1979, \pasj, 31, 87

\bibitem[{{Stello}(2012)}]{Stello2012}
{Stello}, D. 2012, in Astronomical Society of the Pacific Conference Series,
  Vol. 462, Progress in Solar/Stellar Physics with Helio- and Asteroseismology,
  ed. H.~{Shibahashi}, M.~{Takata}, \& A.~E. {Lynas-Gray}, 200

\bibitem[{{Takata}(2016)}]{Takata2016}
{Takata}, M. 2016, \pasj, 68, 91

\bibitem[{{Tassoul}(1980)}]{TASSOUL80}
{Tassoul}, M. 1980, \apj S, {{\bf 43}}, 469

\bibitem[{{Thi} {et~al.}(2010){Thi}, {van Dishoeck}, {Pontoppidan}, \&
  {Dartois}}]{Thi2010}
{Thi}, W.~F., {van Dishoeck}, E.~F., {Pontoppidan}, K.~M., \& {Dartois}, E.
  2010, \mnras, 406, 1409

\bibitem[{{Tian} {et~al.}(2015){Tian}, {Bi}, {Bedding}, \& {Yang}}]{Tian2015}
{Tian}, Z., {Bi}, S., {Bedding}, T.~R., \& {Yang}, W. 2015, \aap, 580, A44

\bibitem[{{Unno} {et~al.}(1989){Unno}, {Osaki}, {Ando}, {Saio}, \&
  {Shibahashi}}]{Unno89}
{Unno}, W., {Osaki}, Y., {Ando}, H., {Saio}, H., \& {Shibahashi}, H. 1989,
  {Nonradial oscillations of stars} (University of Tokyo Press, 1989, 2nd ed.)

\bibitem[{{Vassiliadis} \& {Wood}(1993)}]{VW93}
{Vassiliadis}, E. \& {Wood}, P.~R. 1993, \apj, 413, 641

\bibitem[{{Vrard} {et~al.}(2016){Vrard}, {Mosser}, \& {Samadi}}]{Vrard2016}
{Vrard}, M., {Mosser}, B., \& {Samadi}, R. 2016, \aap, 588, A87

\bibitem[{{White} {et~al.}(2011){White}, {Bedding}, {Stello},
  {Christensen-Dalsgaard}, {Huber}, \& {Kjeldsen}}]{White2011}
{White}, T.~R., {Bedding}, T.~R., {Stello}, D., {et~al.} 2011, \apj, {\bf 743},
  161

\end{thebibliography}

\Online

\begin{figure*}[!tbp]
\centering
\hbox{
\includegraphics[width=6 cm,angle=90]{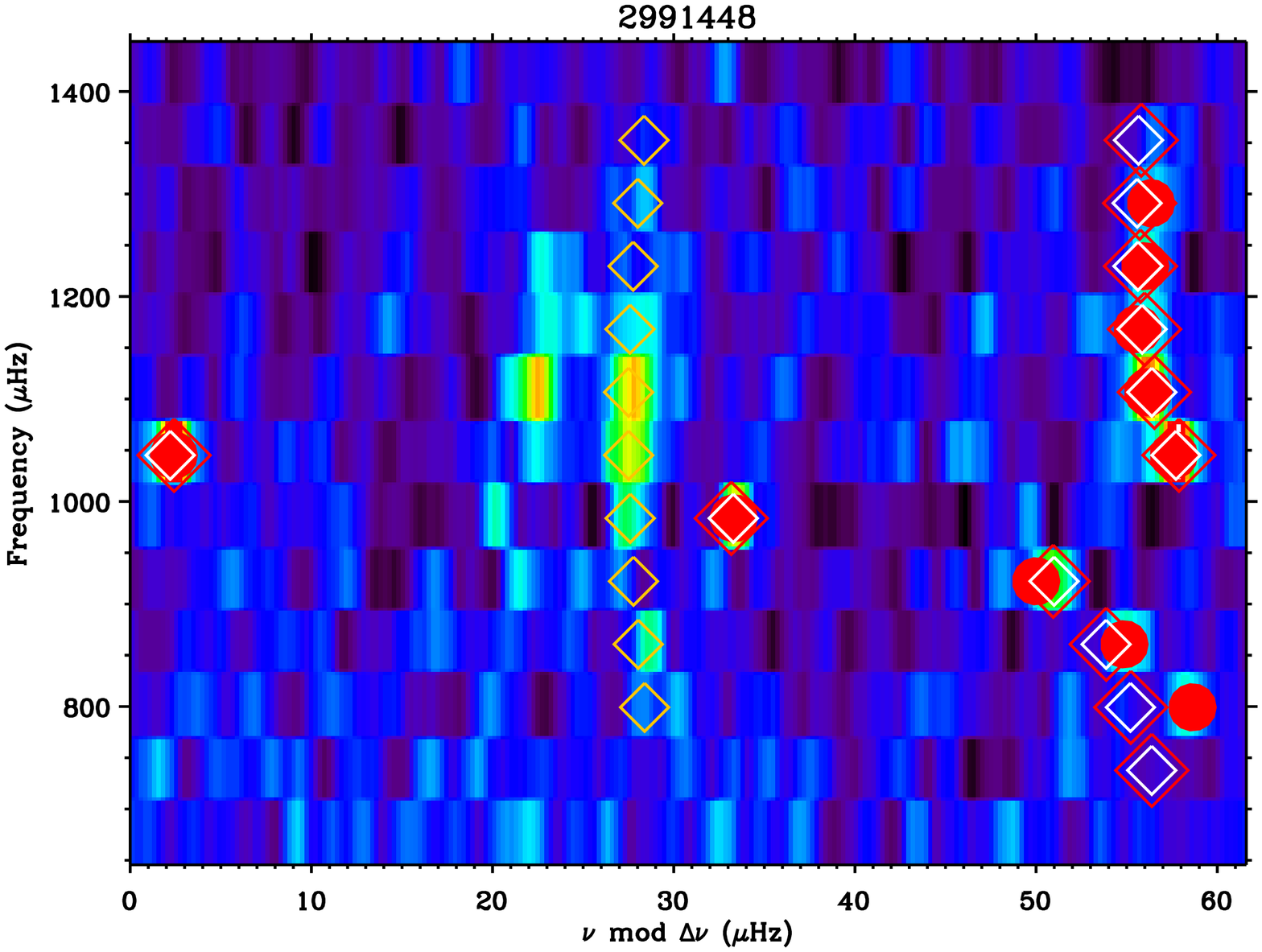}
\includegraphics[width=6 cm,angle=90]{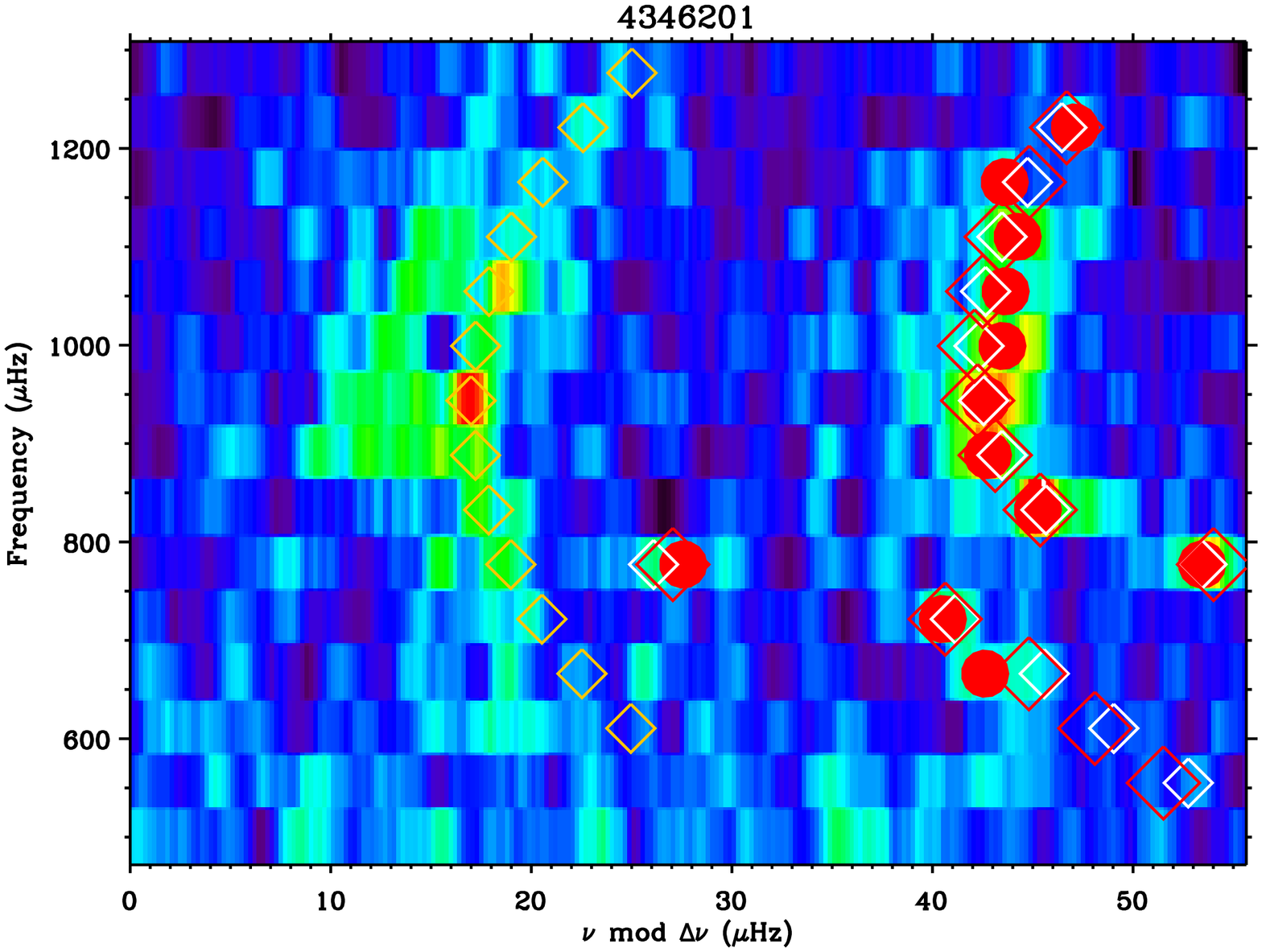}
}
\hbox{
\includegraphics[width=6 cm,angle=90]{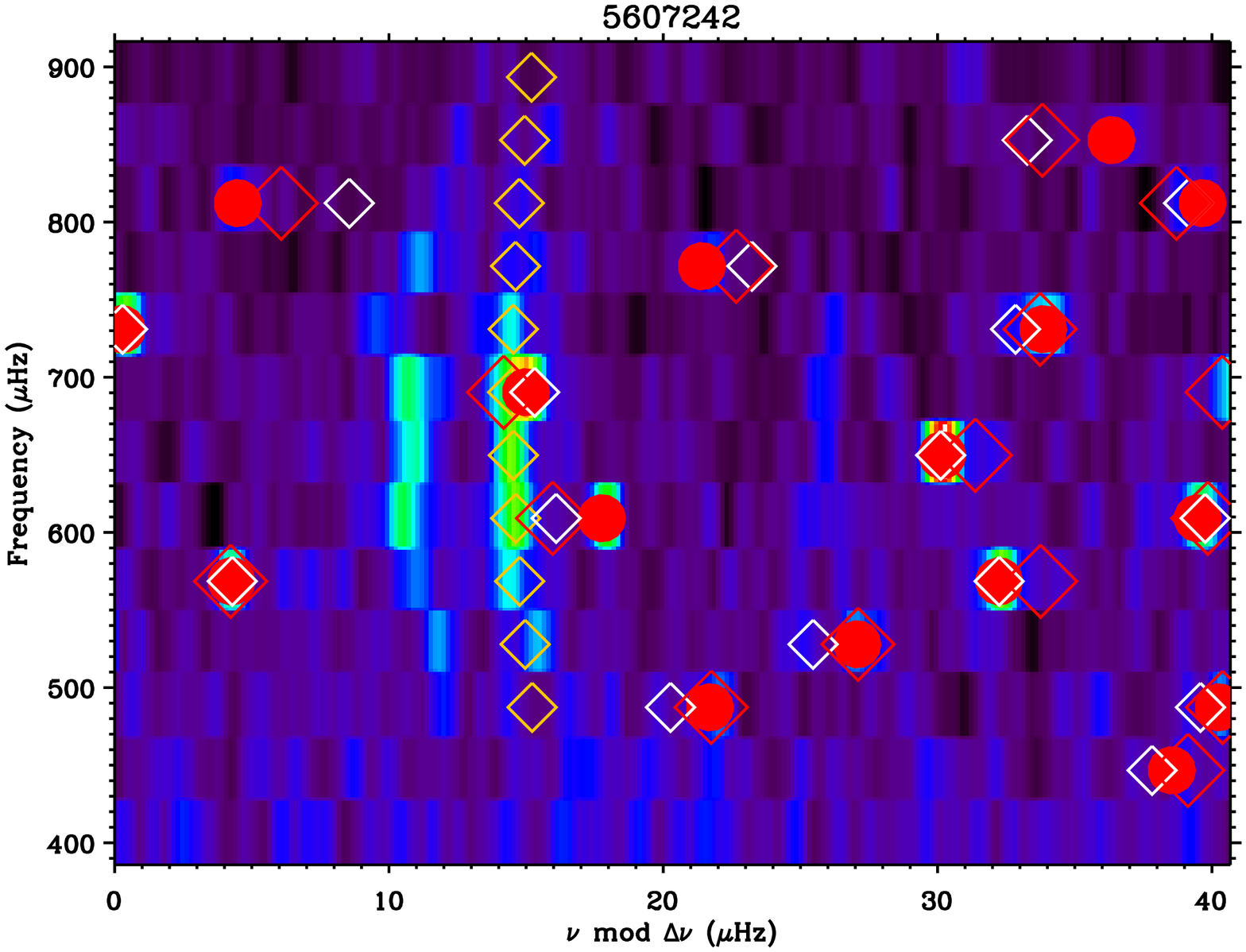}
\includegraphics[width=6 cm,angle=90]{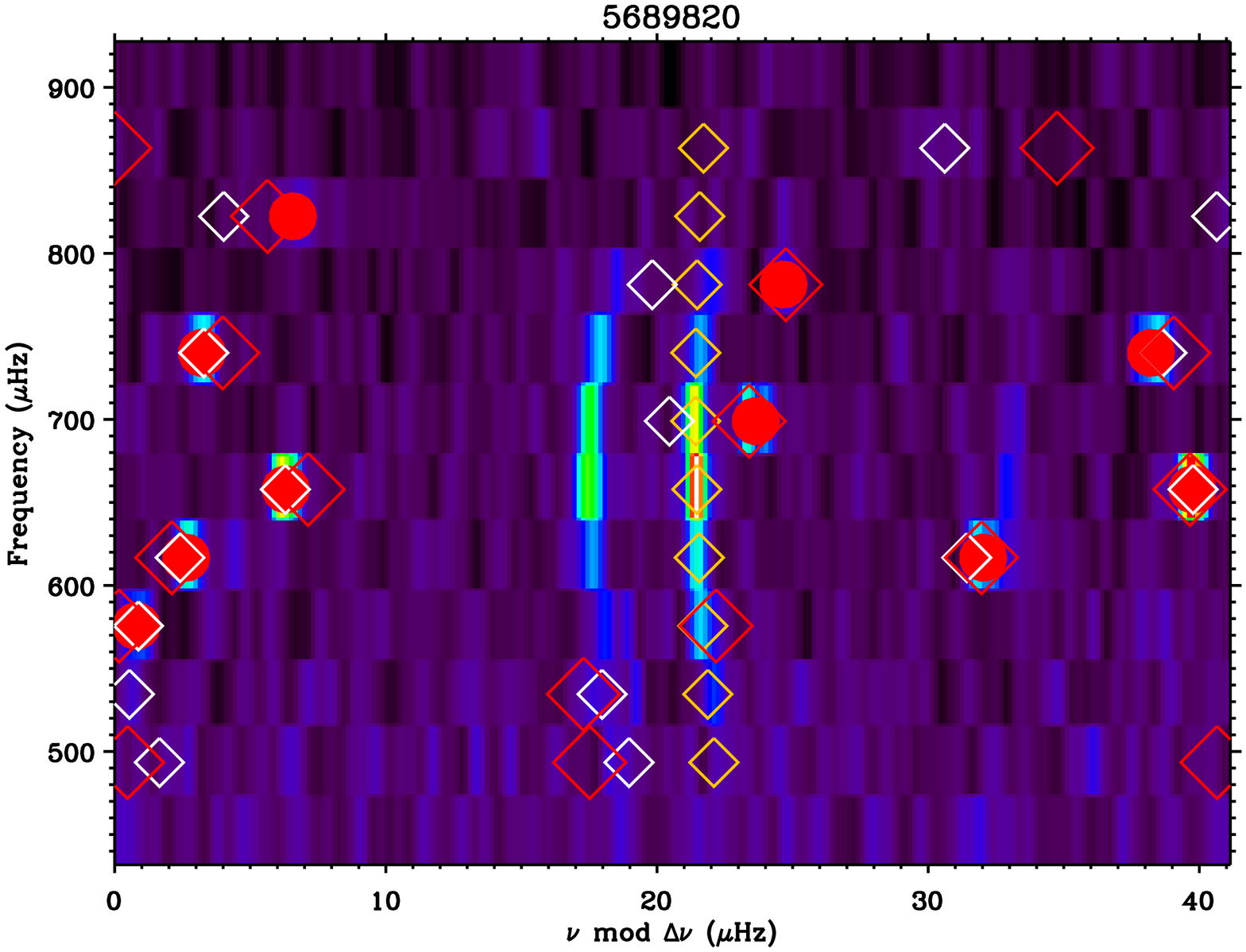}
}
\hbox{
\includegraphics[width=6 cm,angle=90]{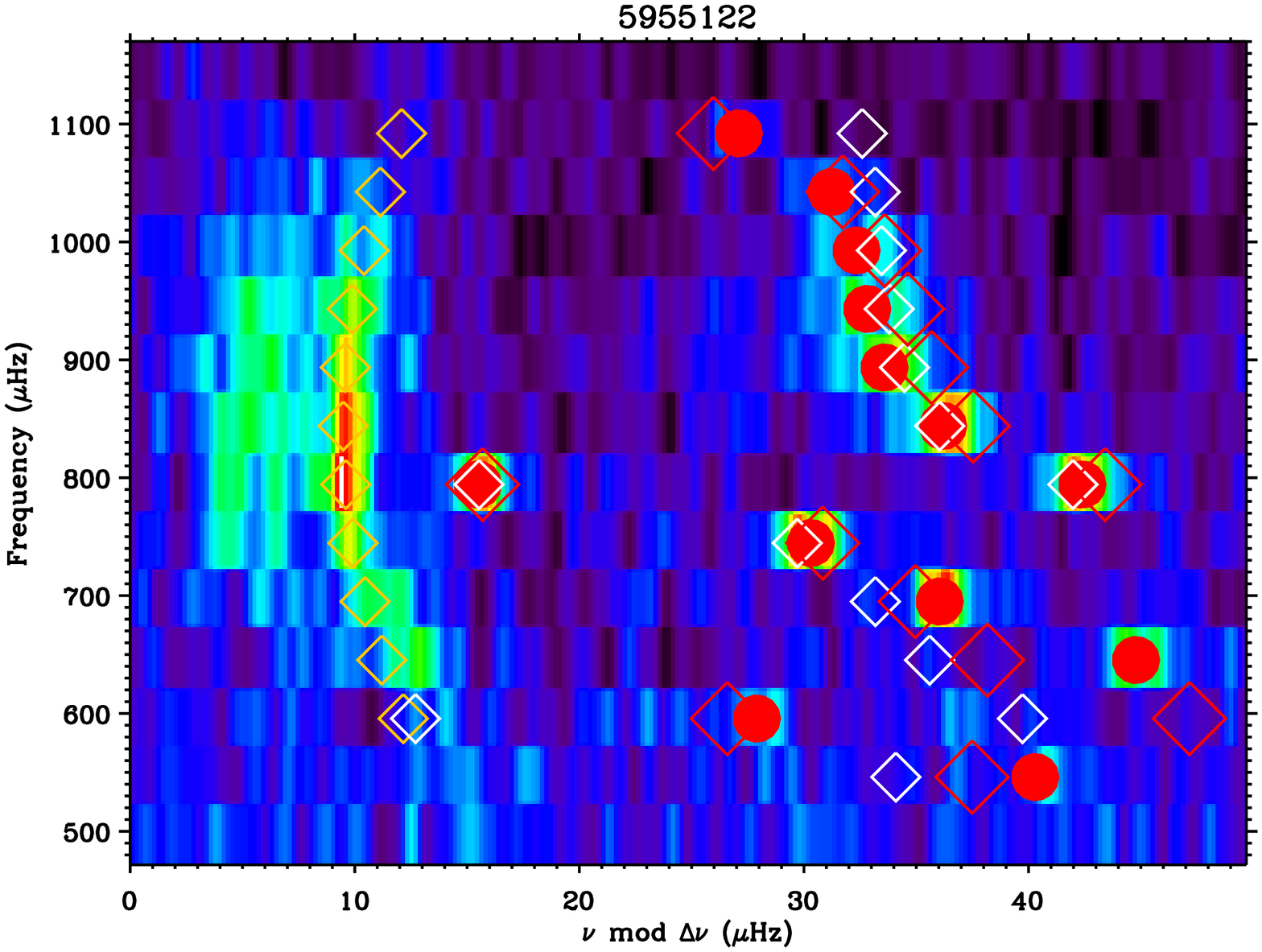}
\includegraphics[width=6 cm,angle=90]{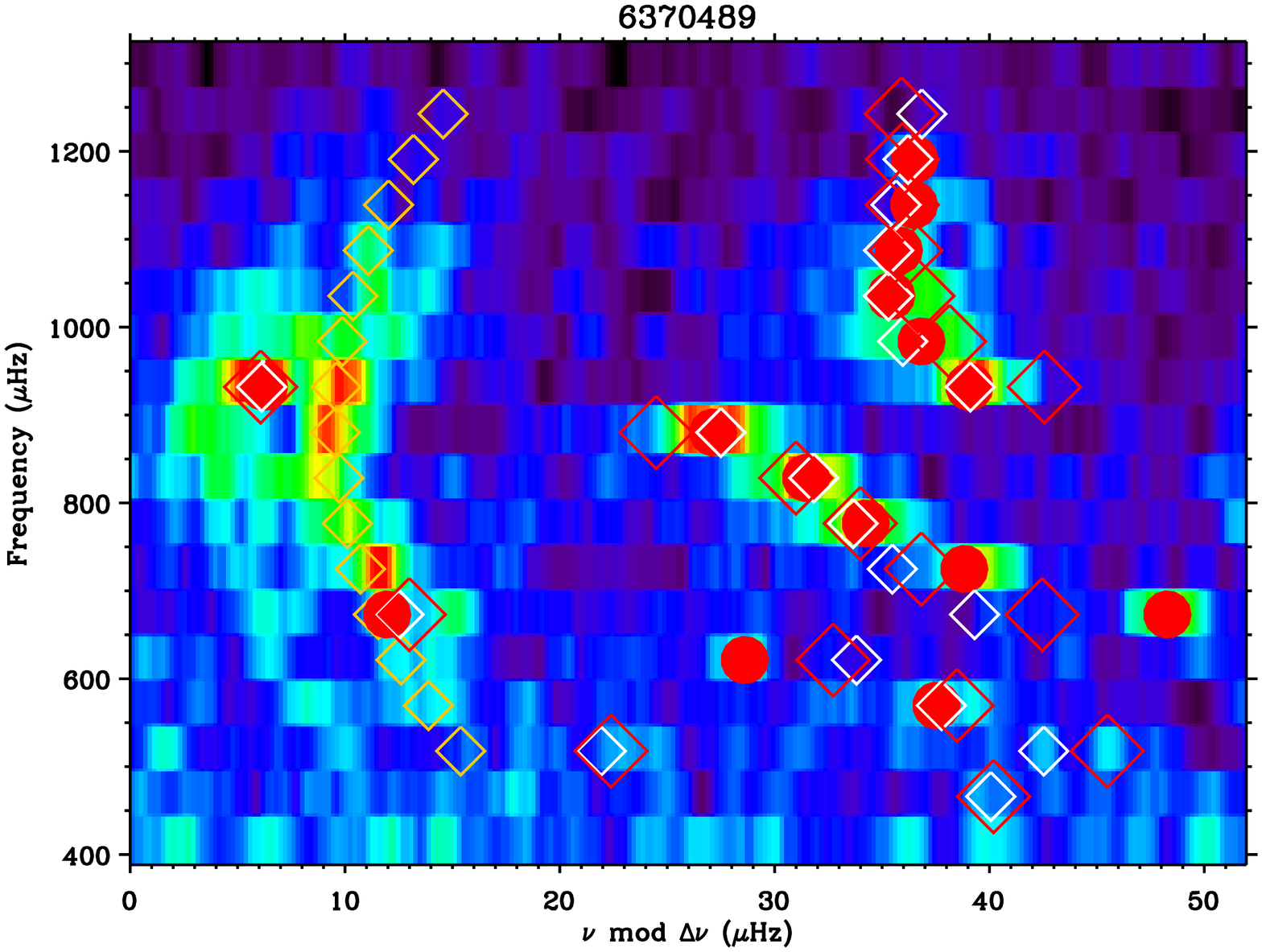}
}
\caption{Echelle diagram for 6 stars.  Frequencies fitted on the power spectra (Red circles), Frequencies from the optimisation: $l=0$ (Orange diamonds), $l=1$ (White diamonds). Frequencies from fitting the asymptotic model on the fitted frequencies (Red diamonds).}
\label{ech_2991448}
\end{figure*}

\begin{figure*}[!tbp]
\centering
\hbox{
\includegraphics[width=6 cm,angle=90]{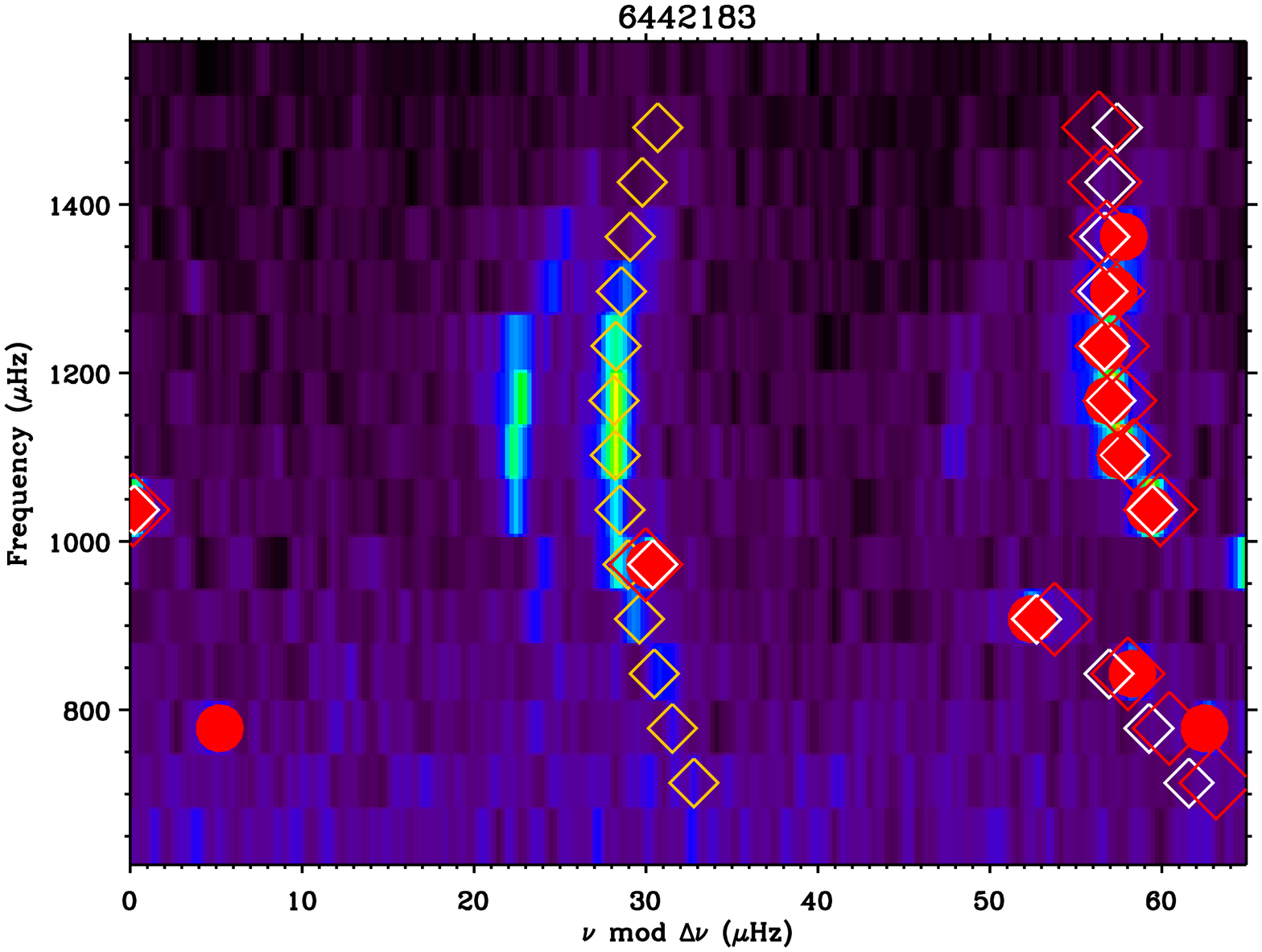}
\includegraphics[width=6 cm,angle=90]{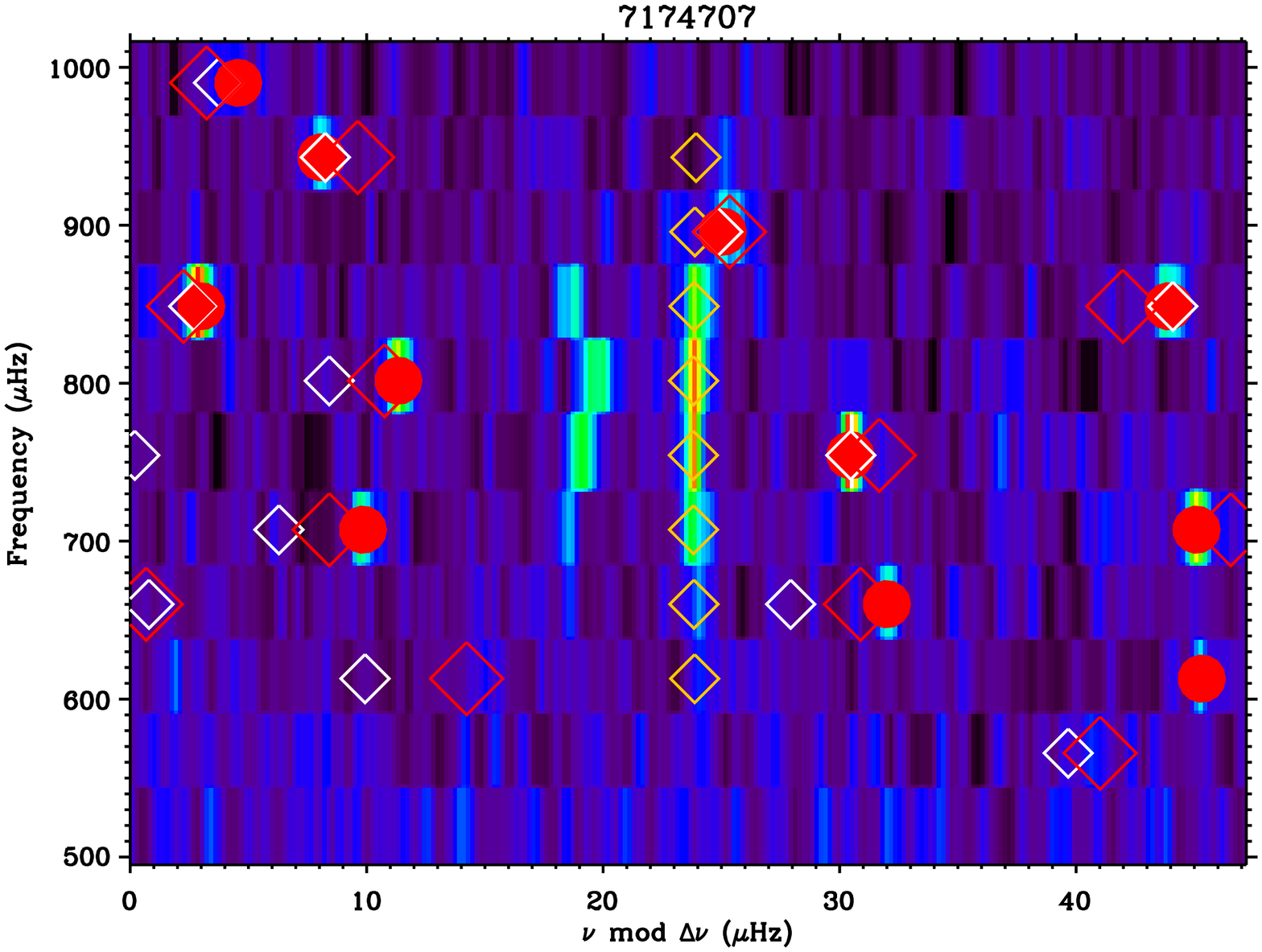}
}
\hbox{

\includegraphics[width=6 cm,angle=90]{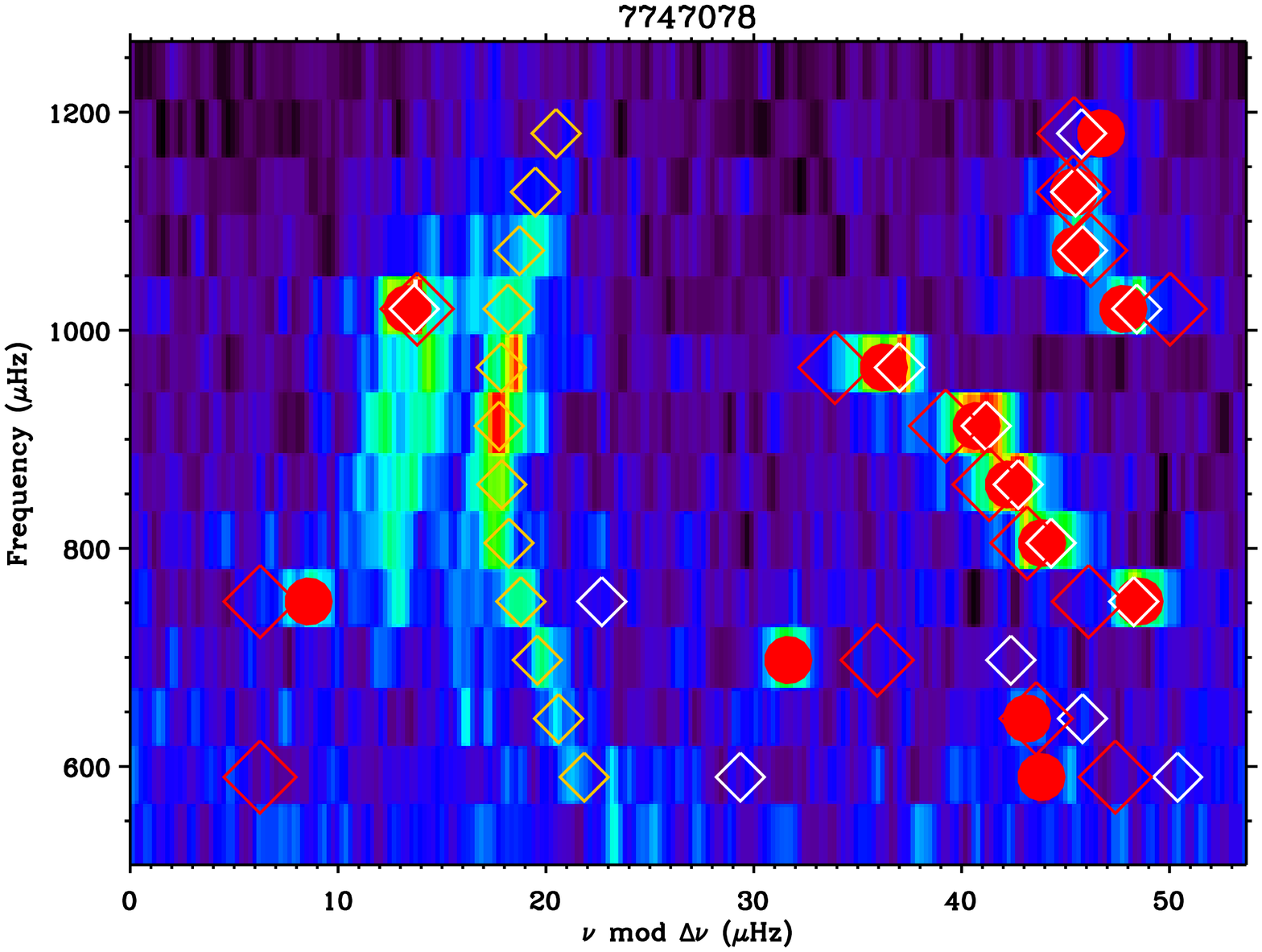}
\includegraphics[width=6 cm,angle=90]{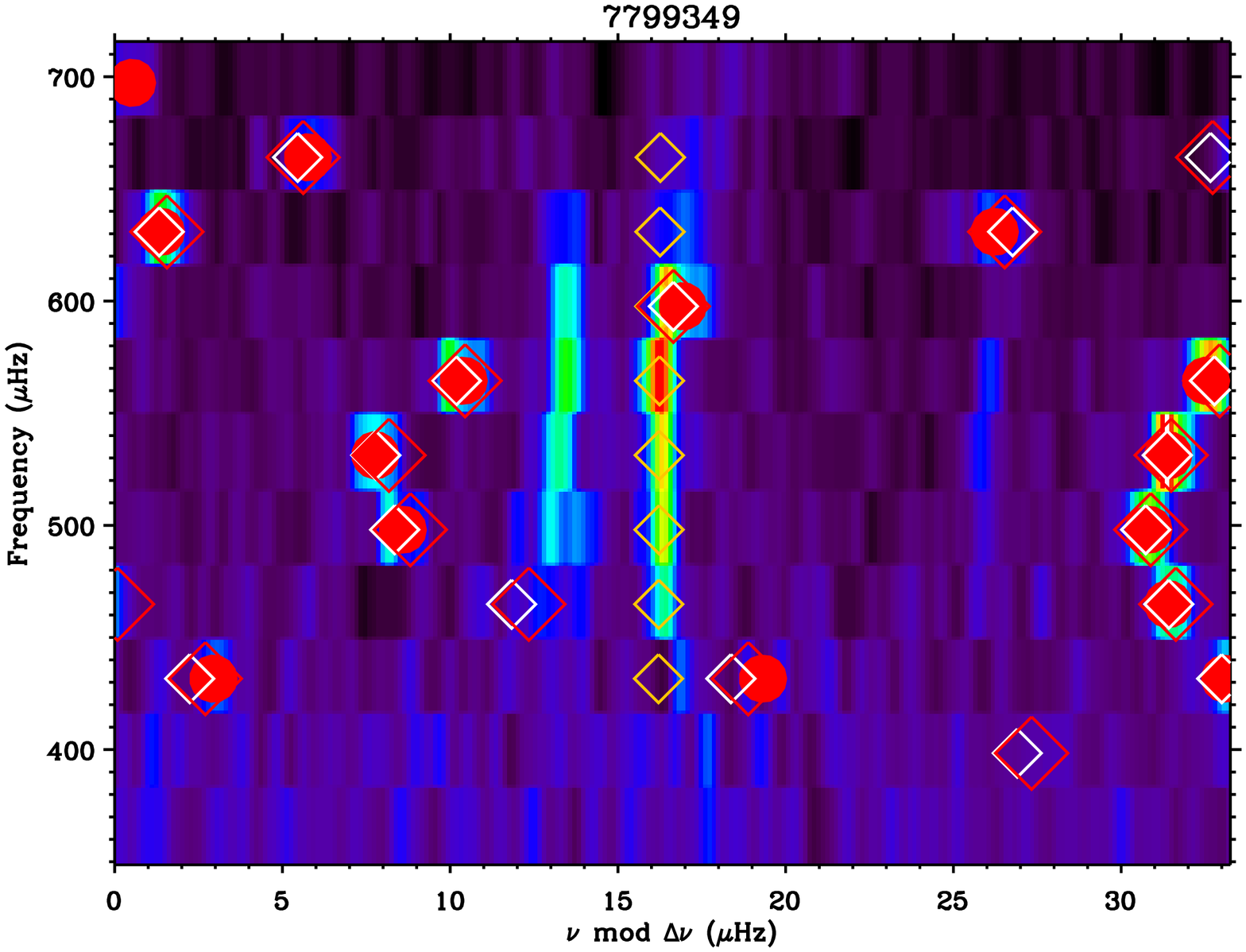}
}
\hbox{

\includegraphics[width=6 cm,angle=90]{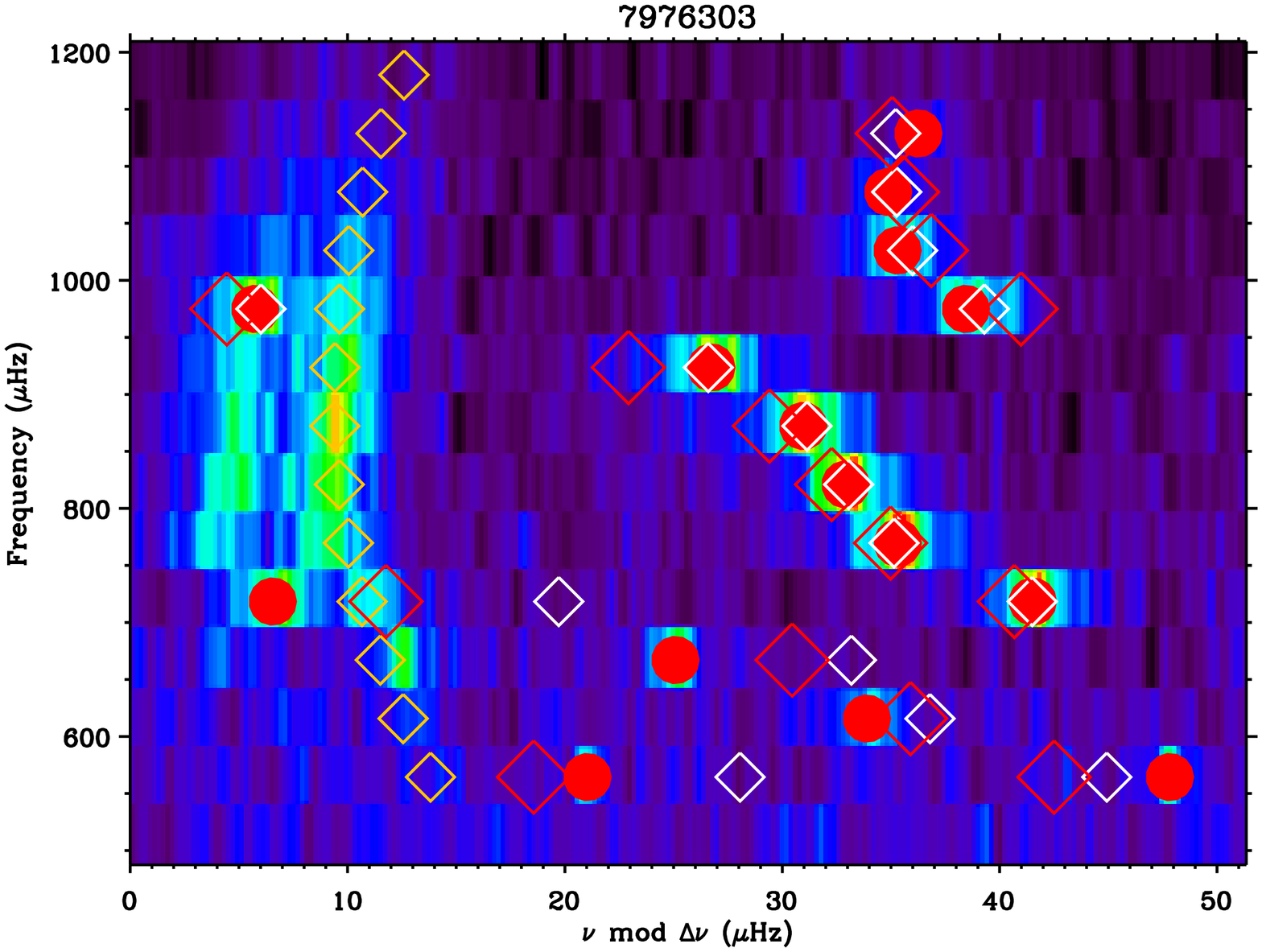}
\includegraphics[width=6 cm,angle=90]{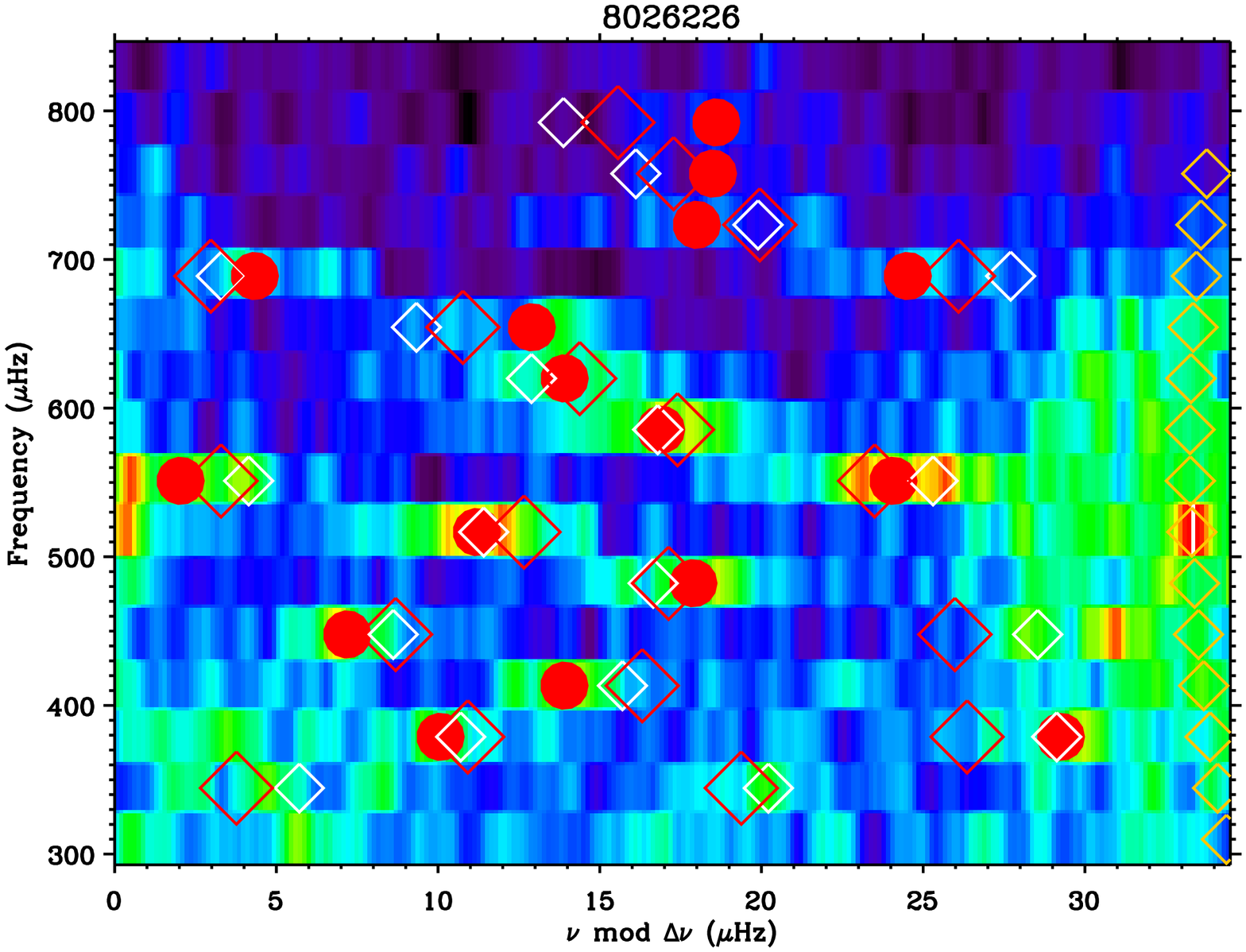}
}
\caption{Echelle diagram for 6 stars.  Frequencies fitted on the power spectra (Red circles), Frequencies from the optimisation: $l=0$ (Orange diamonds), $l=1$ (White diamonds).  Frequencies from fitting the asymptotic model on the fitted frequencies (Red diamonds).}
\label{ech_6442183}
\end{figure*}

\begin{figure*}[!tbp]
\centering
\hbox{

\includegraphics[width=6 cm,angle=90]{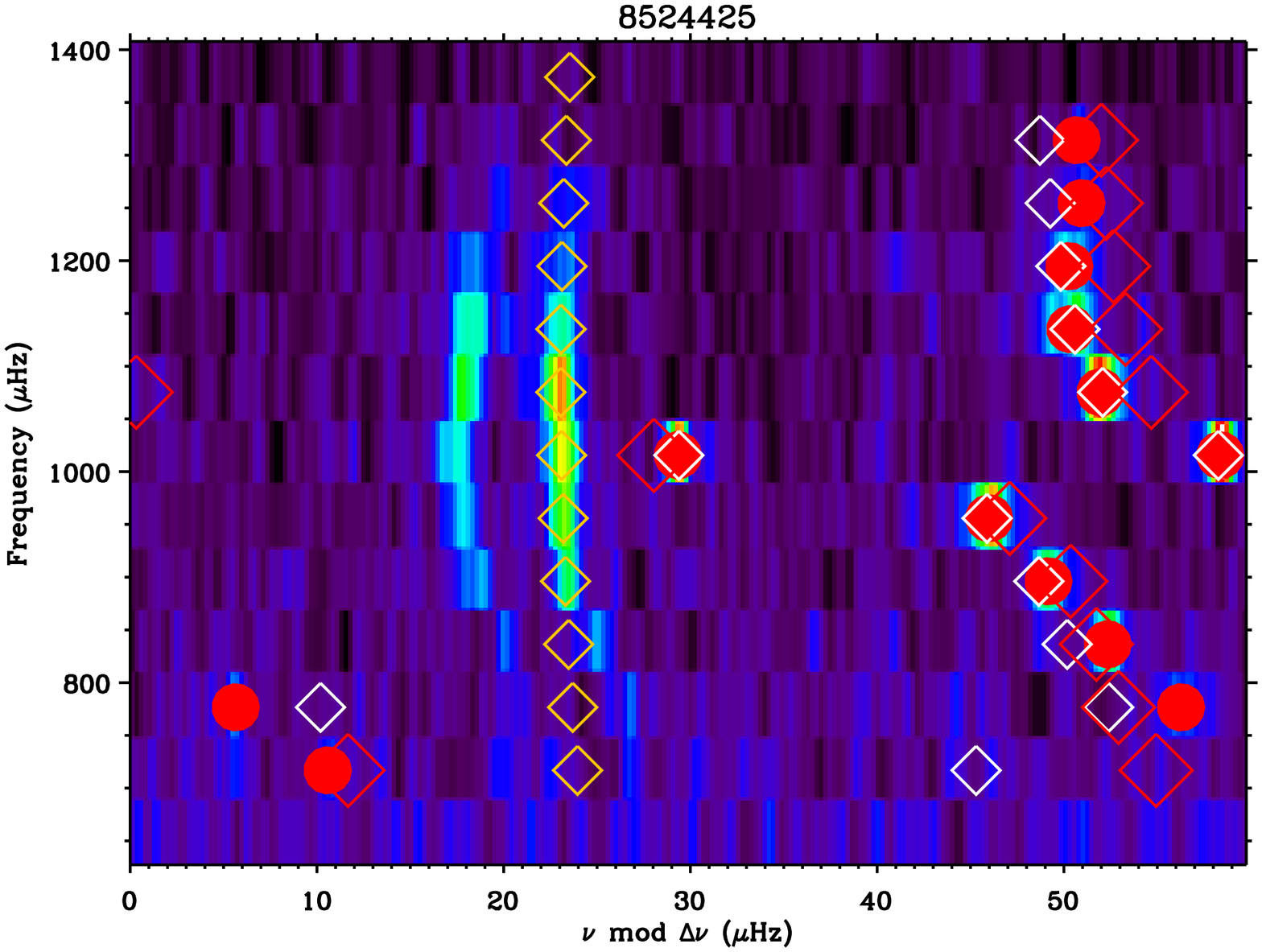}
\includegraphics[width=6 cm,angle=90]{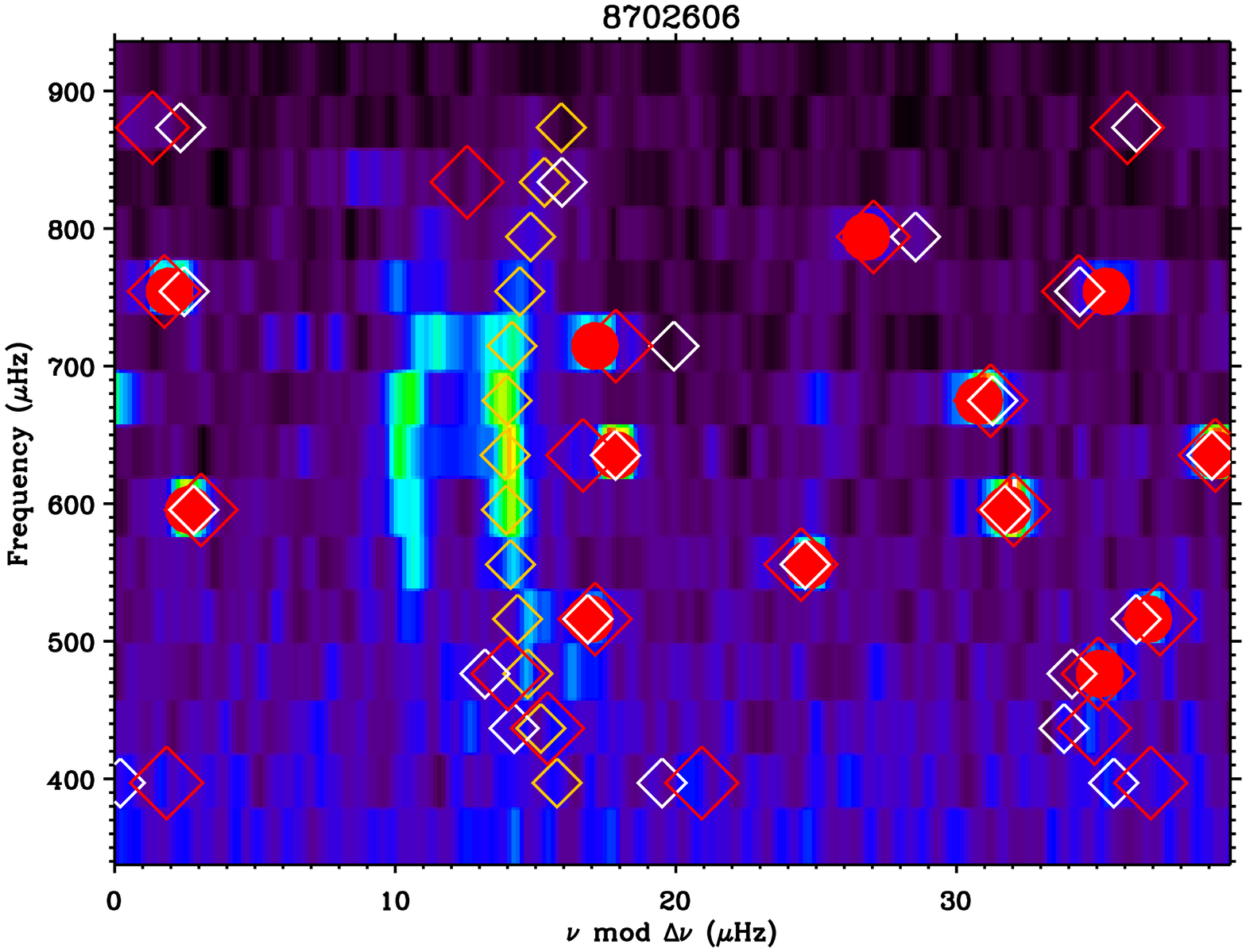}
}
\hbox{

\includegraphics[width=6 cm,angle=90]{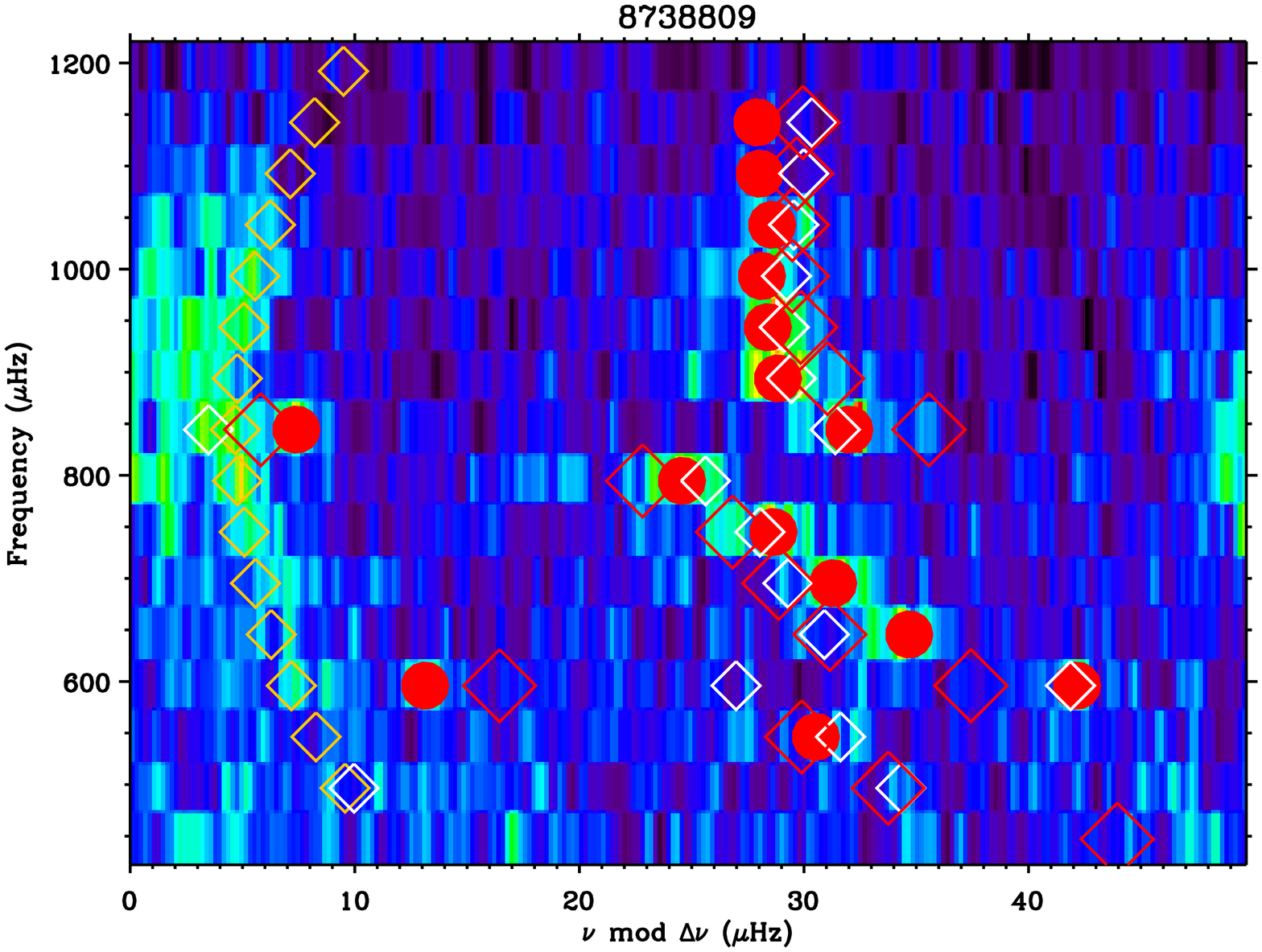}
\includegraphics[width=6 cm,angle=90]{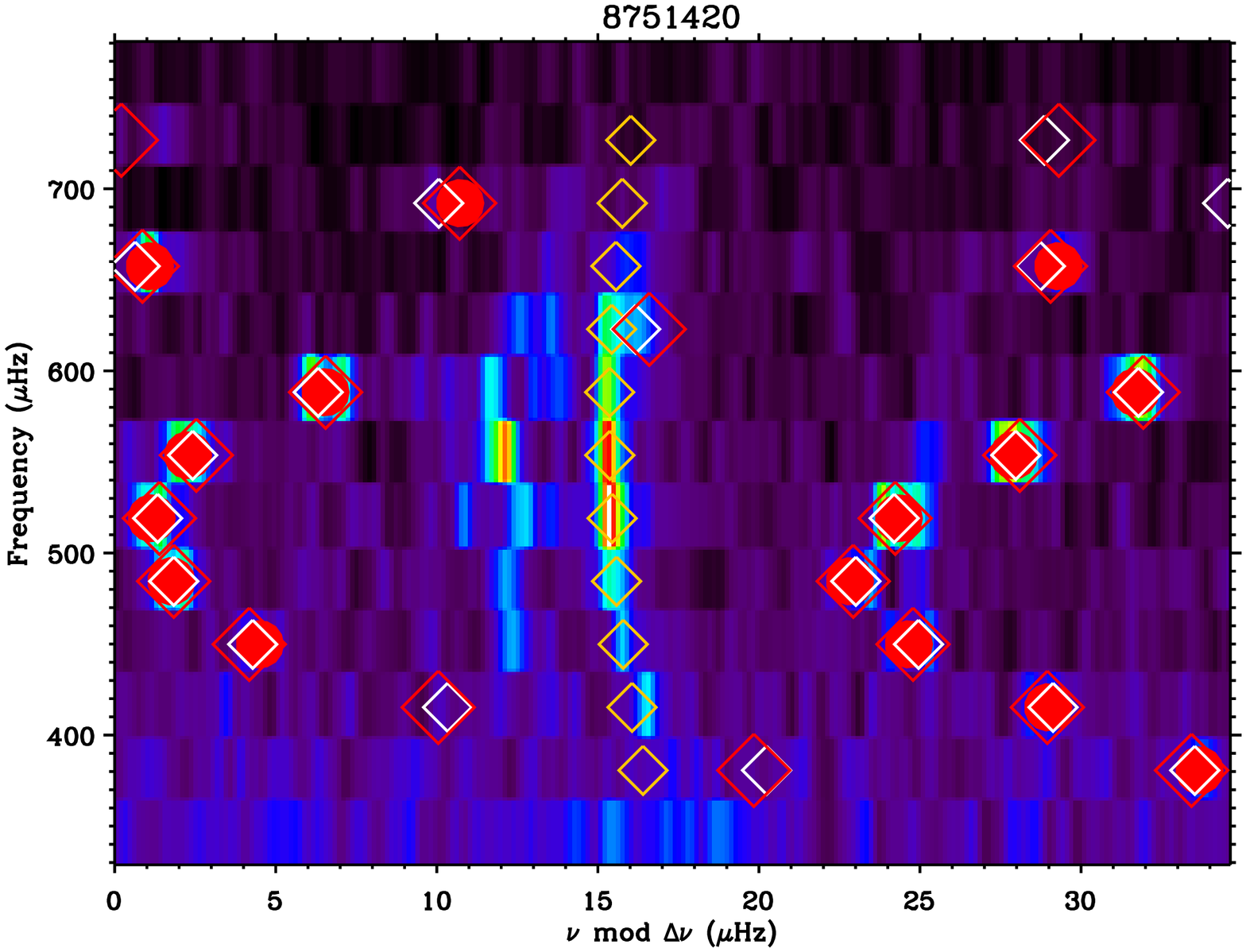}
}
\hbox{

\includegraphics[width=6 cm,angle=90]{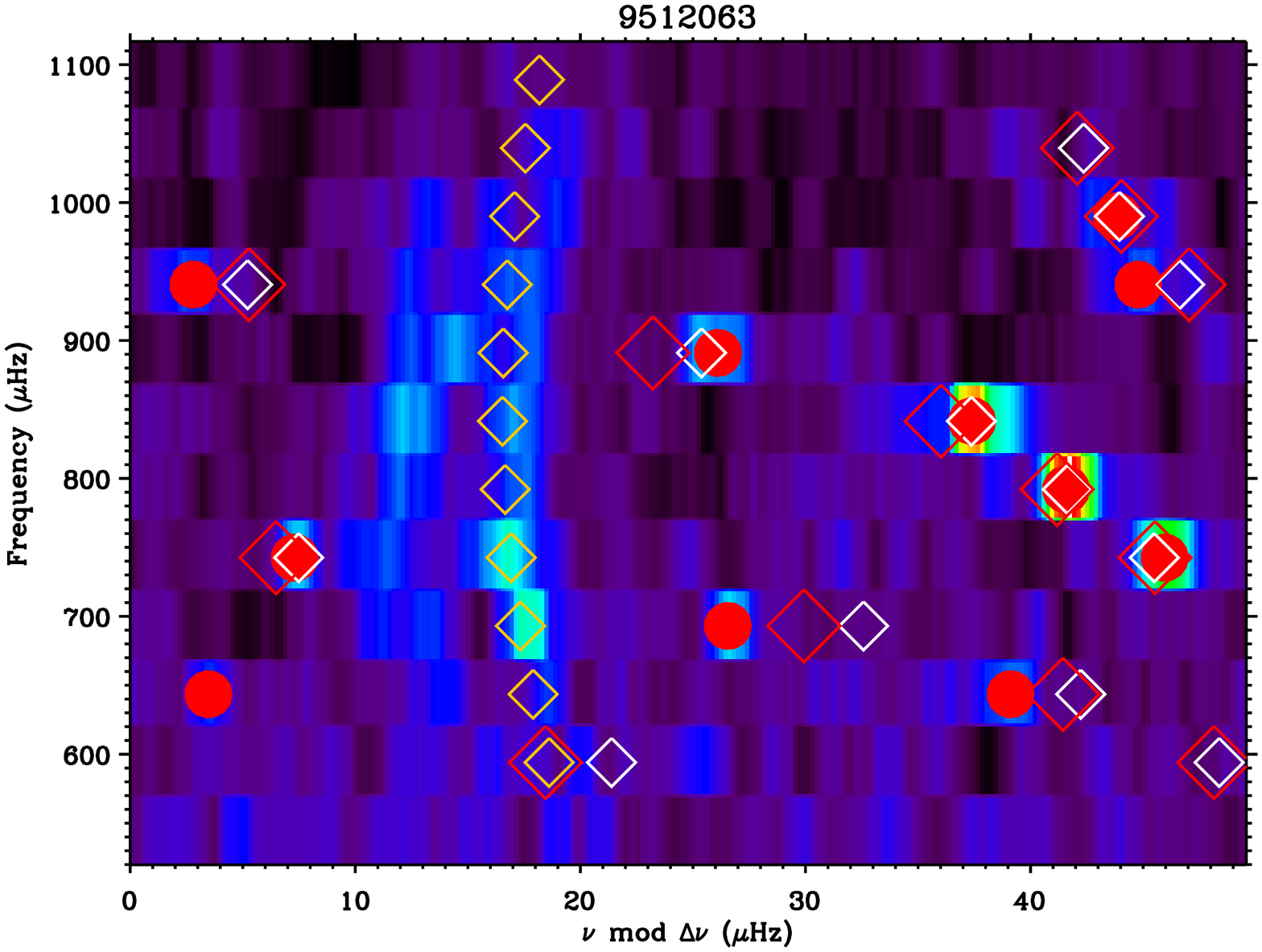}
\includegraphics[width=6 cm,angle=90]{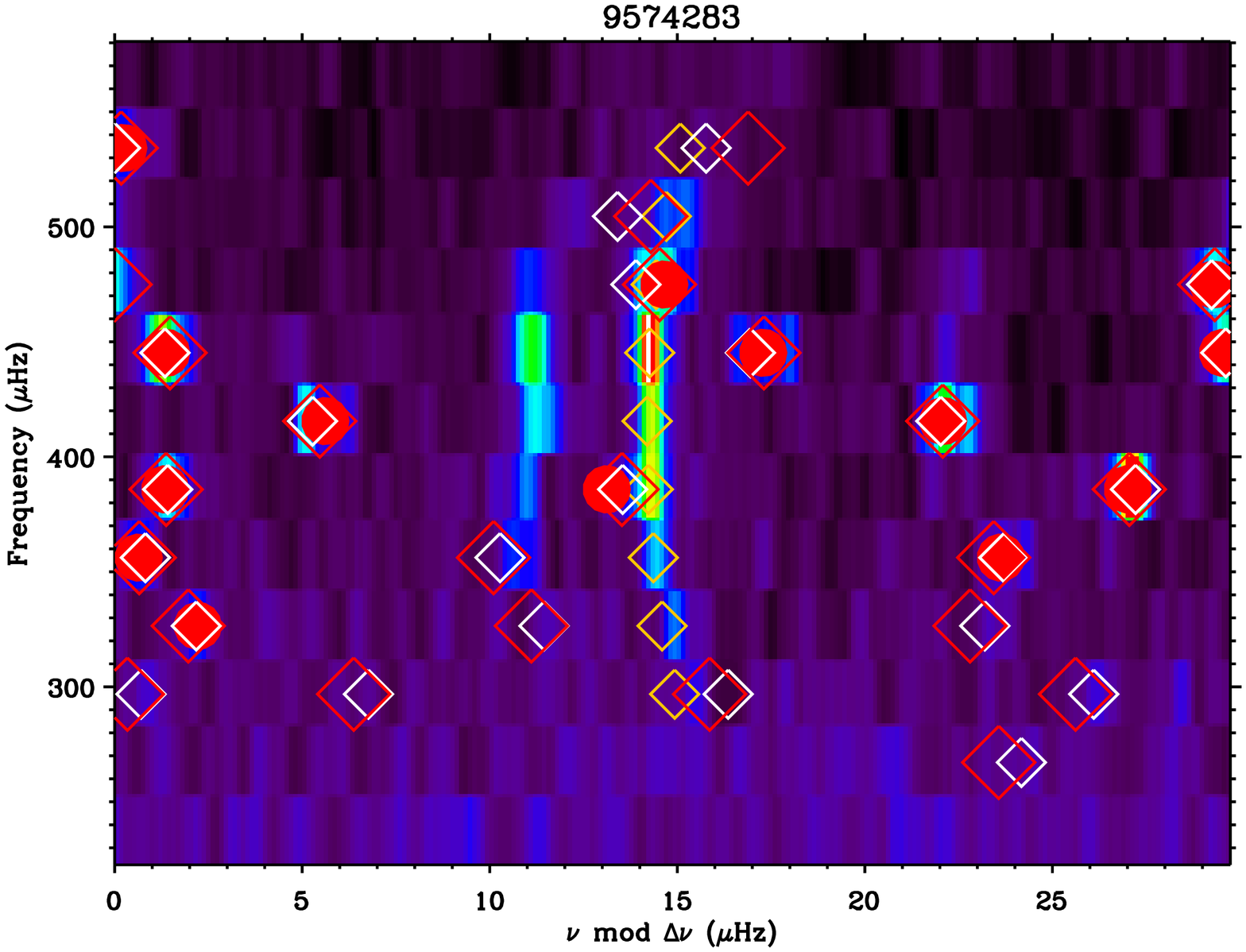}
}
\caption{Echelle diagram for 6 stars.  Frequencies fitted on the power spectra (Red circles), Frequencies from the optimisation: $l=0$ (Orange diamonds), $l=1$ (White diamonds).  Frequencies from fitting the asymptotic model on the fitted frequencies (Red diamonds).}
\label{ech_8524425}
\end{figure*}

\begin{figure*}[!tbp]
\centering
\hbox{

\includegraphics[width=6 cm,angle=90]{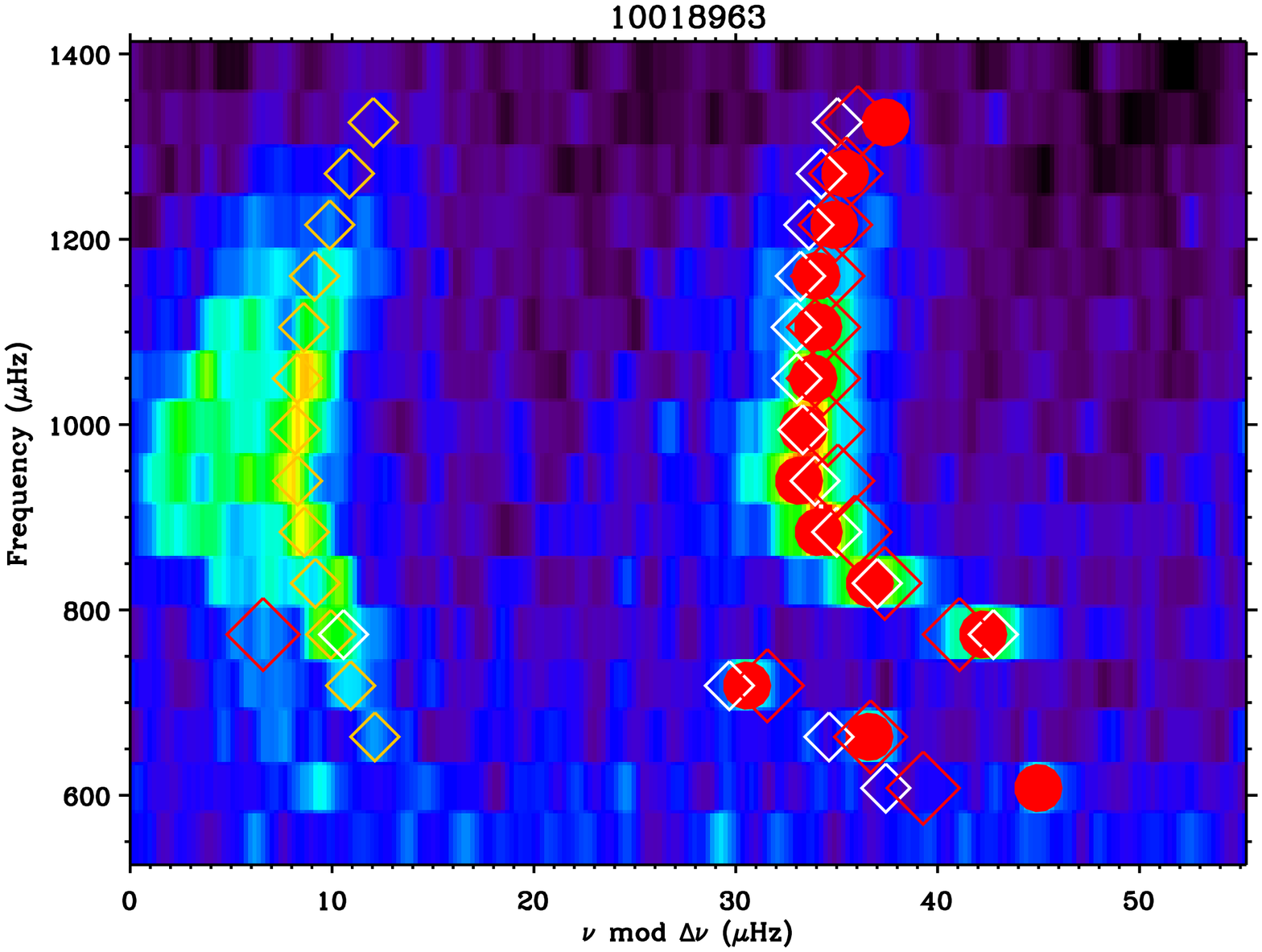}
\includegraphics[width=6 cm,angle=90]{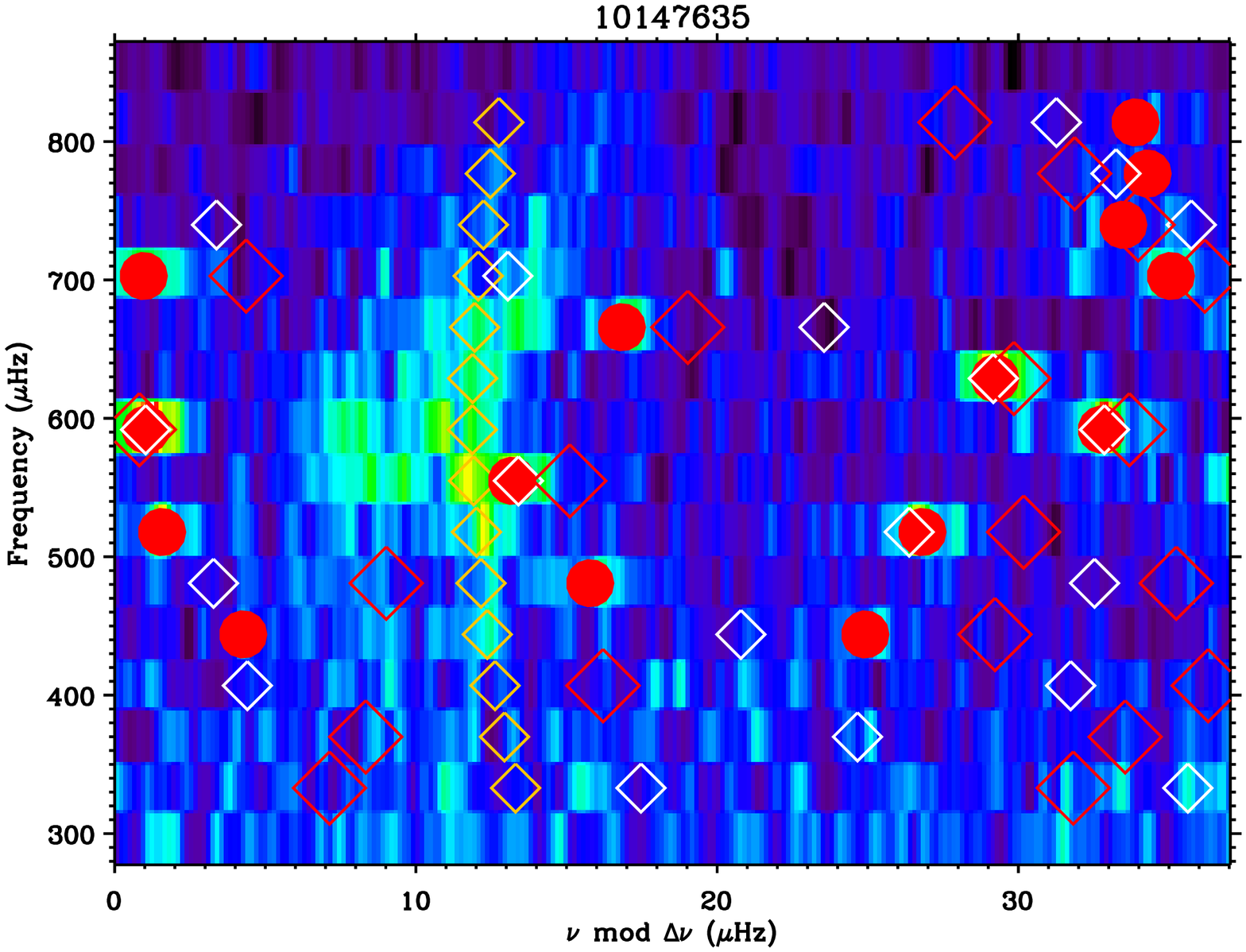}
}
\hbox{

\includegraphics[width=6 cm,angle=90]{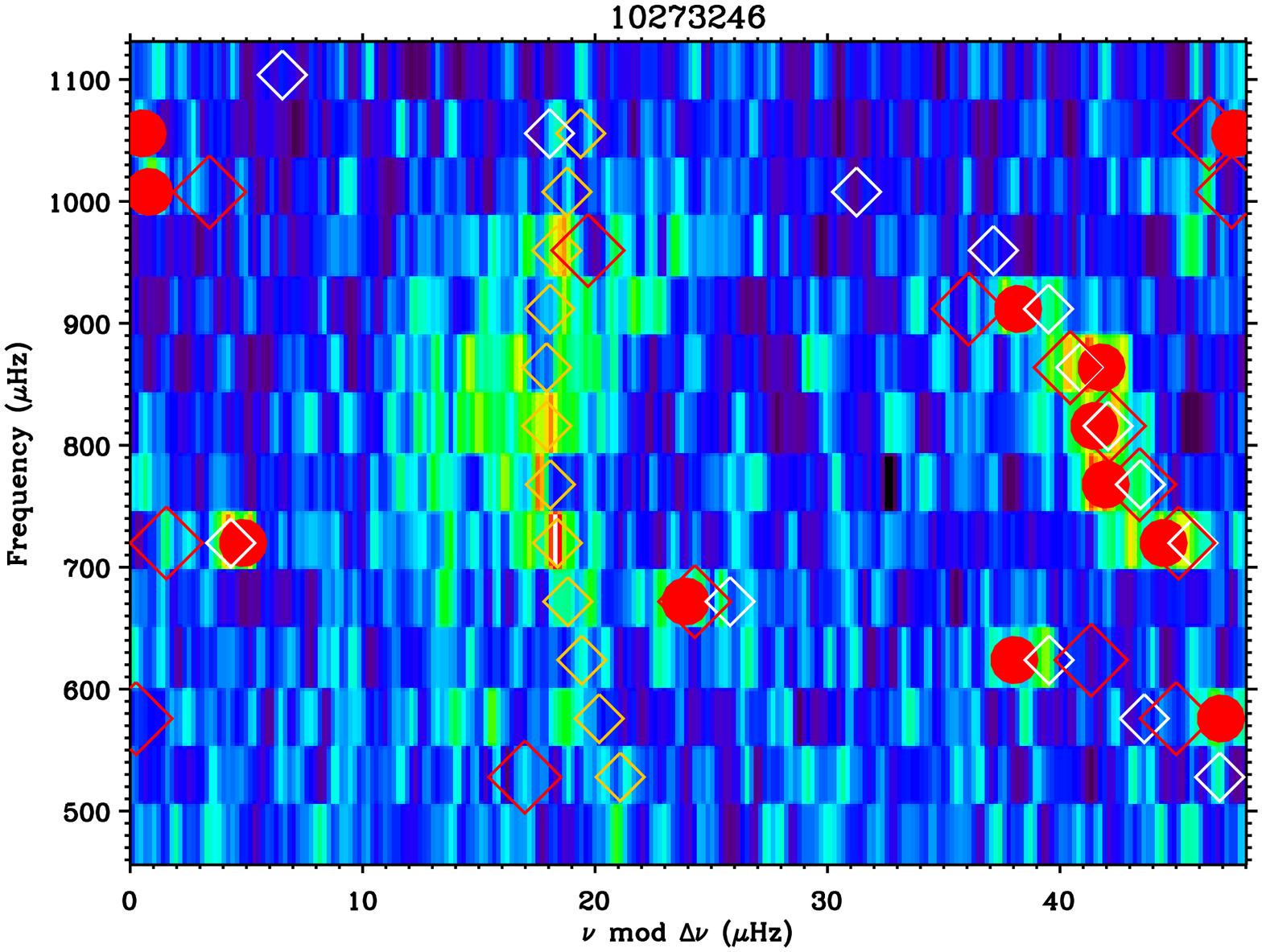}
\includegraphics[width=6 cm,angle=90]{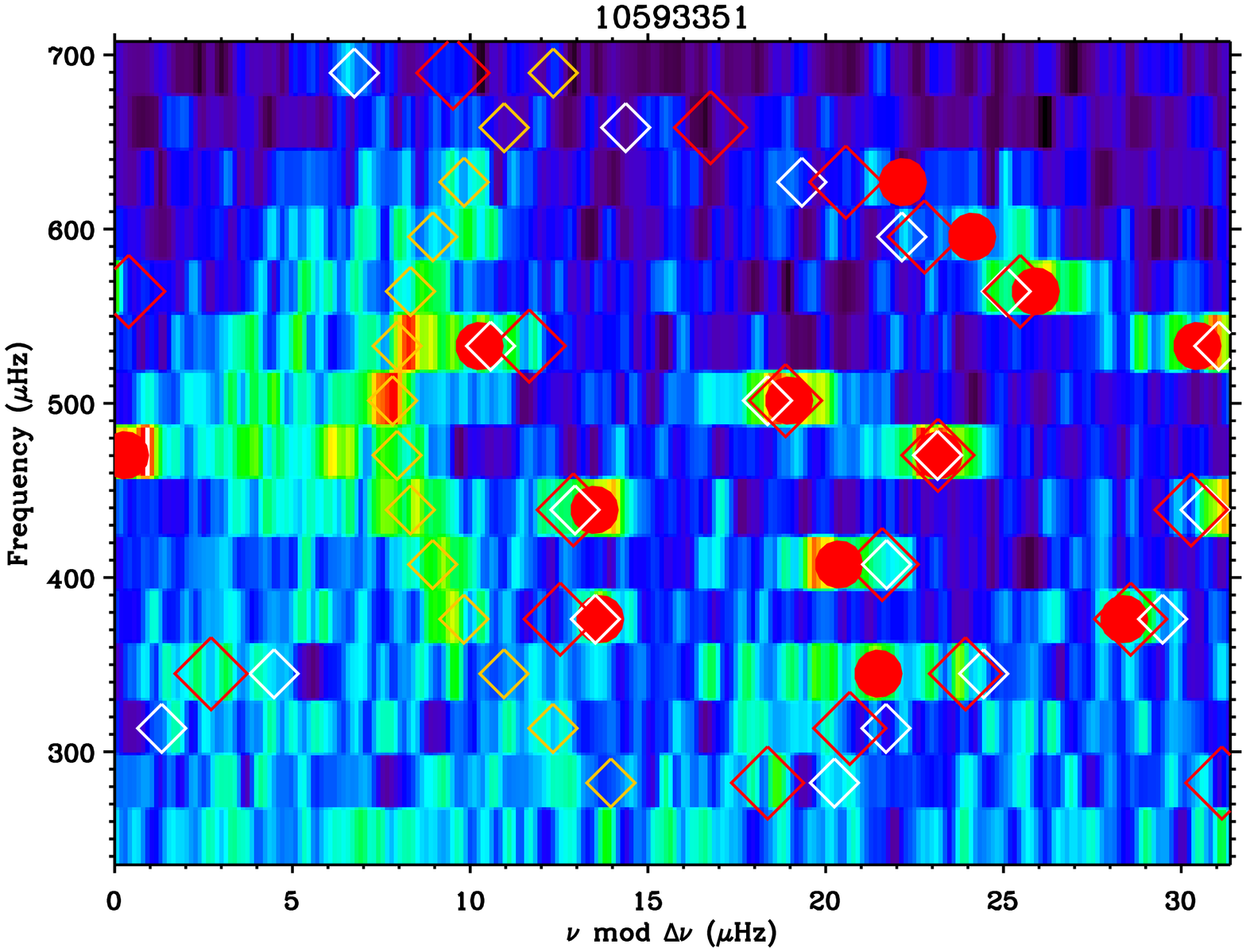}
}
\hbox{

\includegraphics[width=6 cm,angle=90]{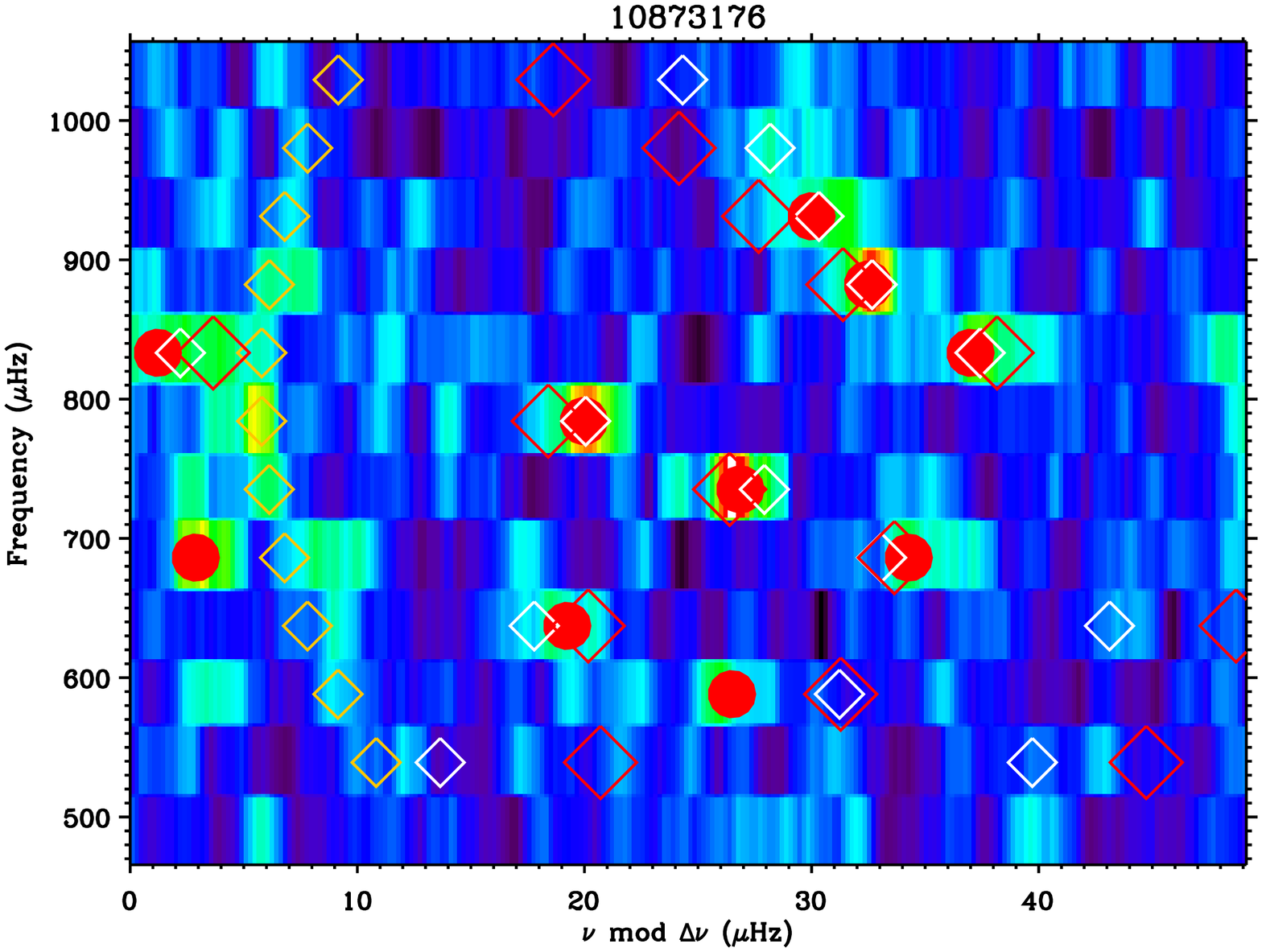}
\includegraphics[width=6 cm,angle=90]{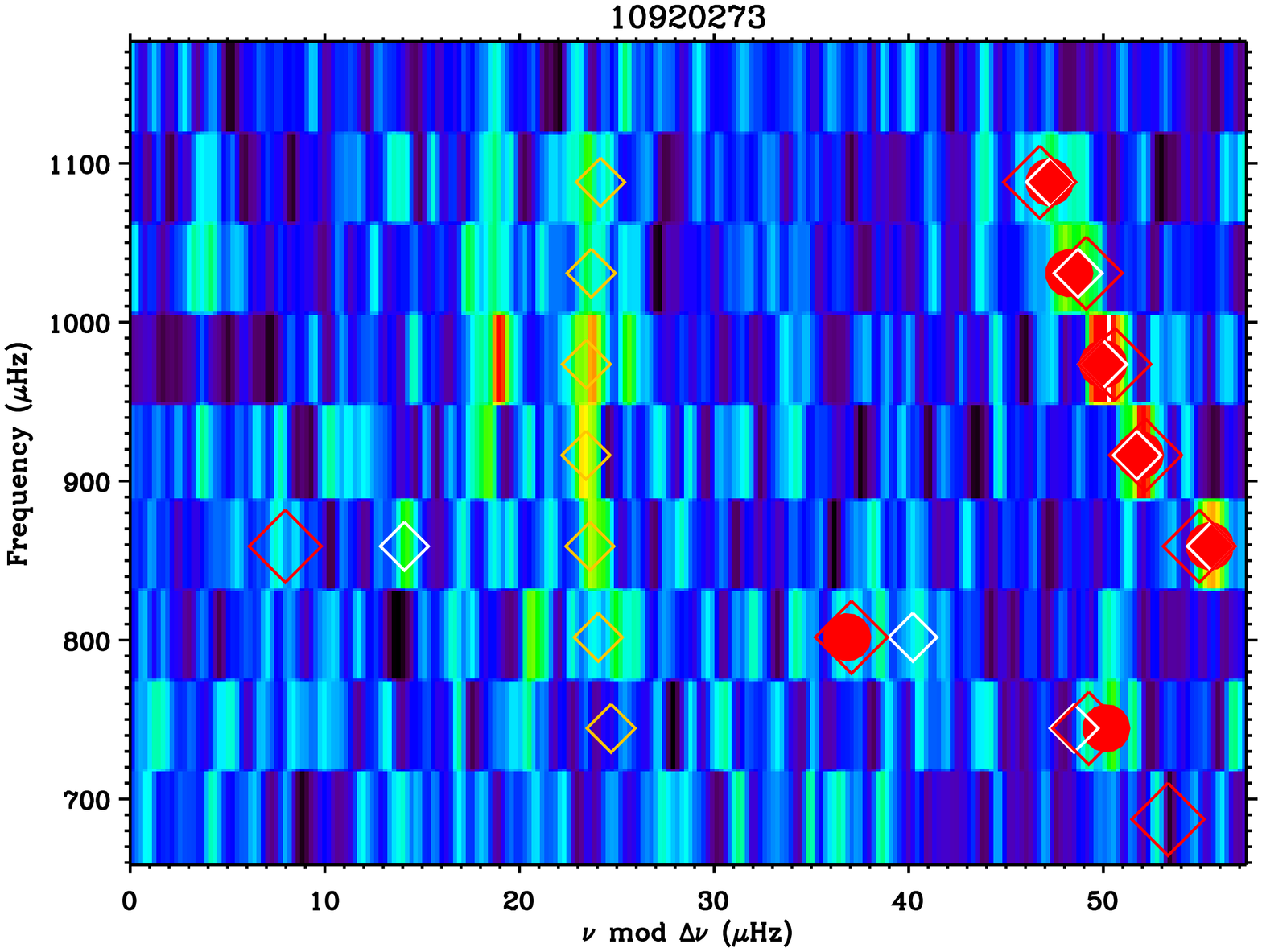}
}
\caption{Echelle diagram for 6 stars.  Frequencies fitted on the power spectra (Red circles), Frequencies from the optimisation: $l=0$ (Orange diamonds), $l=1$ (White diamonds).  Frequencies from fitting the asymptotic model on the fitted frequencies (Red diamonds).}
\label{ech_10018963}
\end{figure*}

\begin{figure*}[!tbp]
\centering
\hbox{

\includegraphics[width=6 cm,angle=90]{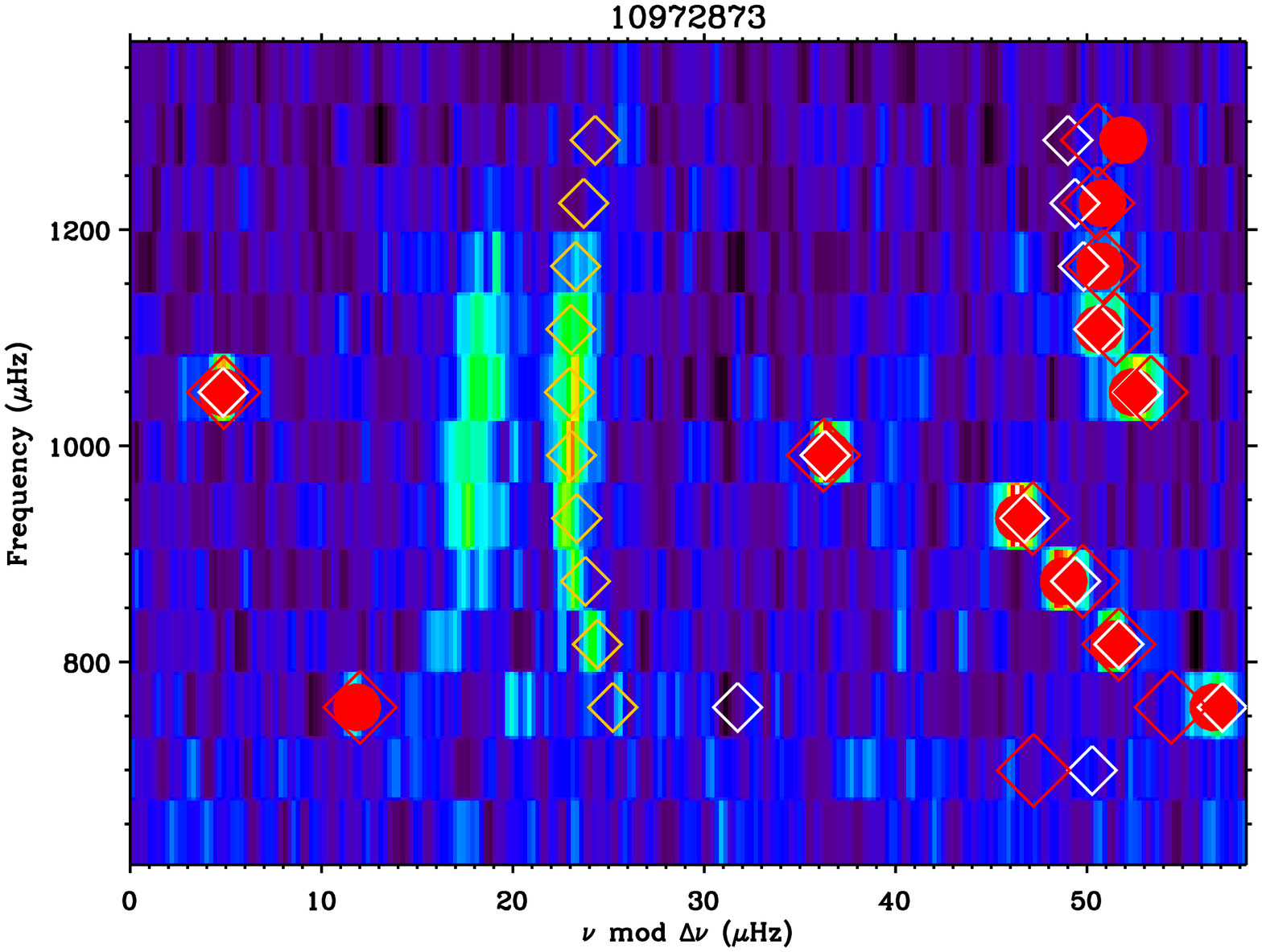}
\includegraphics[width=6 cm,angle=90]{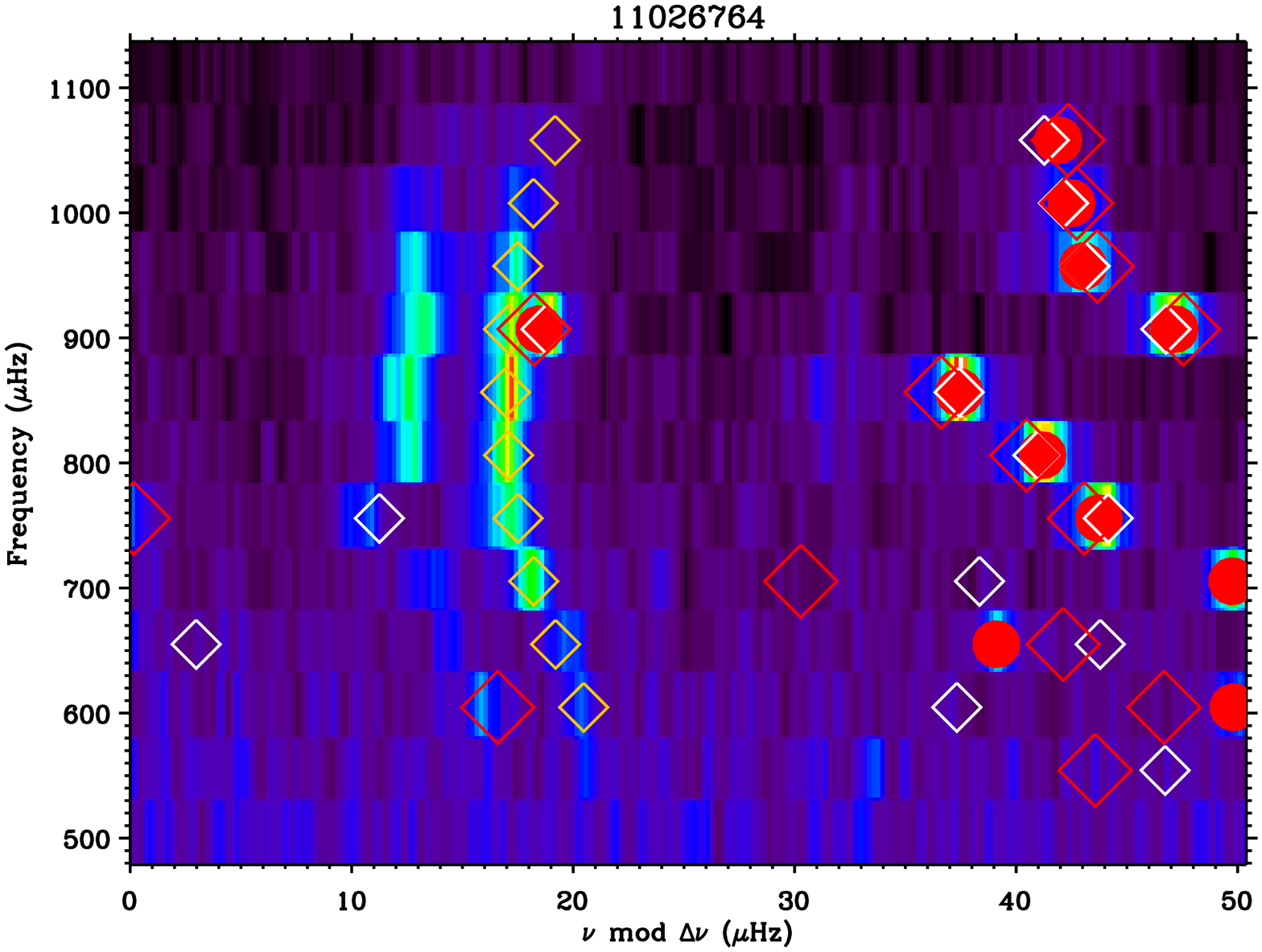}
}
\hbox{

\includegraphics[width=6 cm,angle=90]{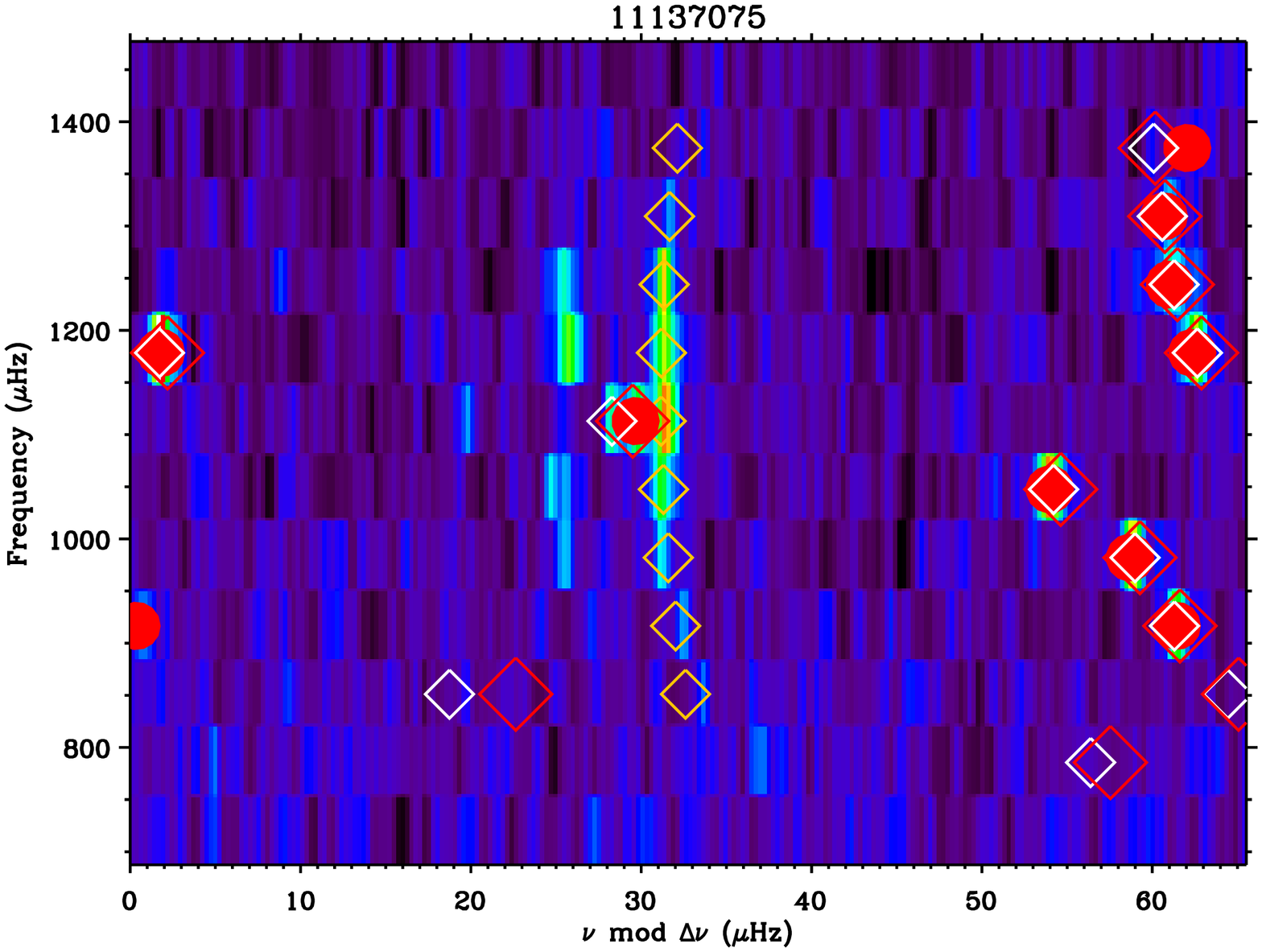}
\includegraphics[width=6 cm,angle=90]{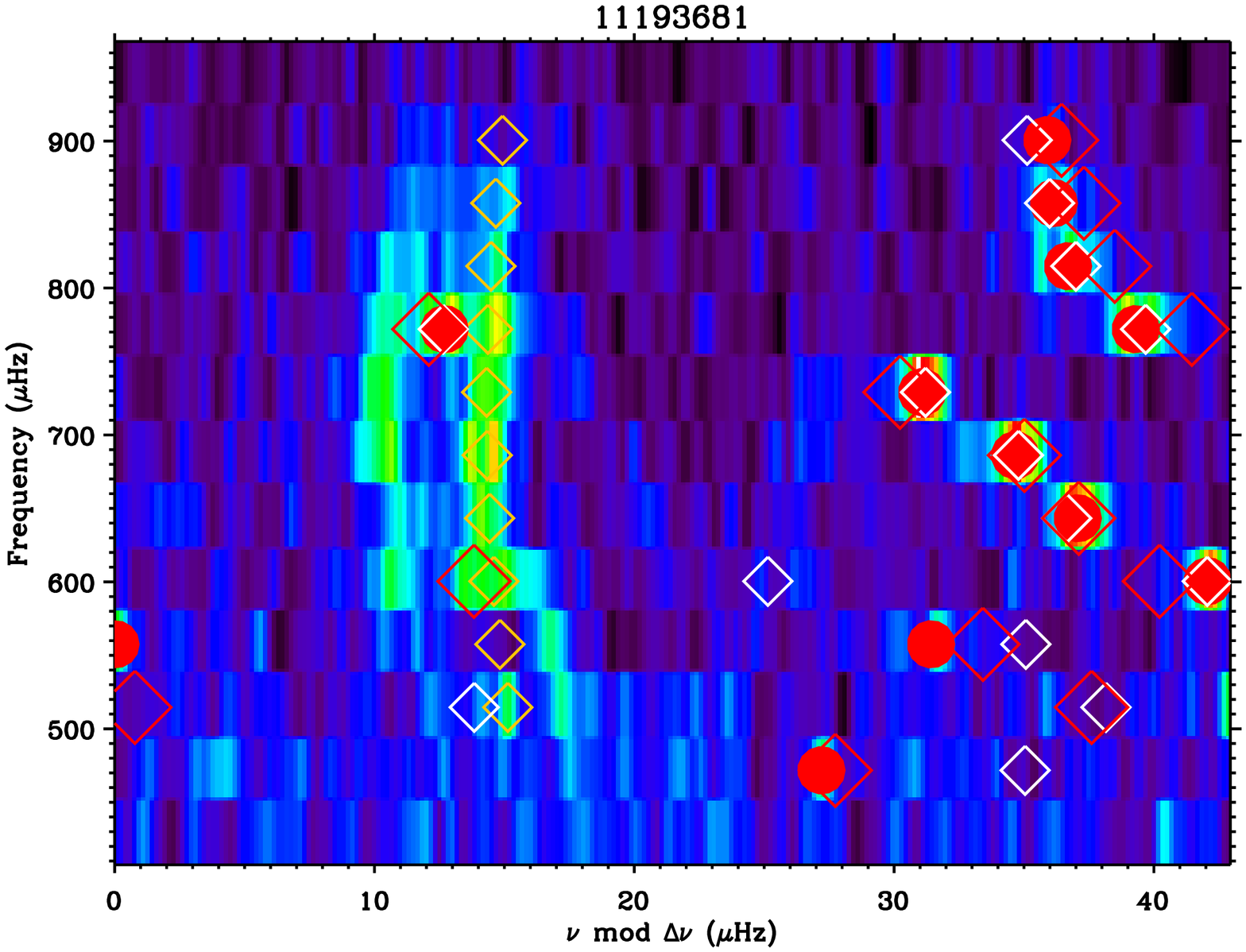}
}
\hbox{

\includegraphics[width=6 cm,angle=90]{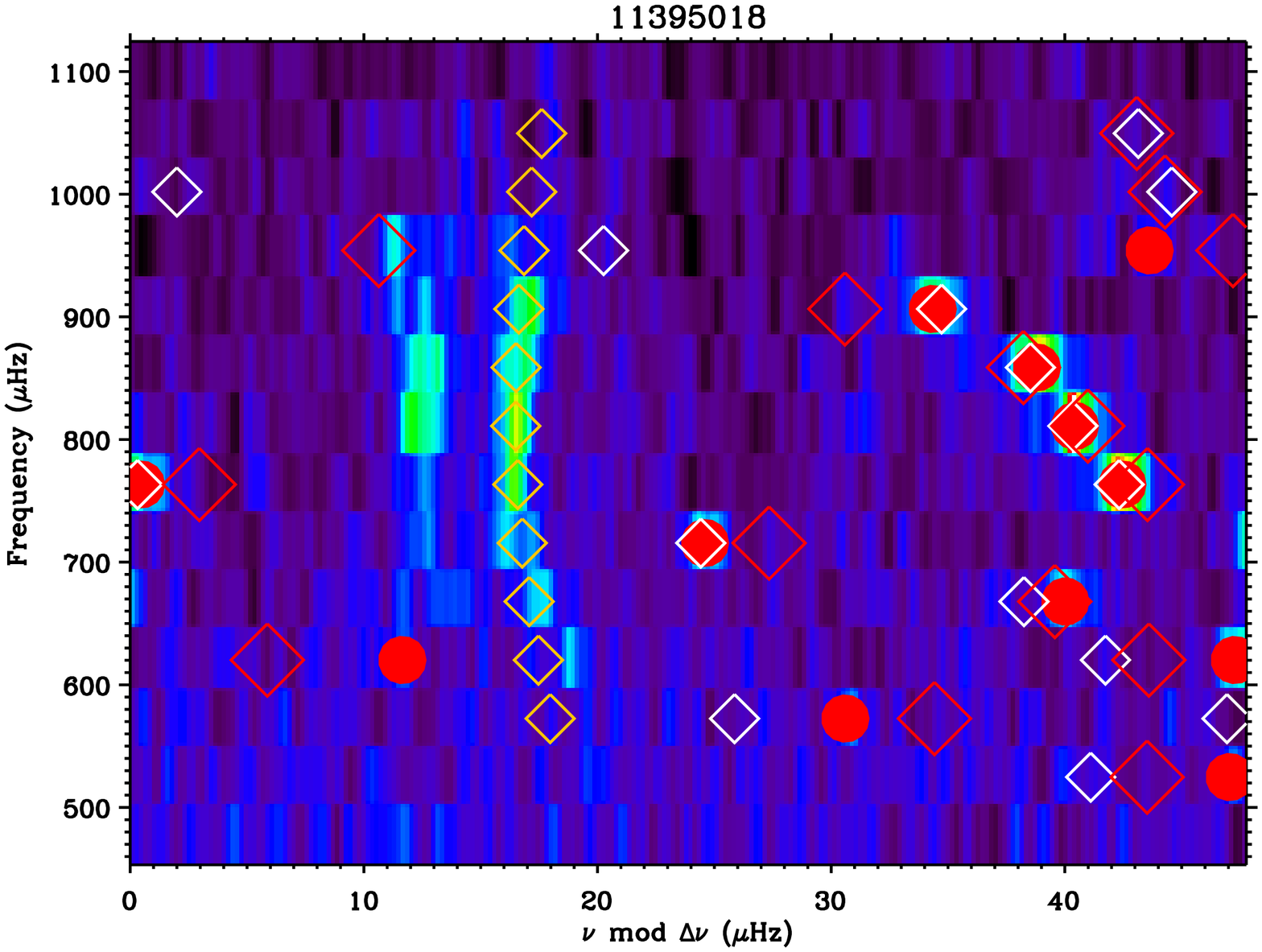}
\includegraphics[width=6 cm,angle=90]{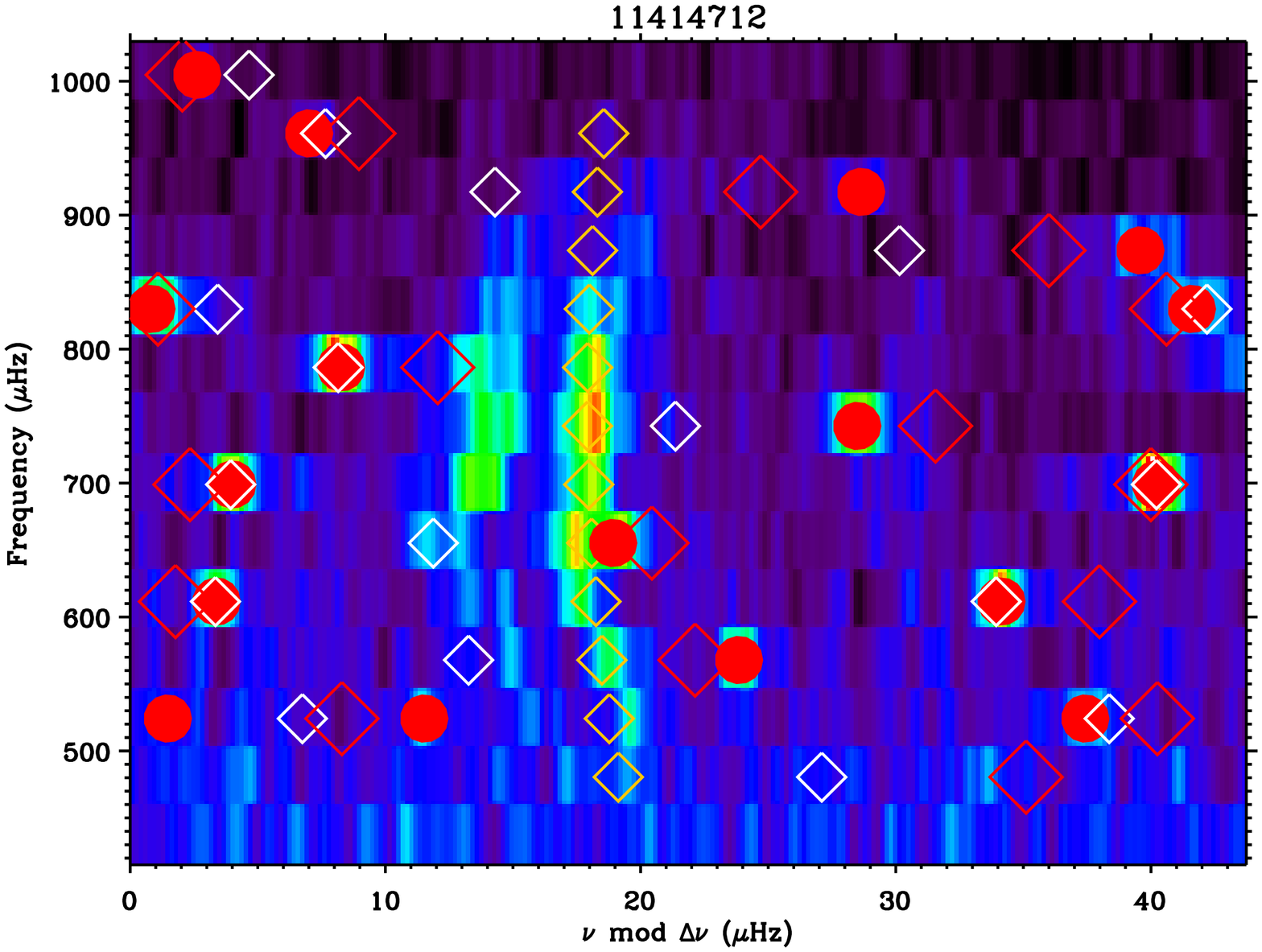}
}
\caption{Echelle diagram for 6 stars.  Frequencies fitted on the power spectra (Red circles), Frequencies from the optimisation: $l=0$ (Orange diamonds), $l=1$ (White diamonds).  Frequencies from fitting the asymptotic model on the fitted frequencies (Red diamonds).}
\label{ech_10972873}
\end{figure*}

\begin{figure*}[!tbp]
\centering
\hbox{

\includegraphics[width=6 cm,angle=90]{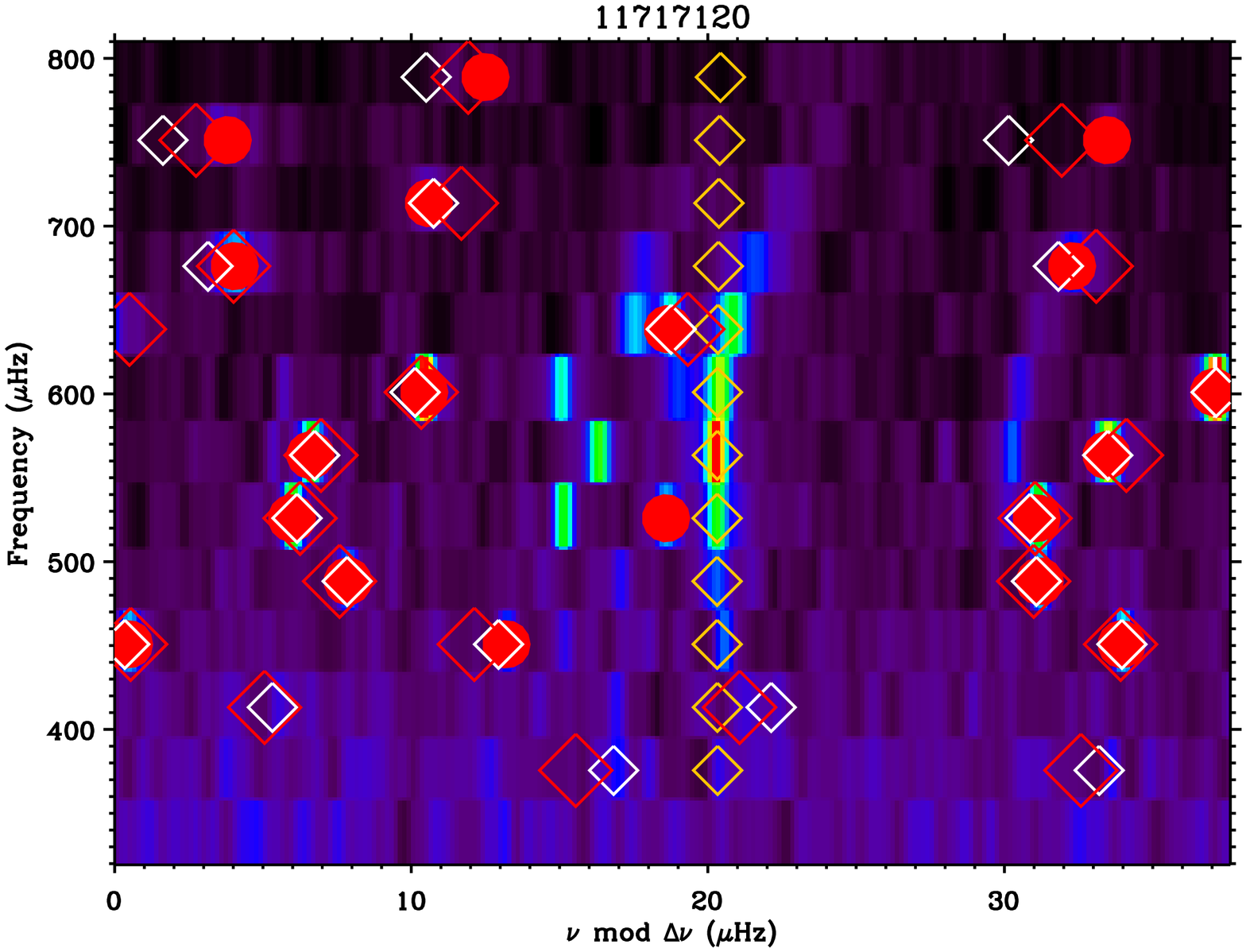}

\includegraphics[width=6 cm,angle=90]{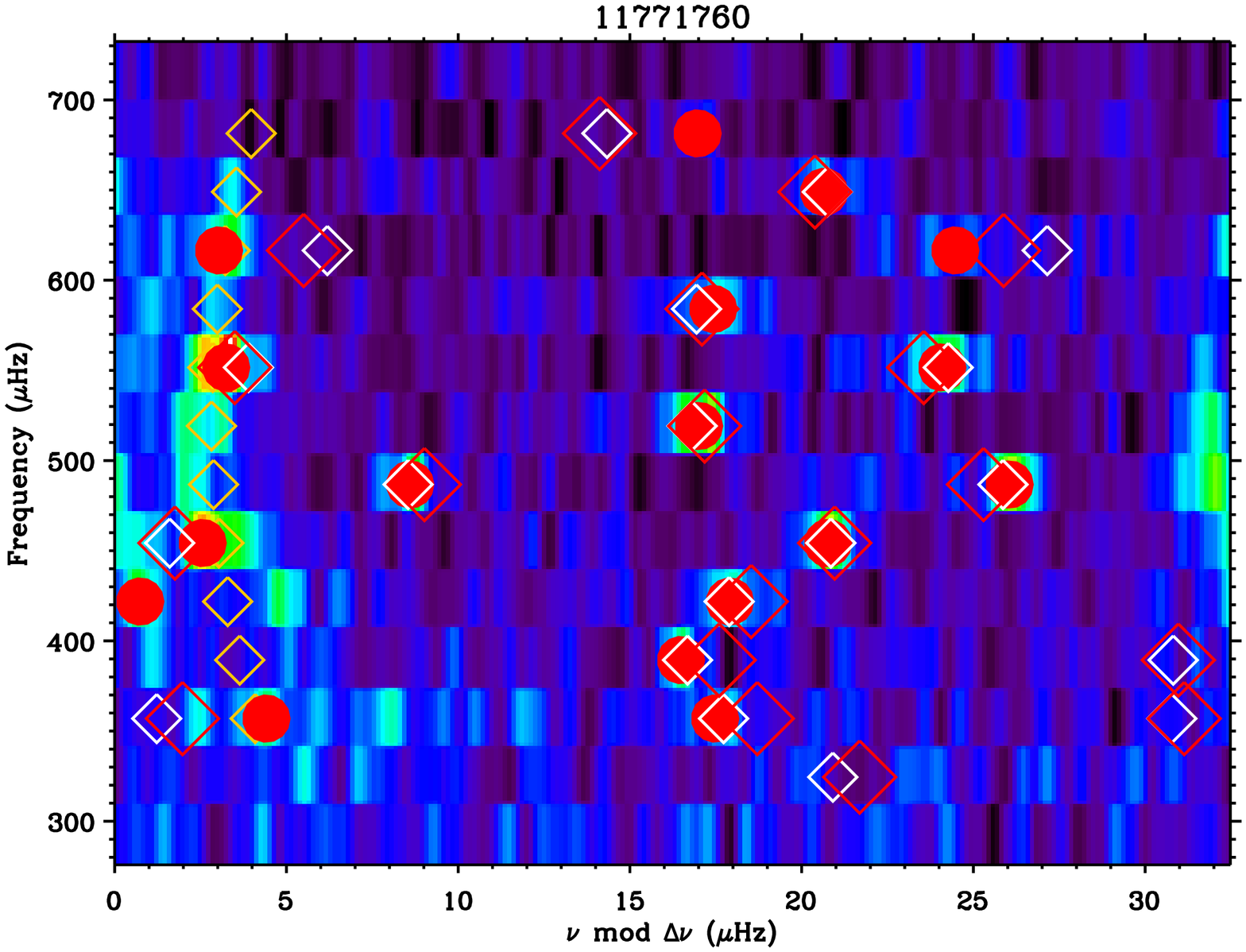}
}

\hbox{
\includegraphics[width=6 cm,angle=90]{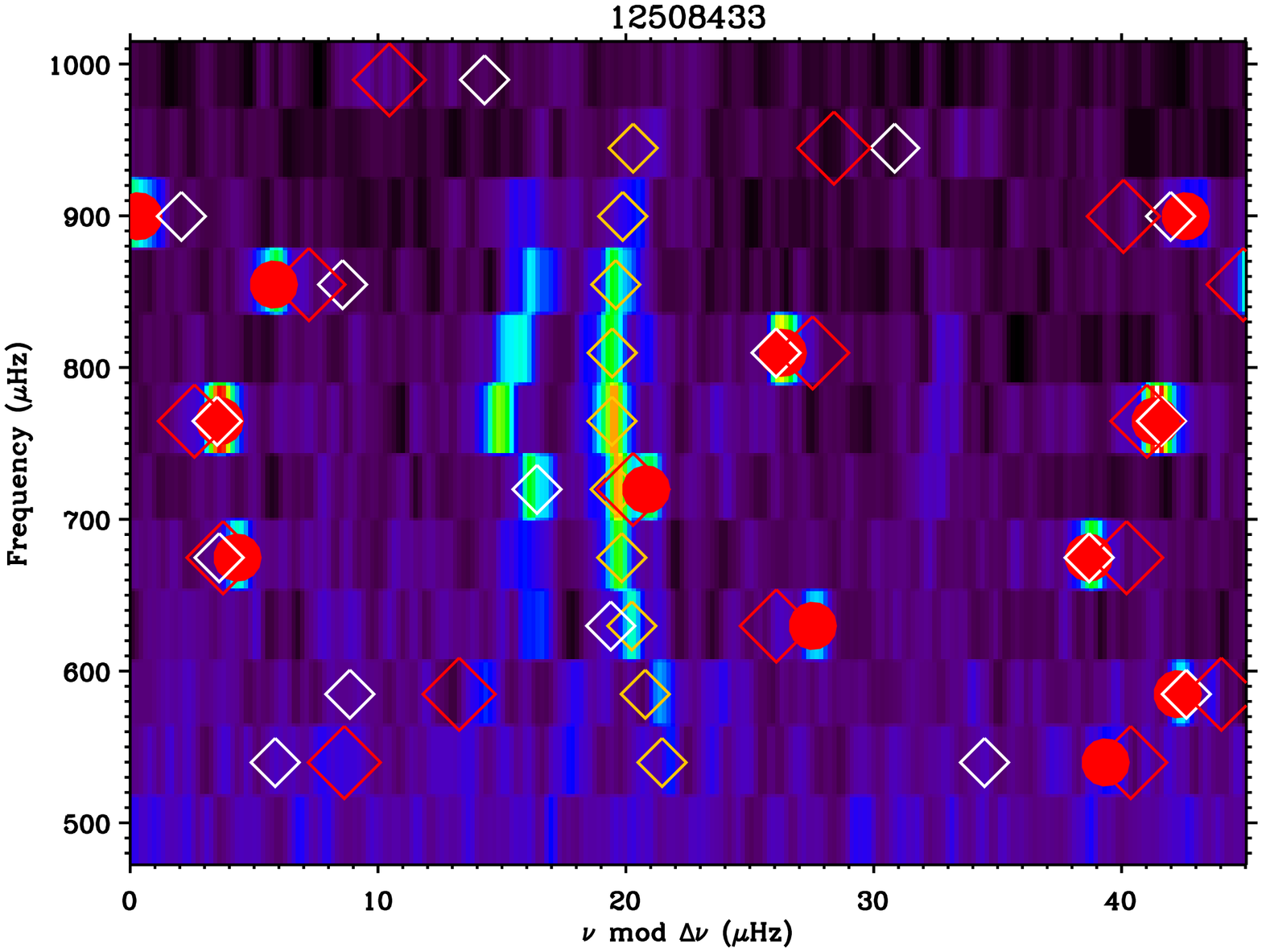}
}
\caption{Echelle diagram for 3 stars.  Frequencies fitted on the power spectra (Red circles), Frequencies from the optimisation: $l=0$ (Orange diamonds), $l=1$ (White diamonds).  Frequencies from fitting the asymptotic model on the fitted frequencies (Red diamonds).}
\label{ech_11717120}
\end{figure*}


\end{document}